\documentclass[article]{aastex631}
\usepackage[utf8]{inputenc}
\usepackage{savesym}
\savesymbol{tablenum}
\usepackage{siunitx}
\restoresymbol{SIX}{tablenum}
\usepackage{amsmath}
\usepackage{natbib}
\usepackage{float}
\usepackage{appendix}
\usepackage{multirow}
\usepackage{booktabs}
\usepackage{hyperref}
\newcommand{\CellWithForceBreak}[2][c]{\begin{tabular}[#1]{@{}c@{}}#2\end{tabular}}

\begin{document}

\title{A High-Resolution Spectroscopic Survey of Directly Imaged Companion Hosts \\
\normalsize {I. Determination of diagnostic stellar abundances for planet formation and composition}}

\author[0000-0003-3708-241X]{Aneesh Baburaj} \affiliation{Department of Astronomy \& Astrophysics, University of California, San Diego, La Jolla, CA 92093, USA} \affiliation{Department of Physics, University of California, San Diego, La Jolla, CA 92093, USA}

\author[0000-0002-9936-6285]{Quinn M. Konopacky} \affiliation{Department of Astronomy \& Astrophysics, University of California, San Diego, La Jolla, CA 92093, USA} 

\author[0000-0002-9807-5435]{Christopher A. Theissen} \affiliation{Department of Astronomy \& Astrophysics, University of California, San Diego, La Jolla, CA 92093, USA} 

\author[0000-0002-1046-025X]{Sarah Peacock} \affiliation{University of Maryland, Baltimore County, MD 21250, USA} \affiliation{NASA Goddard Space Flight Center, Greenbelt, MD 20771, USA}

\author[0009-0009-5477-9375]{Lori Huseby} \affiliation{Lunar and Planetary Laboratory, University of Arizona, Tucson, AZ 85721, USA} 

\author[0000-0003-3504-5316]{Benjamin Fulton} \affiliation{NASA Exoplanet Science Institute/Caltech-IPAC, MC 314-6, 1200 E California Blvd, Pasadena, CA 91125, USA} 

\author[0000-0003-0398-639X]{Roman Gerasimov} \affiliation{Department of Physics and Astronomy, University of Notre Dame, Nieuwland Science Hall, Notre Dame, IN 46556, USA}

\author[0000-0002-7129-3002]{Travis S. Barman} \affiliation{Lunar and Planetary Laboratory, University of Arizona, Tucson, AZ 85721, USA} 

\author[0000-0002-9803-8255]{Kielan K. W. Hoch} \affiliation{Space Telescope Science Institute, 3700 San Martin Drive, Baltimore, MD 21218, USA}


\correspondingauthor{Aneesh Baburaj}
\email{ababuraj@ucsd.edu}


\begin{abstract}
We present the first results of an extensive spectroscopic survey of directly imaged planet host stars. The goal of the survey is the measurement of stellar properties and abundances of 15 elements (including C, O, and S) in these stars.  In this work, we present the analysis procedure and the results for an initial set of five host stars, including some very well-known systems. We obtain C/O ratios using a combination of spectral modeling and equivalent width measurements for all five stars. Our analysis indicates solar C/O ratios for HR 8799 (0.59 $\pm$ 0.11), 51 Eri (0.54 $\pm$ 0.14), HD 984 (0.63 $\pm$ 0.14), and GJ 504 (0.54 $\pm$ 0.14). However, we find a super-solar C/O (0.81 $\pm$ 0.14) for HD 206893 through spectral modeling. The ratios obtained using the equivalent width method agree with those obtained using spectral modeling but have higher uncertainties \textcolor{black}{($\sim$0.3 dex)}.  We also calculate the C/S and O/S ratios, which will help us to better constrain planet formation, especially once planetary sulfur abundances are measured using JWST. Lastly, we find no evidence of highly elevated metallicities or abundances for any of our targets, suggesting that a super metal-rich environment is not a prerequisite for large, widely separated gas planet formation. The measurement of elemental abundances beyond carbon and oxygen also provides access to additional abundance ratios, such as Mg/Si, which could aid in further modeling of their giant companions. 
\end{abstract}

\section{Introduction}
The application of high-contrast imaging techniques to exoplanet research has led to a significant increase in the number of Jovian (Jupiter-like) planets discovered at wide orbits. These planets lie at distances of 9$\--$120 AU from their host stars and have masses ranging from 2$\--$14 $M_\mathrm{Jup}$ (e.g., \citealt{2016PASP..128j2001B}). These planets are fairly young ($\sim$15$\--$200 Myr) compared to planets discovered by other methods as their detectability is enhanced at young ages (e.g., \citealt{2008A&A...482..315B}) when they still retain the heat of their formation. Neither of the two widely accepted models of planet formation $\--$ core (or pebble) accretion or gravitational instability $\--$ can adequately explain the formation of these planets. 
\par
In the core-accretion scenario, heavy elements present in the protoplanetary disk condense into grains, which subsequently form planetesimals. These kilometer-sized planetesimals coalesce to form planetary cores through runaway growth (\citealt{1996Icar..123..180K}; \citealt{2004ApJ...616..567I}). If the core reaches a critical mass ($M_{c,acc}$ $\sim$ several $M_E$) before the depletion of the disk gas, runaway gas accretion occurs, leading to the formation of a gas giant. Theoretical models demonstrate that for gas giants beyond $\sim$ 5 AU, the core accretion scenario is insufficient to explain planet formation due to long core formation timescales (\citealt{2009ApJ...707...79D}). This contradicts observations, where a vast majority of the directly imaged planets discovered have masses above $\sim$ 2$M_J$ and separations $>$5 AU (e.g., \citealt{2016PASP..128j2001B}; \citealt{2019AJ....158...13N}; \citealt{2023AJ....166...85H}).
\par
The gravitational instability model proposes that the initial protoplanetary disk broke up into self-gravitating clumps of gas and dust, which then collapsed further to form gas giants (e.g., \citealt{1997Sci...276.1836B}; \citealt{2004LPI....35.1124B}; \citealt{2013pss3.book....1Y}). This mechanism can explain the formation of Jupiter-sized planets in extremely long orbits ($>$40 AU).  However, it cannot form any planets closer than this distance (\citealt{2009ApJ...707...79D}). These clumps are also more likely to evolve into brown dwarfs or low-mass stellar companions to more massive stars (\citealt{2016ARA&A..54..271K}). While the Backyard Worlds: Planet 9 Citizen Science Project has found some brown dwarf candidates to massive stars (e.g., \citealt{2020ApJ...899..123M}; \citealt{2021ApJ...921..140S}), their numbers are insufficient to entirely validate the gravitational instability model for the formation of wide orbit exoplanets.
\par
\par
Individual elemental abundances are crucial to constrain planet formation. Elemental abundance ratios are particularly informative in this regard. For planets formed by gravitational instability, there is no separation of gas and solids, hence the abundance ratios are expected to match the host star. For a planet formed by core accretion, the abundance ratio for the planet relative to its host star depends on the distribution of the corresponding volatiles through the protoplanetary disk and the acquisition of any solids during the runaway accretion phase. C, N, O, and S abundances and their ratios in exoplanets serve as probes of formation location depending on where the planet formed relative to the snowlines of C/N/O-bearing species and the amount of sulfur refractories accreted (e.g., \citealt{2011ApJ...743L..16O}; \citealt{2023ApJ...956..125O}; \citealt{2023ApJ...952L..18C}). The overall picture is not quite as simple due to post-formation processes such as migration and core erosion (e.g., \citealt{2019ARA&A..57..617M}), but measuring the abundance ratios of a planet can help unlock the mysteries behind its formation. 
\par
Testing predictions regarding the formation mechanism necessitates elemental abundance measurements for the planet as well as its host star. Recent studies have measured carbon and oxygen abundances and constrained the overall metallicity and C/O for directly imaged exoplanets using OSIRIS at Keck (e.g., \citealt{2023AJ....166...85H}), SINFONI at the Very Large Telescope (VLT) (e.g., \citealt{2021Natur.595..370Z}; \citealt{2023A&A...670A..90P}), and JWST (\citealt{2024ApJ...966L..11P}). More recently, transmission spectroscopy of the atmosphere of the hot Jupiter WASP-39b revealed the presence of $\mathrm{SO_2}$, the first discovery of a sulfur-bearing species in an exoplanet atmosphere (\citealt{2023Natur.614..659R}; \citealt{2023Natur.614..664A}; \citealt{2023Natur.617..483T}). Consequently, JWST may soon enable the measurement of sulfur in the atmospheres of directly imaged exoplanets for the first time, facilitating the use of additional formation tracers such as C/S and O/S ratios to better constrain planet formation.
\par
The interpretation of the current and impending measurements of planetary abundances requires their comparison to those of the host star. However, many direct imaging surveys observe young and/or high-mass stars \textcolor{black}{(age $<$ 100 Myr, $M_*$ $>$ 1.2 $M_{\odot}$)} which are often poorly characterized compared to older FGK stars. This is due to their high rotational velocities ($v\sin{i}$ $>$ 30 $\mathrm{kms^{-1}}$), which leads to significant broadening of spectral lines. This leads to the majority of directly imaged exoplanet hosts having \textcolor{black}{either poorly constrained (i.e., lacking uncertainties) or} unknown individual elemental abundances. \textcolor{black}{GJ 504 (\citealt{2017AA...598A..19D}), HR 8799 (\citealt{2020AJ....160..150W}), and more recently HD 984 (\citealt{2024AA...686A.294C}) are the only directly imaged host stars with well-constrained measurements of metallicity and carbon and oxygen abundances}. The increase in the number of planets with measured elemental abundances makes it imperative to have corresponding data on their host stars.
\par
In addition, the composition of the host star encodes the composition of the environment in which the planet forms. Variations in this composition can play a significant role in the occurrence of planets around the host star \textcolor{black}{(e.g., \citealt{2005ApJ...622.1102F}; \citealt{2018A&A...612A..93M}; \citealt{2018AJ....155...68W}; \citealt{2019A&A...624A..94M})}. Recent work has found relationships between planet occurrence and the abundance of individual elements such as Mg, Si, and Ni for the transiting planets discovered by Kepler (\citealt{2022AJ....163..128W}). However, no such study has been conducted among the directly imaged planet population. Measuring the detailed chemical makeup of the host star (in addition to carbon and oxygen) could allow us to investigate some of these population trends among this population of companions. 
\par
In this paper, we present the first results from an optical spectroscopic survey of directly imaged planet host stars, wherein we provide estimates of metallicities, abundances of 15 elements, as well as various elemental abundance ratios for five stars with well-studied companions. In Section \ref{sec:sample}, we briefly describe our overall sample for this survey before providing a more detailed description of the targets analyzed in this work. Section \ref{sec:observation} discusses our observations, including the telescope and instrument used followed by a brief description of the data reduction process. Section \ref{sec:methods} discusses the methods involved in our analysis; the initial part of this section is dedicated to spectral modeling, while the second half talks about equivalent width measurements. In Section \ref{sec:results} we report the results of the analysis of the five stars using the methods described in the previous section. This is followed by the discussion section (Section \ref{sec:discussion}), where we compare our findings with existing literature and also discuss the implications of our results. Lastly, our results and discussion are summarized in the conclusion (Section \ref{sec:conclusion}).

\section{Sample Selection}
\label{sec:sample}
The targets chosen for the overall program are all among the 130 primaries to the directly imaged companions cataloged by the NASA Exoplanet Science Institute. We then restrict our sample to stars brighter than $V\sim12.5$, as they can be observed with SNR $\geq$ \textcolor{black}{150} within a time frame of $\sim$6\,hrs on our choice of telescopes and instruments. This magnitude selection gives us a sample of 65 host stars, distributed over both hemispheres. 

While several directly imaged planet hosts have been observed with optical spectroscopy in other surveys (e.g., \citealt{2011A&A...530A.138C}), we seek to obtain additional data in this program for several reasons. First, most of the datasets from those large surveys are not publicly available, and we intend to make our spectra widely accessible.  The inaccessibility means that we cannot reanalyze them for individual abundance measurements that are not present in the available catalogs. Second, many datasets are from lower-resolution spectrographs ($R<20,000$), \textcolor{black}{we require $R>50,000$ in order to clearly distinguish the telluric lines from the stellar lines. This is especially important for the CI lines at 5052 and 6587 $\mathrm{\AA}$ (e.g., Figure \ref{fig:testco}), and the forbidden [OI] line at 6300 $\mathrm{\AA}$.} Finally, even though several of these sources have estimates of [Fe/H] or [M/H] from large surveys, these do not include individual elemental abundances, and when they do they often do not have uncertainty estimates (e.g., HR 8799, \citealt{2006PASJ...58.1023S}). By using a uniformly analyzed data set, we will obtain unbiased \textcolor{black}{(i.e., unaffected by instrument-to-instrument systematics)} abundance measurements for all the stars in our sample, so that they can be used to constrain formation scenarios. The spectra in the optical regime (380$\--$800 nm) \textcolor{black}{at $R>50,000$ and SNR $\sim150$} can be used to constrain the abundances of 15 elements (C, O, Na, Mg, Si, S, Ca, Sc, Ti, Cr, Mn, Fe, Ni, Zn, Y) to an accuracy of \textcolor{black}{0.1$\--$0.2} dex.  \textcolor{black}{The elements chosen were those with atomic lines in the optical that were reasonably unblended (to measure their equivalent widths more accurately). For instance, Co was not included as most of the lines in the optical had significant blending. Ultimately, we restricted ourselves to species whose abundance ratios might have implications in determining the star-planet connection. An abundance precision of 0.1$\--$0.2 dex gives uncertainties in C/O, C/S, and O/S ratios for these targets up to $<$20\%. This would be a major improvement on the currently unknown abundance ratios of these host stars.} As mentioned previously, the planet C/O for formation by core accretion depends on formation location relative to the various icelines. \textcolor{black}{However, a number of theoretical works suggest that formation via core/pebble accretion would lead to the planet C/O ratios differing from stellar C/O ratios by $>$0.1 dex (\citealt{2011ApJ...743L..16O}; \citealt{2016ApJ...833..203P}). Thus, a precision of $\sim$0.1 dex in C/O ratios could be sufficient to distinguish between formation at least some formation scenarios.}


We choose an initial set of 5 targets to develop and test our analysis and abundance measurement pipeline. The choice of targets is limited to F and G spectral types as those are most conducive to individual elemental abundance measurements. B-type stars have high rotational velocities \textcolor{black}{($v\sin{i}$ $>$ 100 $\mathrm{kms^{-1}}$,} hence extremely broad spectral lines) and a relative lack of strong spectral features for elements other than hydrogen and helium. \textcolor{black}{K and M-type targets are included in this work due to being fainter in the optical, leading to spectra with lower SNR. While data for stars with these spectral types has already been obtained, we reserve their analysis for future work.} Additionally, among the F and G spectral types, the chosen targets have companions that are of great interest among the exoplanet community: \textit{James Webb Space Telescope} programs have been approved for further atmospheric characterization of the companions around 4 of our 5 targets. These include all the companions of the HR 8799 system (GTO 1188; PI Hodapp), GJ 504 b (GTO 2778; PI Perrin), 51 Eridani b (GO 3522; PI Ruffio), and HD 206893 B (GO 5485; PI Baburaj). The companion to the last target (HD 984) has previous measurements of both dynamical (\citealt{2022AJ....163...50F}) and evolutionary masses (\citealt{2015MNRAS.453.2378M}; \citealt{2017AJ....153..190J}).

\subsection{Science Targets for This Work}
\label{sub:targets}
\subsubsection{51 Eri}
51 Eri is a young F0 star ($\sim$ 23 Myr), belonging to the $\beta$ Pictoris moving group (\citealt{2014MNRAS.445.2169M}). A companion to the star, 51 Eri b, was discovered in 2014 as part of the GPIES campaign at GPI South (\citealt{doi:10.1126/science.aac5891}). The companion has a dynamical mass measurement of $\leq$ 11 $M_\mathrm{Jup}$ with a semi-major axis of 11.1$^{+4.2}_{-1.3}$ AU, obtained using $Gaia$ astrometry (\citealt{2020AJ....159....1D}, \citealt{2022MNRAS.509.4411D}). Evolutionary mass estimates range from 2$\--$9 $M_\mathrm{Jup}$ (\citealt{doi:10.1126/science.aac5891}; \citealt{doi:10.1051/0004-6361/201629767}; \citealt{2019AJ....158...13N}; \citealt{doi:10.1051/0004-6361/202244826}; \citealt{2023MNRAS.525.1375W}; \citealt{2024arXiv240101468E}). Recently, \cite{doi:10.1051/0004-6361/202244826} obtained a $\mathrm{C/O}$ = 0.38 $\pm$ 0.09 for 51 Eri b using VLT/Sphere data and atmospheric retrievals.

\subsubsection{HR 8799}
HR 8799 is a young ($\sim$ 30 Myr) $\lambda$ Bootis star (\citealt{1999AJ....118.2993G}) belonging to the Columba association (\citealt{2011ApJ...732...61Z}, \citealt{2013ApJ...762...88M}). $\lambda$ Bootis stars have solar abundances of C, N, O, and S in their atmospheres, but are highly sub-solar in the Fe-peak elements. HR 8799 has previous abundance estimates in the literature by \cite{2006PASJ...58.1023S} which obtained the atmospheric parameters and elemental abundances using high-resolution spectra from the Elodie spectrograph ($R$ = 42,000). More recently, \cite{2020AJ....160..150W} calculated the [Fe/H], carbon and oxygen abundances, and hence the C/O ratio using high-resolution ($R$ $\sim$ 115,000) HARPS data. HR 8799 is one of the best-known multi-planetary systems, with four directly imaged companions discovered to date (\citealt{2008Sci...322.1348M}; \citealt{2010Natur.468.1080M}). All four companions have mass estimates of between 5.8 and 9.8 Jupiter masses with separations ranging from around 17 AU (HR 8799 e) to around 68 AU (HR 8799 b), obtained using relative astrometry and stability arguments (\citealt{2022AJ....163...52S}; \citealt{doi:10.1051/0004-6361/202243862}). \cite{2022A&A...657A...7K} even obtained a dynamical mass as high as 12 $\pm$ 3.5 $M_\mathrm{Jup}$ for HR 8799 e using Gaia EDR3 astrometry. In addition, all the companions also have C/O ratio estimates in the literature. \cite{2021AJ....162..290R} obtained $\mathrm{[C/O]_b}$ = $0.578^{+0.004}_{-0.005}$, $\mathrm{[C/O]_c}$ = 0.562 $\pm$ 0.004, $\mathrm{[C/O]_d}$ = $0.551^{+0.005}_{-0.004}$ for HR 8799 b, c and d respectively. Meanwhile, \cite{doi:10.1051/0004-6361/202038325} obtained $\mathrm{C/O}$ = $0.60^{+0.07}_{-0.08}$ for HR 8799 e. 

\subsubsection{HD 984}
HD 984 is a young (30$\--$200 Myr) F7 star (e.g., \citealt{2015MNRAS.453.2378M}; \citealt{2024AA...686A.294C}). A substellar companion to the star, HD 984 B, was discovered in 2015 (\citealt{2015MNRAS.453.2378M}) using data obtained at the Very Large Telescope (VLT). HD 984 B has a proposed dynamical mass of 61 $\pm$ 4 $M_\mathrm{Jup}$ with a semi-major axis of 28$^{+7}_{-4}$ AU, clearly placing it in the brown dwarf mass regime (\citealt{2022AJ....163...50F}). \cite{2024AA...686A.294C} obtained a C/O ratio of 0.50 $\pm$ 0.01 for the companion.

\subsubsection{GJ 504}
GJ 504 is a G0 star with a super solar metallicity (e.g., \citealt{2005ApJS..159..141V}; \citealt{2021AJ....161..134H}). GJ 504 has previous abundance estimates in the literature; \cite{2017AA...598A..19D} measured the abundances of 14 elements (including carbon and oxygen) as part of their spectroscopic study. Their abundances give a C/O of 0.56$^{+0.26}_{-0.18}$ for the star GJ 504A. A companion to the star, GJ 504 b, was discovered using the High Contrast Instrument for the Subaru Next Generation Adaptive Optics (HiCIAO) on the Subaru telescope by \cite{2013ApJ...774...11K}. The GJ 504 system has two different age estimates of 21 $\pm$ 2 Myr and 4.0 $\pm$ 1.8 Gyr. These give two widely different mass estimates of 1.3$^{+0.6}_{-0.3}$ $M_\mathrm{Jup}$ and 23$^{+10}_{-9}$ $M_\mathrm{Jup}$ corresponding to the young and old ages of the companion (\citealt{doi:10.1051/0004-6361/201832942}). Even with the younger age range, it is one of the coldest directly imaged exoplanets ever discovered. In addition, \cite{doi:10.1051/0004-6361/201832942} also estimate a model-dependent C/O ratio of 0.20$^{+0.09}_{-0.06}$ for the companion.

\subsubsection{HD 206893}
HD 206893 is an F5 star with nearly solar metallicity (\citealt{2016ApJ...826..171G}; \citealt{2017MNRAS.469.3042N}) and highly uncertain age estimates, ranging from $\sim$3$\--$300 Myr (\citealt{doi:10.1051/0004-6361/202140749}). The star currently has two known substellar companions; the first one (HD 206893 B) was discovered at a projected orbital separation of $\sim$10 au by \cite{2017A&A...597L...2M} using VLT/SPHERE data. Due to the highly uncertain age of the system, mass estimates for this companion range from $\sim$5 $M_\mathrm{Jup}$ (\citealt{doi:10.1051/0004-6361/202140749}) to 30$\--$40 $M_\mathrm{Jup}$ (\citealt{2017AA...608A..79D}; \citealt{doi:10.1051/0004-6361/202140749}; \citealt{2021AJ....161....5W}). Most recently, \cite{doi:10.1051/0004-6361/202244727} obtained a dynamical mass of 28.0$^{+2.2}_{-2.1}$ $M_\mathrm{Jup}$ for HD 206893 B. The presence of an 8$\--$15 $M_\mathrm{Jup}$ inner companion was hypothesized by unexplained RV drift for the system (\citealt{2019A&A...627L...9G}) as well as an anomaly in the $Gaia$ EDR3 proper motion (\citealt{doi:10.1051/0004-6361/202140749}). \cite{doi:10.1051/0004-6361/202244727} used this data to directly observe the inner companion (HD 206893 c) for the first time using VLTI/GRAVITY. They also derive a mass of 12.7$^{+1.2}_{-1.0}$ $M_\mathrm{Jup}$ and an orbital separation of 3.53$^{+0.08}_{-0.06}$ au for the companion. \cite{doi:10.1051/0004-6361/202140749} obtained a C/O ratio varying from 0.65$\--$0.90 for HD 206893 B depending on the model grid and the retrieval method used.

\section{Observations and Data Reduction}
\label{sec:observation}
The northern hemisphere targets are observed using the Automated Planet Finder (APF) telescope at the Lick Observatory in San Jose, California, while the SMARTS 1.5m telescope at the Cerro Tololo Inter-American Observatory in Chile is used for the southern hemisphere targets. Since all the targets discussed in this paper are from our northern hemisphere sample, we will only elaborate on the observations conducted using the APF telescope. \par
The Automated Planet Finder (APF) is a 2.4m telescope equipped with the Levy optical echelle spectrometer. The spectrometer has high throughput allowing us to easily obtain spectral resolutions above 100,000 and even as high as 150,000. This makes it among the highest-resolution spectrographs in the Northern Hemisphere. The Levy’s fixed spectral coverage between 374$\--$900nm (\citealt{2014PASP..126..359V}) allows us to investigate lines over the entire optical region and even lines in the near-infrared. As we require the highest possible resolution for our targets, we use the $0.5^{\prime{\prime}}\times8.0^{\prime{\prime}}$ slit (decker N) \textcolor{black}{which allows us to obtain spectral resolutions of R$\sim$130,000. In addition, we assume} a seeing of 1.2$^{\prime{\prime}}$ to compute the integration time for all our targets. For targets that required a total integration time greater than 900s to obtain an SNR $\geq$ 150, we requested multiple observation frames with equal integration time per frame. For targets that required a total integration time greater than 900s to obtain an SNR $\geq$ 150, we requested multiple observation frames with equal integration time per frame. However, for some targets that required very high integration times ($>$3600s), the observations were split over multiple epochs due to observing constraints.
\begin{deluxetable}{ccccc}[H]
\tablecaption{Observing schedule for targets in this work\label{tab:obs}}
\tablewidth{0pt}
\tablehead{
\colhead{Target} & \CellWithForceBreak{Number of \\ frames} & \CellWithForceBreak{Integration Time \\ per frame (s)} & \CellWithForceBreak{Observation date \\ (UT)} & \CellWithForceBreak{Total Int. \\ Time (s)} \\
}
\startdata
{HIP 25278 \tablenotemark{a}} & {1} & {148} & {December 27, 2016} & {148} \\ \hline
{51 Eri}  & {1} & {54}  & {August 26, 2015} & {54} \\ \hline
{HR 8799} & {1} & {300} & {July 29, 2015}   & {300} \\ \hline 
{HD 984}  & {1} & {997} & {November 6, 2015} & {997} \\ \hline
{GJ 504}  & {4} & {25}  & {February 27, 2021} & {100} \\ \hline
\multirow{ 4}{*}{HD 206893}   & {1} & {900} & {August 7, 2023} & \multirow{ 4}{*}{3600}\\
                             & {1} & {900} & {August 9, 2023} &  \\
                             & {1} & {900} & {August 25, 2023} & \\
                             & {1} & {900} & {August 26, 2023} & \\
\enddata
\tablenotetext{a}{Test target used to validate analysis procedure. Refer to Appendix \ref{sec:test} for additional details.}
\end{deluxetable}

Data reduction of the CCD images is necessary in order to obtain the one-dimensional spectra needed for further analysis. The APF has an automated data reduction pipeline that handles most of the reduction process, except for the blaze correction.  The reduction pipeline has been described in detail in \cite{2015ApJ...805..175F}.
\par
For the CCD images obtained using APF, the wavelength along a spectral order is aligned in the horizontal “rows”, whereas the vertical “column” within a spectral order is the spatial dimension. These images are first bias subtracted followed by the determination of locations of the spectral order by means of an algorithm that detects the 3-pixel wide ridge present along the middle of each order. One-dimensional spectra are extracted from the 2D images by summing the counts along the spatial direction for each wavelength; the counts for 8-12 pixels (symmetric about the center of the order) are summed, which accounts for about 99.5\% of the seeing profile. The spectra are then flat-fielded using a normalized flat-field (a flat field that retains the differences in signal gain of the pixels but has an average value of unity), followed by the subtraction of cosmic ray events from the images by identifying individual pixels that are 5$\sigma$ outliers from their neighbors. The one-dimensional spectra are then blaze corrected order-by-order using normalized quartz lamp spectra. The resultant reduced spectra \textcolor{black}{retain only the slope due to the stellar blackbody continuum}.

\subsection{Stellar Lines} \label{sub:stellarlines}
 We identified several stellar absorption lines that could be used for the estimation of carbon and oxygen abundances, and subsequently the C/O ratio. For carbon, we use the C \textsc{i} lines at 4772, 4932, 5052, 5380, and 6587 $\mathrm{\AA}$. Obtaining the oxygen abundance is more difficult due to the lack of easily detectable atomic oxygen features in the optical region of the stellar spectrum as well as other complications (e.g., blends, three-dimensional, non-LTE effects) (\citealt{2005ARA&A..43..481A}). In the optical region, the only lines present are the forbidden [O \textsc{i}] line at 6300 $\mathrm{\AA}$ \hspace{1pt} (which is low-excitation and very sensitive to minor changes in the stellar \textcolor{black}{atmospheric temperature}; \citealt{2005ARA&A..43..481A}) and the O \textsc{i} triplet lines at 6155--58 and 7771--75 $\mathrm{\AA}$. These lines are usually present for stars of spectral types BAFGK, which applies to the stars in our dataset (\citealt{2013PASJ...65...65T}). For the remaining 13 elements (Na, Mg, Si, S, Ca, Sc, Ti, Cr, Mn, Fe, Ni, Zn, Y), the list of stellar lines is given in Table \ref{tab:lineabundance}. \textcolor{black}{We restrict ourselves to the optical region of the spectrum (as opposed to moving to the infrared) for abundance measurements as most previous stellar abundance work has been done in the optical, making literature comparisons easier. In addition, we also lack accessibility to infrared spectrographs of similarly high resolution (R$>$50,000)}. Historically, the measurement of stellar abundances using spectral template fitting versus equivalent widths has been widely debated. A combination of both methods has been used to measure elemental abundances for several transiting planet host stars (e.g., \citealt{2011ApJ...732...55S}; \citealt{2013ApJ...778..132T}; \citealt{2014ApJ...788...39T}) for best results. Hence, we too analyze the spectra using both methods.  First, we perform spectral template fitting in which individual abundances are varied (\citealt{2018ApJ...855L...5D}).  Second, we conduct a more traditional equivalent width analysis of identified lines (\citealt{2014ApJ...788...39T}) to estimate the carbon and oxygen abundances. For other elements, the spectral abundance will be determined only via the equivalent width analysis method because of the significant computation time required for spectral template modeling.

 \section{Methods}
 \textcolor{black}{In this section, we introduce the methods for analysis of our science sample of five targets. We also validate these methods using the star HIP 25278 with previously measured carbon, oxygen, and sulfur abundances from \cite{2014AJ....148...54H} in Appendix \ref{sec:test}.}
 \label{sec:methods}
 \subsection{Spectral Modeling}
 \subsubsection{PHOENIX models} \label{subsub:phoenix}
The first step in the spectral modeling process involves determining the basic stellar atmospheric parameters (described in Section \ref{subsub:stellarparam}): temperature ($T_\mathrm{eff}$), surface gravity ($\log{g}$), and metallicity ($\mathrm{[M/H]}$).  To accomplish this, we utilize the grid of stellar atmospheres (\citealt{10.1051/0004-6361/201219058}) computed using the \textit{PHOENIX} atmosphere code (e.g., \citealt{1997ApJ...483..390H}). The range and spacing of this grid are given in Table \ref{tab:grids}. We interpolate this PHOENIX grid onto a model grid with a resolution of 0.02 $\mathrm{\AA}$ and spanning the wavelength range 380--900nm.
\par
For the determination of individual carbon and oxygen abundances (Section \ref{subsub:codetermination}), the generic \textit{PHOENIX} models are insufficient. This is because we require sampling of multiple carbon and oxygen abundances even at the same metallicity for more accurate abundance measurements. Hence, we computed custom models in a similar fashion as the \citet{10.1051/0004-6361/201219058} grid using the \textit{PHOENIX} model framework. The initial set of these custom models had solar metallicity, with carbon and oxygen abundances varying from $-$0.2 to 0.2 dex in steps of 0.1 dex compared to the solar values (Table \ref{tab:grids}). These models are subsequently used to generate grids of synthetic spectra with a wavelength sampling of 0.02\,$\mathrm{\AA}$ exclusively for the spectral orders containing the carbon and oxygen lines described in Section \ref{sub:stellarlines}. Henceforth, these custom grids will be known as \textit{PHOENIX}$\--$\textit{C/O}.

\begin{deluxetable}{ccccccccccc}
\tablecaption{Parameter range and grid spacing for model grids used in this work \label{tab:grids}}
\tablewidth{0pt}
\tablehead{\colhead{Model grid}  & \colhead{$T_\mathrm{eff}$ (K)}  & \colhead{$\Delta T_\mathrm{eff}$ (K)}  & \colhead{$\log{g}$\tablenotemark{a}}  & \colhead{$\Delta\log{g}$\tablenotemark{a}} & \colhead{$\mathrm{[M/H]}$} & \colhead{$\Delta\mathrm{[M/H]}$} & \colhead{$\mathrm{[C/H]}$} & \colhead{$\Delta\mathrm{[C/H]}$} & \colhead{$\mathrm{[O/H]}$} & \colhead{$\Delta\mathrm{[O/H]}$}  } 

\startdata
\textit{PHOENIX}\tablenotemark{b} & (2300, 12000) & 500 & (0.0, 6.0) & 0.5 & (-4.0, 1.0) & 0.5 & - & - & - & - \\ \hline
\textit{PHOENIX}$\--$\textit{C/O} & (5000, 8000) & 500 & (3.5, 5.0) & 0.5 & 0.00\tablenotemark{c} & - & (-0.2, 0.2) & 0.1 & (-0.2, 0.2) & 0.1 \\ \hline
\textit{PHOENIX}$\--$\textit{C/O} & \multirow{2}{*}{(7300, 7400)} & \multirow{2}{*}{100} & \multirow{2}{*}{(4.0, 4.5)} & \multirow{2}{*}{0.5} & \multirow{2}{*}{-0.50\tablenotemark{c}} & \multirow{2}{*}{-} & \multirow{2}{*}{(-0.2, 0.3)} & \multirow{2}{*}{0.1} & \multirow{2}{*}{(-0.2, 0.3)} & \multirow{2}{*}{0.1} \\
(HR 8799) & \\ \hline
\textit{PHOENIX}$\--$\textit{C/O} & \multirow{2}{*}{(5900, 6000)} & \multirow{2}{*}{100} & \multirow{2}{*}{(4.0, 5.0)} & \multirow{2}{*}{0.5} & \multirow{2}{*}{+0.12\tablenotemark{c}} & \multirow{2}{*}{-} & \multirow{2}{*}{(0.0, 0.6)} & \multirow{2}{*}{0.1$\--$0.2} & \multirow{2}{*}{(0.1, 0.7)} & \multirow{2}{*}{0.1$\--$0.2} \\
(GJ 504)\tablenotemark{d} & \\
\enddata
\tablenotetext{a}{Surface gravity ($g$) in cgs units}
\tablenotetext{b}{Standard grid created using stellar atmospheric code from \citet{10.1051/0004-6361/201219058}}
\tablenotetext{c}{Grid metallicity fixed at stated value}
\tablenotetext{d}{Refer to Section \ref{sub:resultsgj504} for additional details on the grid spacing}
\end{deluxetable}

\subsubsection{Telluric Model}
In order to properly model the full APF spectra, we require synthetic spectra that account for the likely telluric features in them. The telluric spectra are generated using the NASA Planetary Spectrum Generator \footnote{\href{https://psg.gsfc.nasa.gov/}{https://psg.gsfc.nasa.gov/}} (PSG; \citealt{2018JQSRT.217...86V}). The PSG tool provides access to a database of pre-computed telluric transmittances at 5 different altitudes (0m, 2600m, 4200m, 14000m, and 35000m) and 4 different atmospheric water vapor levels expressed in units of precipitable millimeters (hence referred to as precipitable water vapor or $pwv$). Thus, a total of 20 pre-computed telluric transmittance tables are available. 

While forward modeling the telluric parameters, the altitude is kept fixed for data from a specific observatory, and the telluric alpha ($\alpha$) and $pwv$ are solved for. The telluric alpha is an exponential parameter that is used to strengthen/weaken the telluric features (lines). If $p(\lambda)$ is the pixel-to-wavelength mapping, the telluric model is of the form:
\begin{equation}
M_{telluric}[p] = T\Big[p(\lambda), altitude, pwv\Big]^{\alpha} \end{equation}

\subsubsection{Forward Modeling}
A forward modeling approach along the lines of \cite{2020AJ....160..207W}, \cite{2021ApJS..257...45H}, and \cite{2022ApJ...926..141T} is used to find the best-fit models for the stellar spectra. This approach uses the Markov Chain Monte Carlo (MCMC) method built into the emcee package (\citealt{2010CAMCS...5...65G}; \citealt{2013PASP..125..306F}) to traverse the parameter space of stellar atmospheric and telluric/instrumental parameters. This analysis is done in two parts. Initially, the MCMC method is used to determine the basic stellar parameters such as effective temperature ($T_\mathrm{eff}$), surface gravity ($\log{g}$), metallicity ($\mathrm{[M/H]}$), rotational velocity ($v\sin{i}$), radial velocity ($RV$), and telluric/instrumental parameters telluric alpha ($\alpha$), precipitable water vapor ($pwv$) and the line-spread function (LSF). \textcolor{black}{We do not flux calibrate or continuum normalize the spectrum as either step would simply involve multiplying our PHOENIX templates by a scaling factor to match the flux values, which is not necessary for our analysis}.

Each parameter is forward modeled using a uniform prior range. Initially, a very wide prior range of $T_\mathrm{eff}\in$ [4500K, 9000K], $\log{g}\in$ [2.9, 5.1], and $[M/H]\in$ [$-$3, 1] is used for each of our targets. We use the results of MCMC runs using these priors to condense our prior range depending on the spectral type of the star. Ultimately, we expect the solved parameters to have a normal distribution, hence the log-likelihood function used is of the form:
\begin{equation}
\ln{\mathrm{(likelihood)}} = -0.5 \times \left[\sum{\chi}^2 + \sum\left(\ln(2\pi{N^{*}}^2)\right)\right]
\end{equation}
\noindent
Here, $\chi = \frac{data[p]-D[p]}{N^{*}[p]}$ and $N^{*}$ is the overall noise. $N^{*}$ is defined as $N^{*2} = N^2 + \sigma^2$, where $N$ is a noise factor added in quadrature to the spectral noise ($\sigma$) to account for unknown uncertainties or systematics in our model. The SNR for our data is roughly 100, implying a noise of about 1\%.  We increase this by a factor of 2 to account for any additional systematics in the data or models that aren't accounted for, hence the noise is fixed at 2\% of the median flux value for a specific order. The uncertainty is the difference between the 84th and the 50th percentile as the upper bound and the difference between the 50th and the 16th percentile as the lower bound. The equation used to forward-model the data is: 
\begin{equation}
D[p] = C \times \Bigg[\bigg(M\Big[p(\lambda[1+\frac{RV}{c}]), T_\mathrm{eff}, \log{g}, [M/H]\Big] \otimes \kappa_R(v\sin{i})\bigg) \times T\Big[p(\lambda), altitude, pwv\Big]^{\alpha} \Bigg] \otimes \kappa_R(\Delta\nu_{inst}) + C_{flux} 
\end{equation}

\noindent
Here, $p(\lambda)$ is the pixel-to-wavelength mapping, $M\big[p(\lambda)\big]$ is the \textit{PHOENIX} stellar atmosphere model parameterized by the effective temperature ($T_\mathrm{eff}$), surface gravity ($\log{g}$), and metallicity ($\mathrm{[M/H]}$), $C$ is a constant multiplicative flux parameter fit in each MCMC step, $\kappa_R$ is the line broadening due to stellar rotation ($\kappa_R(v\sin{i})$) and instrumental LSF ($\kappa_R(\Delta\nu_{inst})$), $C_{flux}$ is the flux offset and the last term is the telluric contribution as detailed in the previous section. RV is the radial velocity without the barycentric correction, which was corrected later. We also include a wavelength offset to account for linear offsets in our initial wavelength calibration. \textcolor{black}{In most cases, the flux offset is 2$\--$3 orders of magnitude lower than the flux values. The scaling factor that we multiply the PHOENIX templates with is enough to match the data. However, the flux offset was just included to be conservative in case the scaling was not sufficient, but it was sufficient in all cases.}

\subsubsection{Determining Stellar Parameters: Temperature, Gravity, and Metallicity} \label{subsub:stellarparam}
The high-resolution APF data for each star has 50 orders which can be individually forward modeled. However, attempting to forward model all orders simultaneously is computationally intensive. Out of these 50 orders, 20 orders are particularly noisy at the edges and are consequently discarded from further analysis. For the remaining 30 orders, we conduct forward modeling of single spectral orders for each star. The forward modeling process involves two MCMC runs, each performed with 100 walkers and 500 steps and a burn-in of 400 steps. The priors used for each target are given in Table \ref{tab:priors}. In the first run, we obtain an approximate best-fit model to the stellar spectrum. The residuals of this best-fit model and the data are then used to generate a mask that eliminates outliers at a 3$\sigma$ level. Subsequently, a second MCMC run is conducted after outlier rejection to obtain the final best-fit values for the parameters and their uncertainties for that specific order. Based on how well the parameters converge for each order, we identified a subset of orders with sufficient tellurics (needed for telluric parameters and anchoring the rest wavelength frame) and/or strong metal lines (for determining $\mathrm{[M/H]}$); ten such orders have been identified as ``ideal", and used for further analysis.
\par

\begin{deluxetable}{cccccccccc}
\tablecaption{Forward modeling prior ranges for stellar parameter determination \label{tab:priors}}
\tablewidth{0pt}
\tablehead{\colhead{Target}    & \colhead{$T_\mathrm{eff}$ (K)}    & \colhead{$\log{g}$ (cgs)}      & \colhead{$\mathrm{[M/H]}$}     & \colhead{$v\sin{i}$\tablenotemark{*} (km\,s$^{-1}$)}    & \colhead{$RV$\tablenotemark{*} (km\,s$^{-1}$)} & \colhead{$\alpha$\tablenotemark{*}} & \colhead{$pwv$\tablenotemark{*}} & \colhead{LSF\tablenotemark{*}} & \colhead{Noise factor\tablenotemark{*} (N)} }
\startdata
51 Eri & (6500, 8000) & (3.9, 5.0) & (-1.0, 0.5) & \multirow{5}{*}{(0,100)} & \multirow{5}{*}{(-100,100)} & \multirow{5}{*}{(0.0, 2.0)} & \multirow{5}{*}{(0.5, 5.0)} & \multirow{5}{*}{(0.1, 3.0)} & \multirow{5}{*}{(0, 7.5e6)} \\
HR 8799 & (6500, 8500) & (3.9, 4.9) & (-1.5, 0.5) & & & & & &  \\
HD 984 & (5500, 7000) & (3.9, 4.9) & (-1.0, 1.0) & & & & & &  \\
GJ 504 & (5000, 7000) & (3.9, 4.9) & (-0.6, 0.5) & & & & & &  \\
HD 206893 & (6000, 7500) & (3.8, 5.0) & (-0.6, 0.5) & & & & & &  \\
\enddata
\tablenotetext{*}{All five targets use the same priors for these parameters}
\end{deluxetable}

\begin{figure}
\centering
\includegraphics[width=0.52\linewidth]{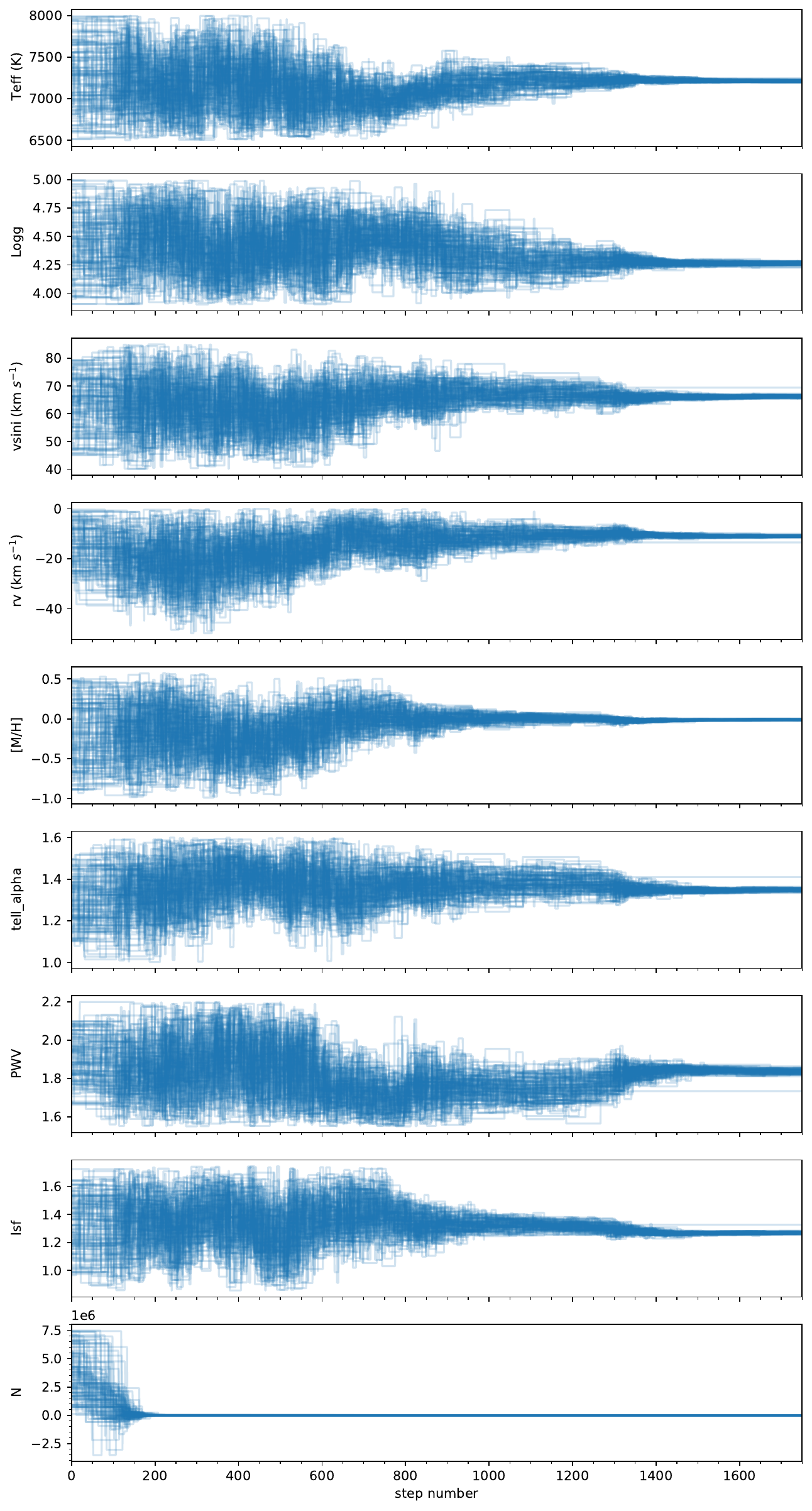}
\caption{Walker plot from an MCMC run for the determination of stellar atmospheric parameters for the host star 51 Eri. The MCMC was run with 100 walkers for 1750 steps, with the first 1250 discarded as burn-in. Convergence was achieved in this run, as seen by the behavior of the walkers, which begin to converge to a definite value around step $\sim$ 1200 for all parameters. The corner plot corresponding to this run is shown in Figure \ref{fig:51ericorner}.}
\label{fig:walkers}
\end{figure}

In order to determine the final best-fit stellar parameters, we fit all of the ideal ten orders at once. During this multiple-order fitting process, all parameters other than $T_\mathrm{eff}$, $\log{g}$, and $\mathrm{[M/H]}$ are constrained to a narrow range determined from the single-order fitting results. Priors for $T_\mathrm{eff}$, $\log{g}$, and $\mathrm{[M/H]}$ are kept the same as defined in Table \ref{tab:priors}. The forward modeling procedure is akin to the single-order fits, with two MCMC runs of 1500--2500 steps and 100 walkers each, with the first 1000--2000 steps discarded as burn-in. For each target, a few initial runs were used to determine how many steps were needed for the walkers to converge. Figure \ref{fig:walkers} shows an example of a walker plot for an MCMC run in which we obtain convergence. Model fitting for the spectra of different host stars required different numbers of steps to converge. Hence, keeping the same number of runs for each target would have led to a waste of our finite computation resources if convergence was achieved in fewer steps. After outlier rejection, the second MCMC run yields the best-fit values for the parameters and their uncertainties for the multi-order run. 

Unlike single-order fits, multi-order MCMC runs often converge at different values of $T_\mathrm{eff}$, $\log{g}$, and $\mathrm{[M/H]}$ for different runs. Due to the large amounts of data (10 high-resolution spectral orders) we are fitting for in a multi-order run, the walkers of the MCMC run \textcolor{black}{often converge in a region of high log-probability (i.e., a local minima of the probability space). Once the walkers are in that region, they often continue to sample within that region. Occasionally the walkers do find a new region of high log-probability, in which case they try and move towards that region of probability space. On such occasions, the walkers do get out but are often not able to converge to a new set of parameters before the MCMC run reaches the step limit.} The walkers could also get stuck in a local minimum where the model fit is especially poor for some of the spectral orders, or the value of $T_\mathrm{eff}$, $\log{g}$, or $\mathrm{[M/H]}$ approaches the edge of the prior range, or one of the three parameters converges at a significantly different value (at a $>$3$\sigma$ level) compared to other multi-order runs. We do not consider such runs while calculating the mean and uncertainties of our stellar parameters. Hence, we need to do multiple runs to effectively sample multiple local minima states and get a more accurate estimate of the mean and uncertainties of the stellar parameters. We continued to do these multi-order fitting runs until we had a minimum of 10 runs overall and at least seven runs that met our criteria. In case fewer than seven runs met our criteria from the initial 10, we continued to do more multi-order fits until we reached our goal of seven eligible ones. The median and standard deviation of the $T_\mathrm{eff}$, $\log{g}$, $\mathrm{[M/H]}$ values from these eligible runs were adopted as the best-fit and the uncertainty respectively.

\subsubsection{Determining Carbon and Oxygen Abundances}\label{subsub:codetermination}
\begin{figure}
\centering
\plottwo{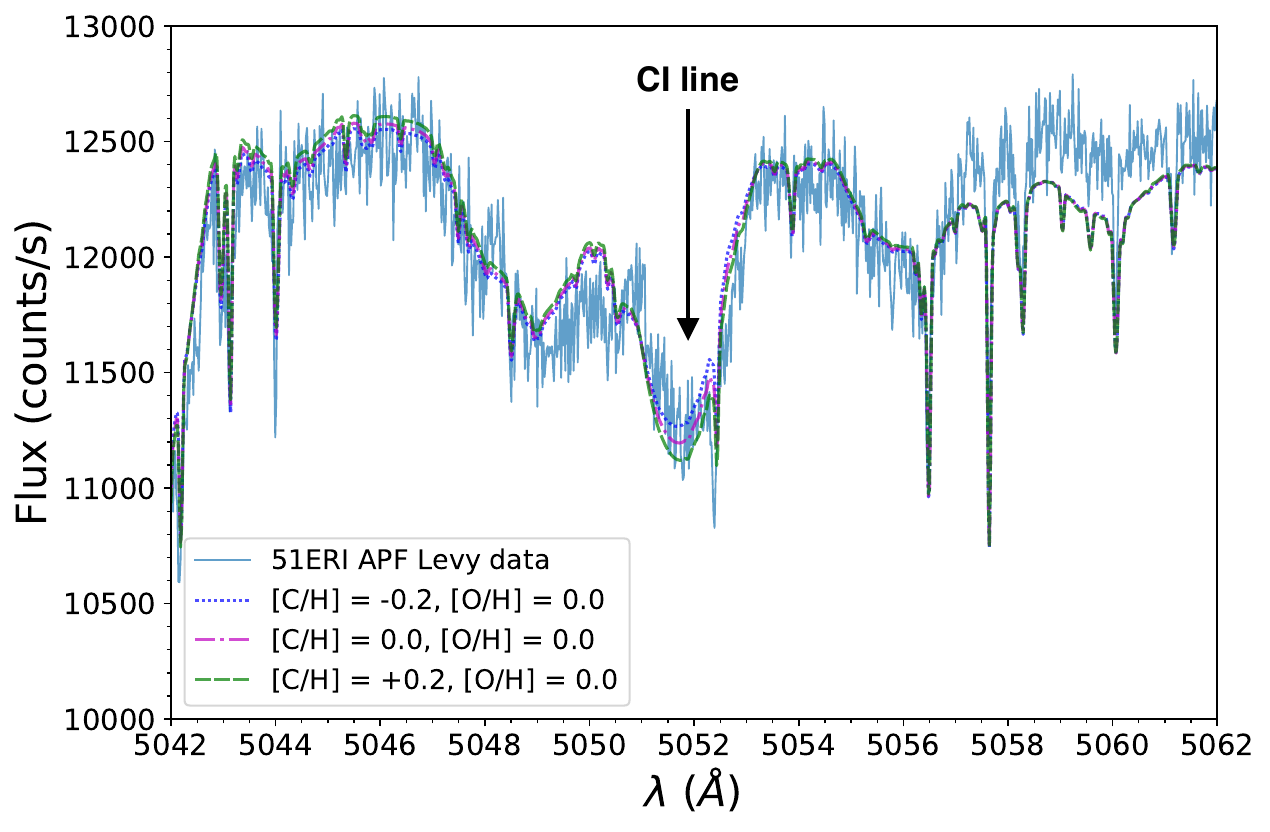}{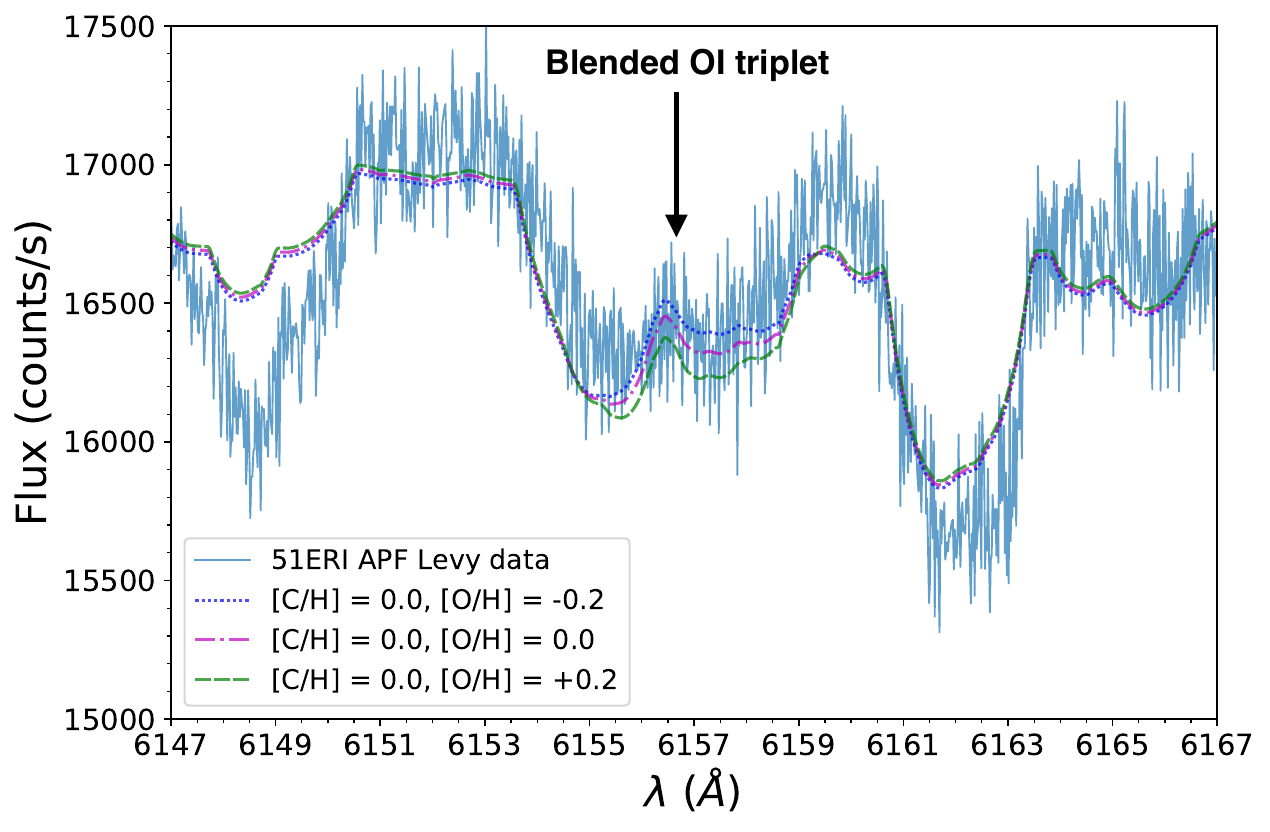}
\caption{(Left) APF spectra of the directly imaged companion host star 51 Eri (cyan) around the C \textsc{i} line at 5052 $\mathrm{\AA}$. Overplotted are \textit{PHOENIX}$\--$\textit{C/O} models at the best-fit $T_\mathrm{eff}$ and $\log{g}$ of 51 Eri, but carbon abundances ranging from $-$0.2 dex (blue, dotted) to solar (magenta, dash-dotted) to +0.2 dex (green, dashed). The oxygen abundance is kept constant at solar values for all models. The \textit{PHOENIX}$\--$\textit{C/O} model is a good fit for our data and this grid can resolve the differences in flux due to varying carbon abundance. (Right) Same as the figure on the left, but for varying oxygen abundances around the O \textsc{i} triplet at 6155$\--$58 $\mathrm{\AA}$. The carbon is kept constant at solar abundance for all models. \textcolor{black}{While the models seem to have offsets from the science spectra in certain wavelength pockets (e.g, 6147$\--$51 $\mathrm{\AA}$, these offsets do not coincide with carbon and oxygen lines of interest and are consistent across different carbon and oxygen abundances, Hence, they are probably due to an issue in the models that does not impact the carbon and/or oxygen abundances.}}
\label{fig:testco}
\end{figure}

Once the best-fit $T_\mathrm{eff}$, $\log{g}$ and $\mathrm{[M/H]}$ have been obtained, we perform a refined fit for the carbon and oxygen abundances using the \textit{PHOENIX}$\--$\textit{C/O} grid mentioned in Section \ref{subsub:phoenix}. The priors are set to (best-fit $-$  uncertainty, best-fit $+$ uncertainty) for all the stellar and telluric parameters (except the metallicity, which is fixed) as determined in the forward modeling using the standard \textit{PHOENIX} grid. The priors are set to the entire abundance range supported by the custom grid for carbon and oxygen abundances. Before using the grids for abundance forward modeling, we checked whether the \textit{PHOENIX}$\--$\textit{C/O} models showed differences in the strengths of the carbon and oxygen atomic line(s) at temperatures and gravities similar to those of the targets in our current host star sample. Figure \ref{fig:testco} shows a clear difference between models with different carbon and oxygen abundances, respectively.

\par
The \textit{PHOENIX}$\--$\textit{C/O} grid was exclusively used to forward model the orders with relatively strong carbon and oxygen lines. Among the orders with good S/N, we identified five (5) orders with strong carbon lines and 2 orders with strong oxygen lines (line wavelengths given in Table \ref{tab:lineabundance}). We perform multi-order fits on the orders with the strong carbon lines using the \textit{PHOENIX}$\--$\textit{C/O} grid to obtain an estimate of the carbon abundance. \textcolor{black}{For obtaining the abundance estimates, we consider only those runs with a reduced chi-square within 10\% of the lowest chi-square value obtained for that particular spectral line. This large range of reduced chi-square allows us to sufficiently account for spectral noise and model systematics.} As discussed previously, oxygen is more difficult to obtain the abundance for, due to the small number of strong lines in the optical region. Out of the available lines, the triplet at 7771$\--$75 $\mathrm{\AA}$ is in a noisy spectral order and, hence cannot be fit reliably using MCMC. This leaves only the forbidden line at 6300 $\mathrm{\AA}$ and the triplet at 6155--58 $\mathrm{\AA}$ as the lines to use for fitting. To determine the oxygen abundance, we perform multi-order fits over the two (2) oxygen orders. \textcolor{black}{For faster rotators, fitting over the two oxygen orders has a large uncertainty due to the significant rotational broadening This issue is overcome on a case-by-case basis.}
\subsection{Equivalent Width Determination}
The equivalent width method is used to determine the abundance of carbon, oxygen, and 13 other elements (Na, Mg, Si, S, Ca, Sc, Ti, Cr, Mn, Fe, Ni, Zn, Y). For carbon and oxygen, the abundances, and hence the C/O ratio obtained using this method are compared to that obtained through spectral modeling. As mentioned previously, this is to ensure a secondary check to our spectral modeling abundances as well as due to a lack of consensus in the literature regarding the superior method of determining abundances. The procedure for the measurement of equivalent widths and subsequent abundance estimations is different for carbon and oxygen and for the 13 other elements; we shall go over both of them in detail.
\\
\subsubsection{Carbon and Oxygen}
 Due to the high rotation velocities of all five targets, all carbon and oxygen absorption lines are blended with the surrounding lines. In this scenario, it is not straightforward to obtain the equivalent width of the individual carbon and oxygen lines. Hence, we continuum-normalize the region around the blended feature \textcolor{black}{by fitting a first-order curve to 2$\--$3 flat regions of the spectrum (with at least one on either side of the feature). This is followed by measuring the equivalent width of the entire blend using trapezoidal integration.} Using the wavelength range of the blended line and the NIST database, we identify all the absorption lines present in the blend. The spectral analysis software MOOG (\citealt{1973ApJ...184..839S}) is then used to obtain the abundances from the equivalent widths.  Kurucz ATLAS9 models are generated using stellar atmospheric parameters determined in the initial forward modeling. Using these models and the line values, i.e., the wavelength, excitation potential ($\chi_{ep}$), and the oscillator strength ($\log{gf}$) for all the lines present in the blend from NIST, we use the `blend' driver in MOOG to obtain the abundance of the element (carbon or oxygen) with an absorption feature present in the blend. This procedure is used for each carbon absorption line that is part of a blended feature. The abundance value adopted for carbon is the average of the abundances measured corresponding to each absorption line. \textcolor{black}{The error in the abundance estimates includes the standard deviation of the abundances obtained from each C \textsc{i} line and stellar atmospheric parameter uncertainties. The two sources of uncertainty are combined in quadrature.}\par
\begin{figure}
    \centering
    \includegraphics[width=1.0\linewidth]{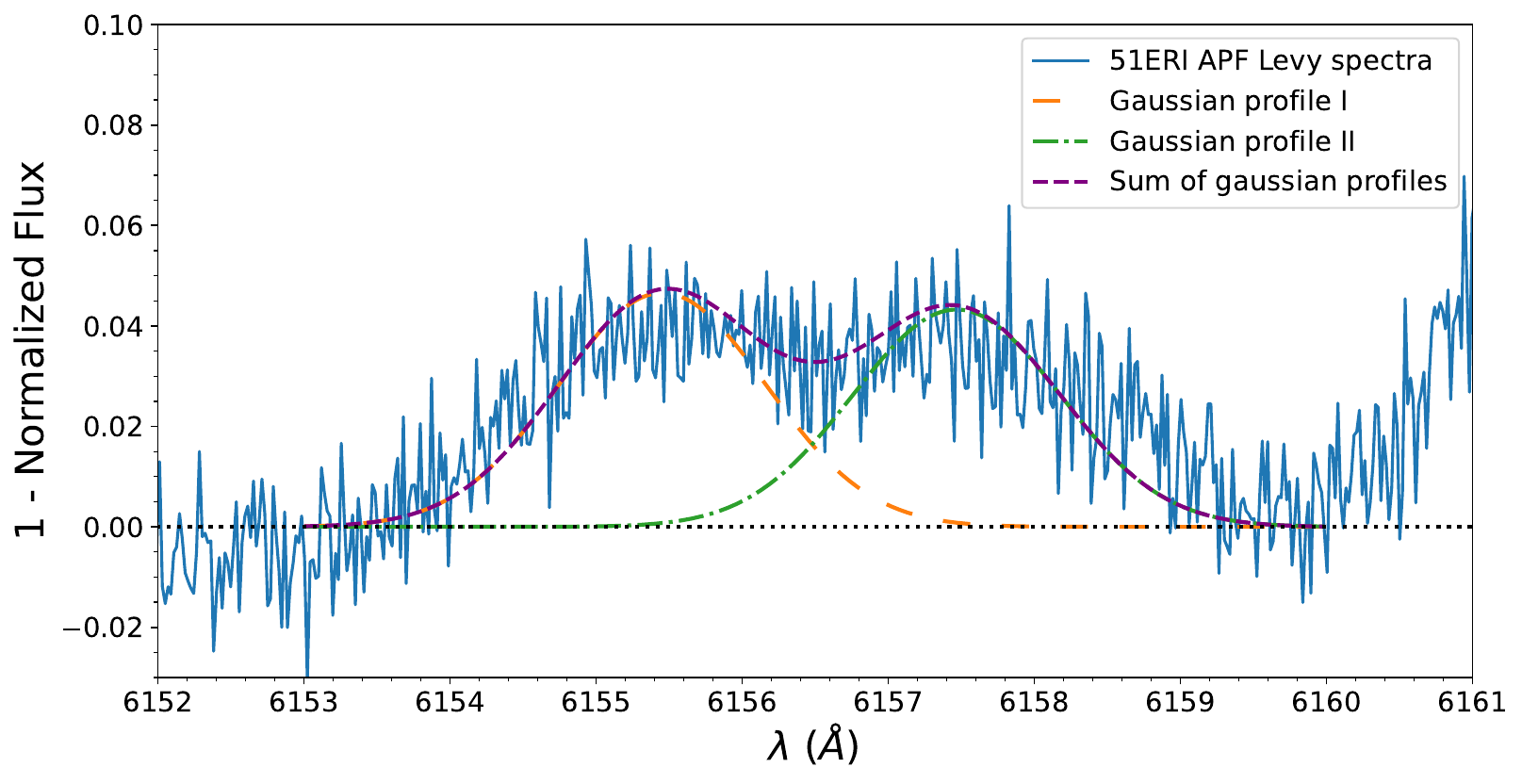}
    \caption{Continuum-normalized spectra of the host star 51 Eri (blue) illustrating the procedure for measurement of the equivalent width of the oxygen triplet feature at 6155$\--$58 $\mathrm{\AA}$. We fit two independent Gaussian profiles (orange dashed and green dash-dotted lines) to the blended feature and adjust the parameters of the two Gaussians such that the sum of the two profiles (purple dashed line) fits the observed spectral feature. Subsequently, trapezoidal integration on the area under the Gaussian blend gives the equivalent width of the blended feature. The black dotted line at the bottom shows the continuum level.}
    \label{fig:eqwidthoxy}
\end{figure}

 Measuring the equivalent width abundance for oxygen is a more difficult process, primarily because of the difficulties in measuring the equivalent width of each of its spectral features. The order with the triplet at 7771$\--$75 $\mathrm{\AA}$ is too noisy to reliably determine the continuum and measure the equivalent widths. The forbidden line at 6300 $\mathrm{\AA}$ is too weak and often indiscernible in the spectra of the faster rotators due to rotational broadening. \textcolor{black}{While we do use the forbidden line for oxygen abundance measurement for GJ 504, we ignore NLTE effects, as the uncertainty in oxygen abundance would be dominated by the variance between the various oxygen lines rather than the non-application of NLTE effects. For the faster rotators, equivalent width measurements are possible only for the oxygen triplet at 6155$\--$58 $\mathrm{\AA}$. However, this triplet appears as one single blended feature for all of our targets except GJ 504. For the latter, we can see two distinct spectral features in that wavelength region: one is a blend of the OI lines at 6155.98 and 6156.78 $\mathrm{\AA}$, and the other is the singular OI line at 6158.18 $\mathrm{\AA}$.} Thus, it is only possible to measure the abundance corresponding to the entire blended feature (i.e., for all the oxygen lines in the triplet), rather than for each of the lines individually. Since 4 out of our 5 targets only have a single oxygen feature with a measurable equivalent width, we came up with a slightly different procedure to measure the equivalent width abundance and its errors. We continuum-normalize the region around the blended feature, followed by fitting two independent Gaussian profiles to the overall blended feature (Figure \ref{fig:eqwidthoxy}). Trapezoidal integration on the area under the double Gaussian fit is used to estimate the equivalent width. The parameters of the two Gaussian profiles are tweaked to slightly alter the fit to the absorption feature. In this manner, we can obtain a mean value and standard deviation for the equivalent width, which can then be translated to an abundance value and uncertainty using MOOG.\textcolor{black}{ Using MOOG, we also investigate the uncertainty in oxygen abundance due to the uncertainties in the stellar atmospheric parameters.}
 \\
 
 \subsection{Other elements}
 Most other elements have a large number of absorption lines present in the optical region of the stellar spectrum. Hence, for these elements, we use only those lines for abundance measurements that are not blended with any neighboring lines. We only continuum-normalize a 20$\mathrm{\AA}$ region around each line, followed by measuring the equivalent width of the line using trapezoidal integration. Using the Kurucz ATLAS9 models and the line values for the absorption line from NIST, we use the `abfind' driver in MOOG to obtain the abundance of the element corresponding to the specific absorption line. The abundance value adopted for each species is the average of the abundances measured corresponding to each absorption line. \textcolor{black}{The error in the abundance estimates is computed by adding the standard deviation of the abundances obtained from each individual line for a given species and the contribution due to the uncertainties in the stellar atmospheric parameters in quadrature.} We make two exceptions to this method, for S \textsc{i} and Ti \textsc{i}, due to the lack of non-blended lines for these species for some of our targets. These two species are analyzed using a similar procedure to the carbon lines. The lines used for each species and the equivalent widths for each star are given in Table \ref{tab:lineabundance} in the appendix. The abundance corresponding to each ionic species is given in Table \ref{tab:ewabundance}.

\section{Results}
\label{sec:results}

\subsection{51 Eri}
\begin{figure}
    \centering
    \includegraphics[width=1.0\linewidth]{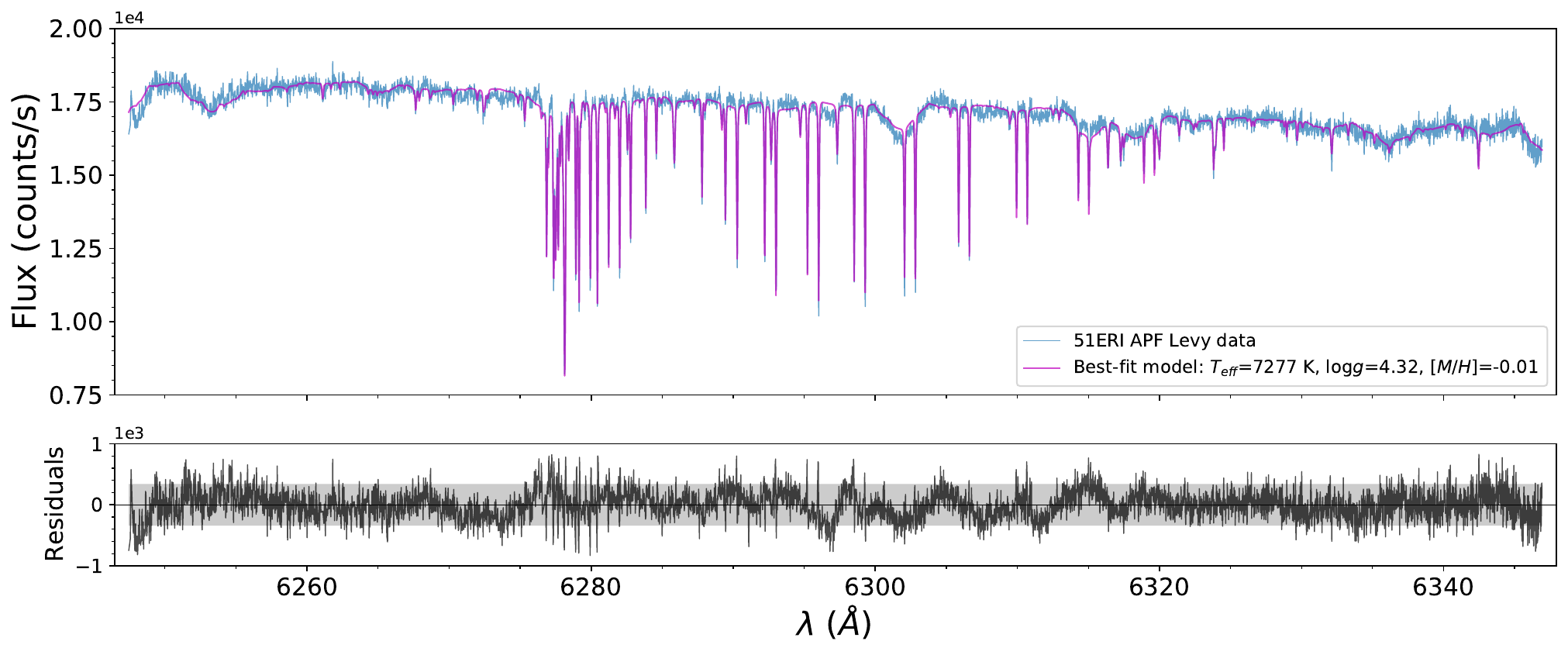}
    \caption{Best-fit PHOENIX model to the APF spectrum for the target 51 Eri (cyan), shown for APF order \#74. The broader absorption features correspond to stellar features while the sharper ones (those similar to spikes) correspond to telluric features. The parameters of the best-fit PHOENIX model are estimated by computing the median for the effective temperature ($T_\mathrm{eff}$), surface gravity ($\log{g}$), and the metallicity ($\mathrm{[M/H]}$) over nine retained runs after performing multiple multi-order MCMC runs. This model has $T_\mathrm{eff}$ = 7277 K, $\log{g}$ = 4.32, $\mathrm{[M/H]}$ = -0.01 (magenta). The residuals between the data and the model are plotted in black and other noise limits are shown in grey.}
    \label{fig:51erispec74}
\end{figure}

\begin{figure}
    \centering
    \includegraphics[width=1.0\linewidth]{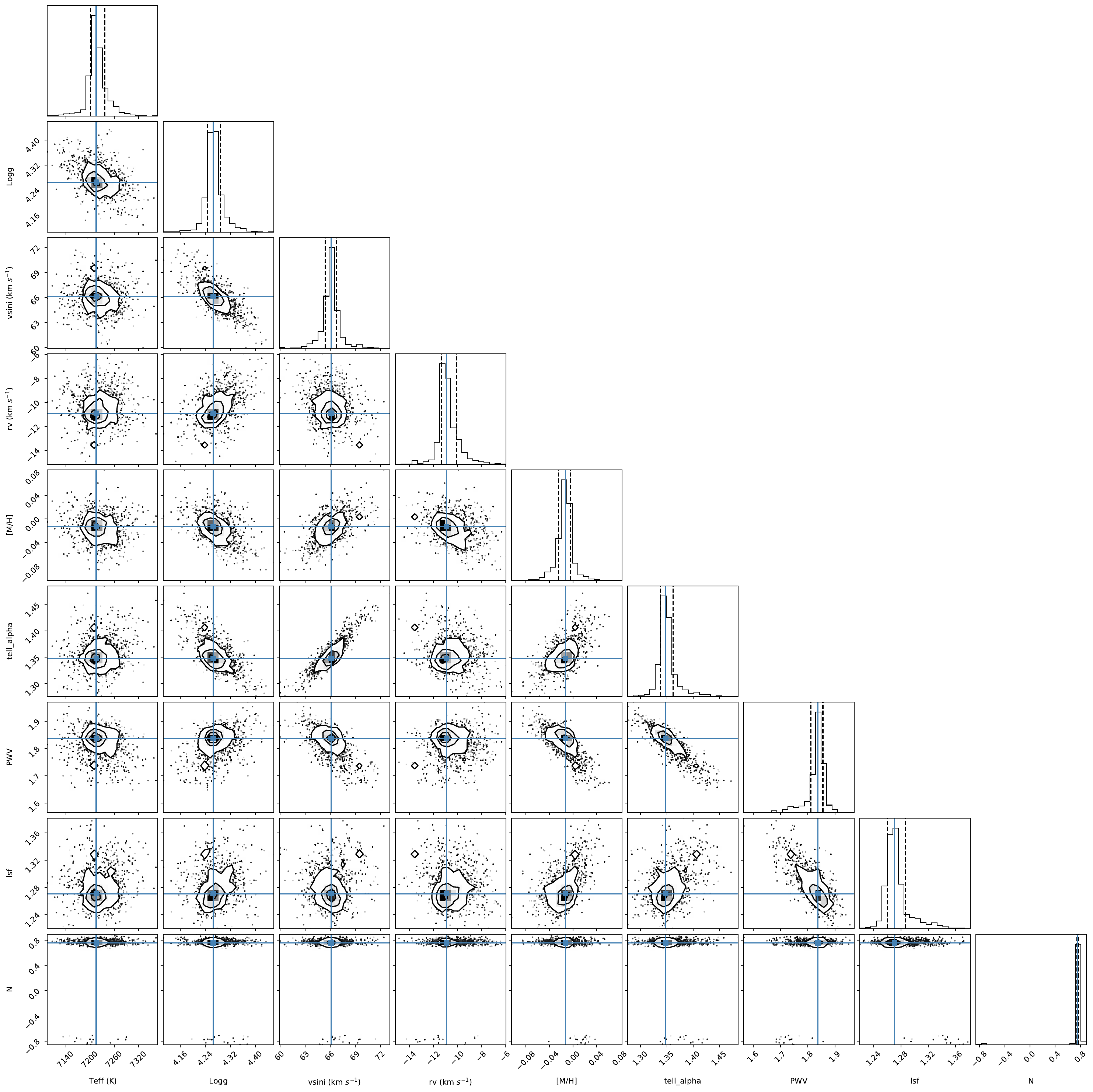}
    \caption{Corner plot for one of the retained runs for a multi-order fit of the PHOENIX grid to the spectrum of 51 Eri. The marginalized posteriors are shown along the diagonal. The blue lines represent the 50 percentile, and the dotted lines represent the 16 and 84 percentiles. The subsequent covariances between all the parameters are in the corresponding 2-D histograms. This run gives a best-fit $T_\mathrm{eff}$ = 7215 K, $\log{g}$ = 4.27, $\mathrm{[M/H]}$ = -0.01.}
    \label{fig:51ericorner}
\end{figure}

The determination of atmospheric parameters for 51 Eri involved multi-order fitting with 15 two-part MCMC runs, each with 100 walkers and 1750 steps, and the first 1250 discarded as burn-in. Out of these, two runs were discarded due to poor fits to some of the spectral orders. Another four runs were discarded as the values of some of the fitted parameters ($T_\mathrm{eff}$, $\log{g}$ and $\mathrm{[M/H]}$ in particular) hit the edge of the prior range. For the nine runs that were retained, the median and standard deviation were computed for all the fitted parameters, this gave $T_\mathrm{eff}$ = 7277 $\pm$ 164 K, $\log{g}$ = 4.32 $\pm$ 0.23, $\mathrm{[M/H]}$ = -0.01 $\pm$ 0.11 as the stellar atmospheric parameters. Figures \ref{fig:51erispec74} and \ref{fig:51ericorner} show the best-fit model to our spectral data for one of the APF orders and an example corner plot for one of the atmospheric runs, respectively. \par 

\begin{figure}
    \centering
    \includegraphics[width=1.0\linewidth]{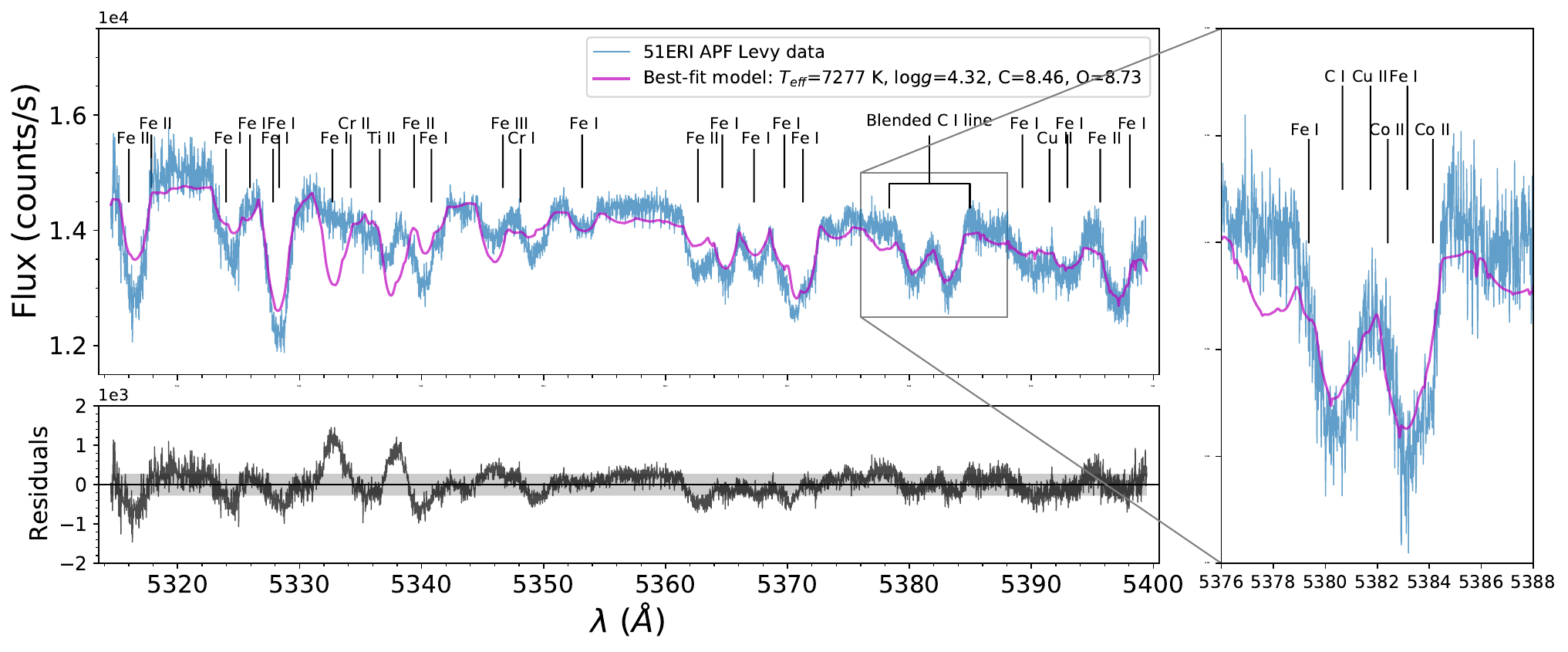}
    \caption{Best-fit PHOENIX-C/O model to the APF spectrum for the target 51 Eri (cyan), shown for order \#87 containing a blended C I line at 5380 $\mathrm{\AA}$. Some significant spectral absorption lines in this order have been labeled. The zoomed-in subplot on the right shows a $\sim$10$\mathrm{\AA}$ region around the blended line, with the lines considered in the C I blend marked individually. The best-fit PHOENIX-C/O model has $T_\mathrm{eff}$ = 7277 K, $\log{g}$ = 4.32, $\log{\epsilon_C}$ = 8.46, $\log{\epsilon_O}$ = 8.73 (magenta). The residuals between the data and the model are plotted in black and other noise limits are shown in grey. \textcolor{black}{While the model fit is excellent around the blended C I line, it does not fit well around some of the other lines, probably due to the use of the solar metallicity PHOENIX-C/O grid and or other unknown systematics the stellar atmospheric models. Due to these irregularities between the model and the spectra, we fit for only a small region around the spectral line of interest while determining carbon and oxygen abundances.}}
    \label{fig:51erispec87}
\end{figure}

\begin{figure}
    \centering
    \includegraphics[width=1.0\linewidth]{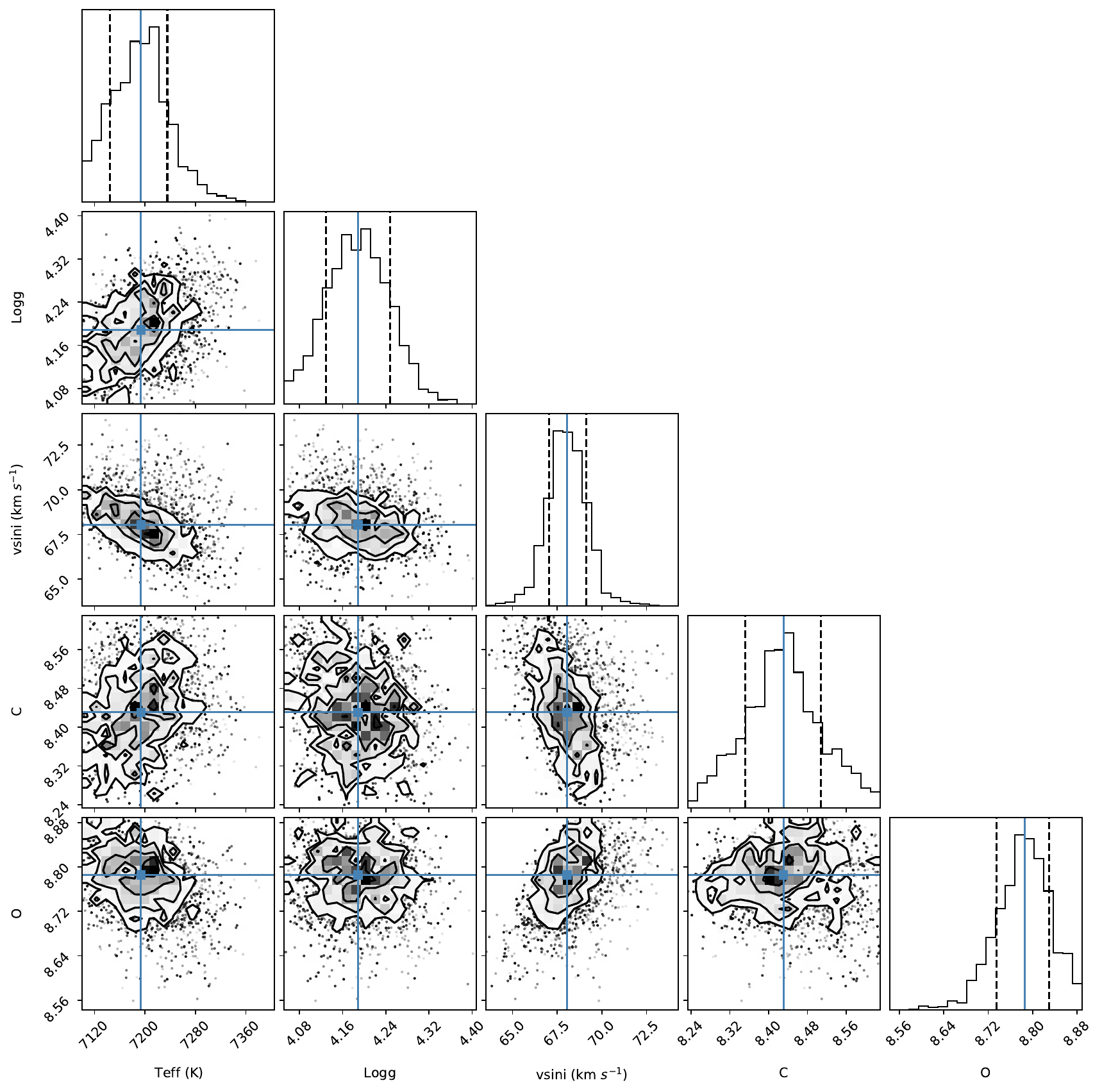}
    \caption{Corner plot for one of the retained runs for the PHOENIX-C/O grid fit to the spectrum of 51 Eri. The marginalized posteriors are shown along the diagonal. The blue lines represent the 50 percentile, and the dotted lines represent the 16 and 84 percentiles. The subsequent covariances between all the parameters are in the corresponding 2-D histograms. This run gives a best-fit $\log{\epsilon_C}$ = 8.43, $\log{\epsilon_O}$ = 8.79.}
    \label{fig:51ericornerco}
\end{figure}

As the metallicity of this star is consistent with solar, we use the \textit{PHOENIX}$\--$\textit{C/O} grid to estimate the carbon and oxygen abundances. We first spliced a $\sim$40$\mathrm{\AA}$ region around the carbon and oxygen spectral lines and fitted all of the carbon and oxygen orders simultaneously. Sixteen multi-order runs were performed with four runs discarded due to poor convergences and two runs discarded due to poor fit to one of the orders. Computing the median and standard deviation of the fitted parameters over the 10 runs, we obtain [C/H] = 0.03 $\pm$ 0.08 and [O/H] = 0.04 $\pm$ 0.08. These values give us a spectral fit C/O = 0.54 $\pm$ 0.14. Figures \ref{fig:51erispec87} and \ref{fig:51ericornerco} show the best-fit model to our spectral data for the order with the C \textsc{i} line at 5380 $\mathrm{\AA}$ and an example corner plot for one of the C/O runs, respectively.\par 

Carbon and oxygen abundance measurements using the equivalent width method give us [C/H] = $-$0.02 $\pm$ 0.16, and [O/H] = $-$0.01 $\pm$ 0.17, giving a C/O = 0.54 $\pm$ 0.29. These values are in excellent agreement with the corresponding abundances obtained using the spectral fitting method. The C/O ratios in both cases are effectively solar ($\sim$0.55). The sulfur abundance is also measured and comes out at [S/H] = $-$0.01 $\pm$ 0.12, giving C/S = 19.95 $\pm$ 9.19 and O/S = 37.15 $\pm$ 17.80. For the other 12 elements, the abundances obtained align with solar values up to 1$\sigma$ for all species except yttrium (Y). Y is found to be super-solar at the 1$\sigma$ level.

\subsection{HR 8799}
HR 8799 is a $\lambda$ Bootis star, which implies that the stellar atmosphere is deficient in iron peak elements. \textcolor{black}{The deficiency of heavier elements (like iron) in $\lambda$ Bootis stars is possibly due to the accretion of a thin outer gas envelope from a companion or circumstellar object (\citealt{2015AJ....150..166J})}. Hence, the spectrum does not have many Fe lines, making the forward modeling take fewer steps for convergence than other targets. The atmospheric parameter multi-order fitting for HR 8799 involved 10 two-part MCMC runs, each with 100 walkers and 1500 steps, with the first 1000 discarded as burn-in. Out of these, three runs were discarded as the walkers did not converge sufficiently within the 1500 steps. Computing the median and standard deviation of the fitted parameters over the remaining seven runs, we obtain $T_\mathrm{eff}$ = 7317 $\pm$ 176 K, $\log{g}$ = 4.32 $\pm$ 0.20, $\mathrm{[M/H]}$ = $-$0.57 $\pm$ 0.08 as the values of the stellar atmosphere parameters. These values of the stellar parameters were then used to generate a custom grid to fit for the carbon and oxygen abundance. \par

Due to the significant sub-solar metallicity, we cannot use the \textit{PHOENIX}$\--$\textit{C/O} grid mentioned previously. We instead compute a customized grid with fixed $\mathrm{[M/H]}$ = $-$0.5 with varying [C/H] $\in$ [$-$0.2,0.3], and [O/H] $\in$ [$-$0.2,0.3]. The carbon and oxygen abundances have a grid step of 0.1 dex, similar to the \textit{PHOENIX}$\--$\textit{C/O} grid. The grid was used to estimate the abundances by first performing multi-order fits on all carbon orders simultaneously. Ten multi-order runs were performed and we did not discard the results of any of the runs. Computing the median and standard deviation of the carbon abundance from the 10 carbon-only runs gave [C/H] = 0.13 $\pm$ 0.04. After this, we fit the three carbon and the two oxygen orders together but restricted the carbon priors according to the carbon abundance obtained previously ($\in$ [0.09, 0.17]). We performed 16 runs, with six runs discarded due to poor convergence of walkers. From the 10 retained runs, we get [O/H] = 0.10 $\pm$ 0.07. These abundances give C/O = 0.59 $\pm$ 0.11. \par

We also measure the abundances of 15 elements (including carbon and oxygen) using the equivalent width method. Carbon and oxygen abundances obtained using this method are [C/H] = 0.04 $\pm$ 0.19 and [O/H] = 0.11 $\pm$ 0.20, giving a C/O = 0.47 $\pm$ 0.30. These values agree quite well with the corresponding abundances obtained using the spectral fitting method. The C/O ratio obtained using both methods agrees with the solar value ($\sim$0.55; \citealt{2009ARA&A..47..481A}). HR 8799 is also under-abundant in sulfur, with [S/H] = $-$0.22 $\pm$ 0.09. This gives us a C/S = 37.15 $\pm$ 17.99 and O/S = 79.43 $\pm$ 40.11. In addition to sulfur, the star is also significantly under-abundant in the rest of the elements heavier than carbon and oxygen, with the abundance for each of them ranging between 0.22$\--$0.88 dex below solar. 

\subsection{HD 984}
The multi-order fitting to determine the atmospheric parameters for HD 984 involved 19 two-part MCMC runs, each with 100 walkers and 2000 steps, with the first 1500 discarded as burn-in. Of these, five runs were discarded due to poor fits to some of the spectral orders. Four runs were discarded as the $\log{g}$ values hit the edge of the prior range. An additional three runs were discarded as the best-fit value of $\mathrm{[M/H]}$ for them deviated by more than 3$\sigma$ from the median. For the seven runs that were retained, the median and standard deviation were computed for all the fitted parameters, this gave $T_\mathrm{eff}$ = 6401 $\pm$ 177 K, $\log{g}$ = 4.42 $\pm$ 0.16, $\mathrm{[M/H]}$ = $-$0.01 $\pm$ 0.09 as the stellar atmospheric parameters.\par

As the stellar metallicity is nearly solar, we use the \textit{PHOENIX}$\--$\textit{C/O} grid to fit the carbon and oxygen lines. Similar to 51 Eri, estimation of the carbon and oxygen abundance involved splicing a $\sim$40$\mathrm{\AA}$ around the carbon and oxygen spectral lines and fitting all of the carbon and oxygen orders simultaneously. Nineteen multi-order runs were performed with three runs discarded due to poor convergence of carbon and/or oxygen abundance and six runs discarded due to poor fit to one or more spectral orders. Computing the median and standard deviation of the fitted parameters over the remaining 10 runs, we obtain [C/H] = 0.06 $\pm$ 0.07 and [O/H] = 0.00 $\pm$ 0.07, giving a spectral fit C/O = 0.63 $\pm$ 0.14. \par 

Subsequently, the equivalent width method was used to determine the abundances of the 15 elements (including carbon and oxygen). This analysis gives us [C/H] = 0.04 $\pm$ 0.09 and [O/H] = 0.10 $\pm$ 0.24, leading to a C/O = 0.48 $\pm$ 0.28. These values agree with the corresponding abundances obtained using the spectral fitting method. Both the spectral fit and equivalent width C/O ratio agree with the solar C/O = 0.55. We also get [S/H] = 0.09 $\pm$ 0.17, giving C/S = 18.20 $\pm$ 8.06 and O/S = 38.02 $\pm$ 25.75. Only nickel (Ni) deviates by more than 1$\sigma$ from solar values among the remaining elements and is slightly super-solar.

\subsection{GJ 504}\label{sub:resultsgj504}
Atmospheric parameter determination for GJ 504 involved multi-order fitting over 22 two-part MCMC runs, each with 100 walkers and 1750 steps, with the first 1500 discarded as burn-in. Of these, 12 runs were discarded due to poor fits to one or more spectral orders. Two runs were discarded as the $\log{g}$ values hit the edge of the prior range. An additional run was discarded as the best-fit value of $\mathrm{[M/H]}$ for that run had a deviation of more than 3$\sigma$ from the median. The seven runs that were retained, gave $T_\mathrm{eff}$ = 5959 $\pm$ 145 K, $\log{g}$ = 4.65 $\pm$ 0.33, $\mathrm{[M/H]}$ = 0.12 $\pm$ 0.08 as the stellar atmospheric parameters. \par

As GJ 504 is significantly super-solar, we cannot use the solar metallicity \textit{PHOENIX}$\--$\textit{C/O} grid. In addition, the carbon and oxygen measurements from equivalent width were also found to be elevated. Hence we computed a custom grid with fixed $\mathrm{[M/H]}$ = +0.12 and varying [C/H] $\in$ (0.0, 0.2, 0.3, 0.4, 0.6), and [O/H] $\in$ (0.1, 0.3, 0.4, 0.5, 0.6, 0.7). The procedure used to estimate the carbon and oxygen was similar to that done for HR 8799. First, Ten multi-order runs were performed for the carbon orders and we did not discard the results of any of the runs. Computing the median and standard deviation of the carbon abundance from the 10 carbon-only runs gave [C/H] = 0.27 $\pm$ 0.03. For the oxygen abundance, we fit the three carbon and the two oxygen orders together but restricted the carbon priors according to the carbon abundance obtained previously ($\in$ [0.24, 0.30]). We performed 16 runs, with five runs discarded due to poor convergence of the walkers. From the 11 retained runs, we get [O/H] = 0.28 $\pm$ 0.11. These abundances give C/O = 0.54 $\pm$ 0.14. \par

Measurement of the abundances using the equivalent width method gave [C/H] = 0.35 $\pm$ 0.19 and [O/H] = 0.47 $\pm$ 0.16, giving a C/O = 0.42 $\pm$ 0.24. The carbon and oxygen abundances, as well as the C/O ratio, agree with the spectral fit abundance. The C/O ratio obtained using both methods is also solar ($\sim$0.55) within 1$\sigma$. Among the other 13 elements, the average abundance is significantly ($>$0.1 dex) super-solar in all of them except Mg, which is solar. However, the large uncertainties (owing to the large uncertainty in the $\log{g}$ measurement) imply that some of them (Na, Ca, Mn, Fe, and Ni) are solar at the 1$\sigma$ level. Our measurement of [S/H] = 0.42 $\pm$ 0.17, gives us a C/S = 17.38 $\pm$ 10.20 and O/S = 41.69 $\pm$ 22.41.  

\subsection{HD 206893}
Multi-order fitting to determine the atmospheric parameters for HD 206893 involved 20 two-part MCMC runs, each with 100 walkers and 2500 steps, out of which the first 2000 were discarded as burn-in. Of these 20 runs, eight were discarded due to poor fits to certain spectral orders. Two additional runs were discarded as the $\mathrm{[M/H]}$ values hit the edge of the grid. For the 10 runs that were retained, the atmospheric parameters obtained were $T_\mathrm{eff}$ = 6617 $\pm$ 46 K, $\log{g}$ = 4.27 $\pm$ 0.15 and $\mathrm{[M/H]}$ = 0.06 $\pm$ 0.18. This target required runs with a higher number of steps compared to other targets as the lower SNR of the spectra and the rotational velocity ($v\sin{i}$ $\sim$ 35 km\,s$^{-1}$) were not conducive to easier convergence. Despite these changes in the MCMC runs, we still get substantially higher errors in the metallicity estimate ($\delta[M/H]$ $\sim 0.2$ dex) compared to our other targets. \par

Since the star has nearly solar metallicity, we use the \textit{PHOENIX}$\--$\textit{C/O} grid to estimate the carbon and oxygen abundances. The procedure we use to estimate the abundance is similar to that used for HR 8799 and GJ 504. First, multi-order MCMC fits are performed on the three carbon orders to obtain the carbon abundance. Ten runs are performed with 100 walkers and 1000 steps, with the first 500 steps discarded as burn-in. All runs are retained and give us a carbon abundance of [C/H] = 0.14 $\pm$ 0.03. Subsequently, we attempted to determine the oxygen abundance by fitting the three carbon and the two oxygen orders together but restricting the priors according to the carbon abundance obtained previously ($\equiv$ [0.11, 0.17]). Sixteen runs were performed, with five runs discarded due to the oxygen abundance converging near the lower edge of the grid, and one run discarded due to poor convergence of the walkers. The 10 retained runs give us an oxygen abundance [O/H] = $-$0.03 $\pm$ 0.07. These abundances give us a C/O = 0.81 $\pm$ 0.14, which is nearly 2$\sigma$ over the solar value ($\sim$0.55). \par

Measuring the equivalent width abundances for carbon and oxygen yielded [C/H] = 0.12 $\pm$ 0.16 and [O/H] = 0.02 $\pm$ 0.15, giving a C/O = 0.69 $\pm$ 0.35. This value agrees with that obtained by the spectral fit method, as well as the solar C/O. We measure the abundances of the 13 other elements mentioned previously as well, including a sulfur abundance of [S/H] = 0.02 $\pm$ 0.05. This gives us C/S = 25.70 $\pm$ 9.92 and O/S = 37.15 $\pm$ 13.53. We only obtain a super-solar abundance for sodium (Na) among the other elements.

\section{Discussion}
\label{sec:discussion}
We analyzed a sample of five F/G-type directly imaged planet host stars and measured the abundances of 15 elements $\--$ carbon and oxygen using the spectral fit and the equivalent width method, and the remaining 13 elements using just the equivalent width method. In addition, we also calculated the C/O, C/S, and O/S ratios for all five stars. Of this sample of stars, only HR 8799 and GJ 504 had previous abundance and C/O measurements in the literature. However, the abundance measurements for HR 8799 from \cite{2006PASJ...58.1023S} had no attached uncertainty estimates. In this work, we measure elemental abundances for the first time for 51 Eri, HD 984, and HD 206893. In addition, we also measure abundances for two previously analyzed stars to address the lack of uncertainty estimates in the literature, and also to have a uniformly analyzed sample of directly imaged companion host stars.

\subsection{Comparison with Previous Measurements}

\subsubsection{51 Eri}
The atmospheric parameters of 51 Eri have been determined several times in the literature (Table \ref{tab:litcompare}). \cite{2019AA...627A.138A} used medium-resolution ($R$ $\sim$ 9000$\--$11000) spectra from the X-shooter instrument on the VLT, while \cite{2012AA...538A.143K} used medium-resolution ($R$ $\sim$ 1000) spectra from the HST. Our parameter values agree with both of these studies. High-resolution spectral analysis of 51 Eri has also been done by \cite{2017AJ....153...21L} and \cite{2021AA...647A..49S}. \cite{2017AJ....153...21L} used $R$ $\sim$ 60,000 spectra from the Sandiford spectrograph at the McDonalds Observatory for their work. We notice a big discrepancy between our values and the metallicity reported in the work, possibly as they used the equivalent width method to determine the iron abundance (and hence the metallicity), which is subject to additional uncertainty when performed on fast rotators such as 51 Eri (e.g., \citealt{2021AA...647A..49S}). However, the values obtained by \cite{2021AA...647A..49S} using $R$ $\sim$ 115,000 HARPS spectra aligns with our work.

\subsubsection{HR 8799}
The atmospheric parameters obtained from our spectral fitting (Table \ref{tab:litcompare}) are in good agreement with the results from \cite{2006PASJ...58.1023S} and the results from \cite{2020AJ....160..150W} for the combined PEPSI and HARPS spectra. Except the $\mathrm{[M/H]}$, the values are also in close agreement with \cite{2021AA...647A..49S}. All stellar parameters also agree with \cite{2003AJ....126.2048G}. For the elemental abundances, the [C/H], [O/H], and C/O ratio (from both spectral fitting and equivalent width) are aligned with those reported in \cite{2020AJ....160..150W}: [C/H] = 0.11 $\pm$ 0.12, [O/H] = 0.12 $\pm$ 0.14, and C/O = $0.54^{+0.12}_{-0.09}$. All three values also agree with those from \cite{2006PASJ...58.1023S} ([C/H] = 0.20, [O/H] = 0.19, and C/O = 0.56).

\begin{deluxetable}{c|c|ccccc}
\tablecaption{Comparison of this work with literature \label{tab:litcompare}}
\tablewidth{0pt}
\tablehead{\colhead{Target}    & \colhead{Work}     & \colhead{$T_\mathrm{eff}$ (K)}    & \colhead{$\log{g}$ (cgs)}     & \colhead{[M/H]}     & \colhead{[C/H]} & \colhead{[O/H]}
}
\startdata
\multirow{6}{*}{51 Eri}  & \multirow{2}{*}{This work} & \multirow{2}{*}{7277 $\pm$ 164} & \multirow{2}{*}{4.32 $\pm$ 0.23} & \multirow{2}{*}{-0.01 $\pm$ 0.11} & 0.03 $\pm$ 0.08 \tablenotemark{a} &  0.04 $\pm$ 0.08 \tablenotemark{a }\\ 
  & & & & & -0.02 $\pm$ 0.16 \tablenotemark{b} & -0.01 $\pm$ 0.17 \tablenotemark{b} \\
  & \cite{2021AA...647A..49S} & 7259 $\pm$ 167 & 4.12 $\pm$ 0.20 & -0.06 $\pm$ 0.10 &  &   \\ 
  & \cite{2019AA...627A.138A} & 7366 $\pm$ 146 & 4.09 $\pm$ 0.21 & 0.09 $\pm$ 0.07 &  &   \\  
  & \cite{2017AJ....153...21L}  & 7146 $\pm$ 62 & 4.23 & 0.24 $\pm$ 0.35 & & \\ 
  & \cite{2012AA...538A.143K}  & 7414 $\pm$ 31 & 4.09 $\pm$ 0.13 & -0.02 $\pm$ 0.08 & & \\ \hline
\multirow{6}{*}{HR 8799}  & \multirow{2}{*}{This work} & \multirow{2}{*}{7317 $\pm$ 176} & \multirow{2}{*}{4.32 $\pm$ 0.20} & \multirow{2}{*}{-0.57 $\pm$ 0.08} & 0.13 $\pm$ 0.04 \tablenotemark{a} &  0.10 $\pm$ 0.07 \tablenotemark{a }\\ 
  & & & & & 0.04 $\pm$ 0.19 \tablenotemark{b} & 0.11 $\pm$ 0.20 \tablenotemark{b} \\
  & \cite{2021AA...647A..49S}  & 7301 $\pm$ 190 & 4.12 $\pm$ 0.23 & -0.70 $\pm$ 0.15 & & \\
  & \cite{2020AJ....160..150W} & 7390 $\pm$ 80 & 4.35 $\pm$ 0.07  & -0.52 $\pm$ 0.08  &  0.11 $\pm$ 0.12 & 0.12 $\pm$ 0.14 \\ 
  & \cite{2006PASJ...58.1023S} & 7250 & 4.30 & -0.50 & 0.20 & 0.19   \\  
  & \cite{2003AJ....126.2048G}  & 7422  & 4.22 & -0.50 & & \\ \hline
\multirow{5}{*}{HD 984}  & \multirow{2}{*}{This work} & \multirow{2}{*}{6401 $\pm$ 177} & \multirow{2}{*}{4.42 $\pm$ 0.16} & \multirow{2}{*}{-0.01 $\pm$ 0.09} & 0.06 $\pm$ 0.07 \tablenotemark{a} &  0.00 $\pm$ 0.07 \tablenotemark{a }\\ 
  & & & & & 0.04 $\pm$ 0.09 \tablenotemark{b} & 0.10 $\pm$ 0.24 \tablenotemark{b} \\
  & \cite{2024AA...686A.294C}  &   &  & -0.01 $\pm$ 0.12 & -0.05 $\pm$ 0.10 & 0.09 $\pm$ 0.20 \\
  & \cite{2020ApJ...898..119R} & 6479 $\pm$ 42 & 4.43 $\pm$ 0.05 & 0.06 $\pm$ 0.02  &  &   \\ 
  & \cite{2018AJ....155..111L} & 6266 $\pm$ 24 & 4.31 & 0.27 &  &   \\  
  & \cite{2005ApJS..159..141V}  & 6490  & 4.83 & -0.05 & & \\ \hline
\multirow{4}{*}{GJ 504\tablenotemark{c}}  & \multirow{2}{*}{This work} & \multirow{2}{*}{5959 $\pm$ 145} & \multirow{2}{*}{4.65 $\pm$ 0.33} & \multirow{2}{*}{0.12 $\pm$ 0.08} & 0.27 $\pm$ 0.03 \tablenotemark{a} &  0.28 $\pm$ 0.11 \tablenotemark{a }\\ 
  & & & & & 0.35 $\pm$ 0.19 \tablenotemark{b} & 0.47 $\pm$ 0.16 \tablenotemark{b} \\
  & \cite{2021AJ....161..134H} & 6080 $\pm$ 100  & 4.3 $\pm$ 0.1 & 0.21 $\pm$ 0.06 &  &   \\ 
  & \cite{2017AA...598A..19D} & 6205 $\pm$ 20 & 4.29 $\pm$ 0.07 & 0.22 $\pm$ 0.04 & -0.004 $\pm$ 0.109 & 0.030 $\pm$ 0.059  \\  \hline
\multirow{5}{*}{HD 206893}  & \multirow{2}{*}{This work} & \multirow{2}{*}{6617 $\pm$ 46} & \multirow{2}{*}{4.27 $\pm$ 0.15} & \multirow{2}{*}{0.06 $\pm$ 0.18} & 0.14 $\pm$ 0.03 \tablenotemark{a} &  -0.03 $\pm$ 0.07 \tablenotemark{a }\\ 
  & & & & & 0.12 $\pm$ 0.16 \tablenotemark{b} & 0.02 $\pm$ 0.15 \tablenotemark{b} \\
  & \cite{2022AA...667A..63Z} & 6680  & 4.34 & 0.08  &  &   \\ 
  & \cite{2017AA...608A..79D} & 6500 $\pm$ 100 & 4.45 $\pm$ 0.15 & 0.04 $\pm$ 0.02 & & \\  
  & \cite{2012AA...541A..40M} &  &  & -0.01 & & \\  \hline
\enddata
\tablenotetext{a}{Abundances obtained using spectral fitting}
\tablenotetext{b}{Abundances obtained using equivalent width}
\tablenotetext{c}{Additional literature comparisons made within text (refer Section \ref{subsub:discgj504})}
\end{deluxetable}

\subsubsection{HD 984}
HD 984 has three measurements of the atmospheric parameters in the literature (Table \ref{tab:litcompare}). The first one was by \cite{2005ApJS..159..141V} using $R$ $\sim$ 70,000 spectra from Keck, Lick, and AAT (Anglo-Australian Telescope). This work uses isochrones to determine the $\log{g}$ of the star to predict a stellar age of 1.2 Gyr. \textcolor{black}{This age estimate was an isochronal age, which inherently has large uncertainties for main-sequence stars like HD 984. In several works since that time, other age indicators have been used to decrease those uncertainties and show that the star is young. Those include activity indicators (X-rays, Ca H\&K), rotation rate, and kinematics. Updated kinematics from Gaia have not shown a definitive association with Columba (\citealt{2018ApJ...856...23G}) as originally suggested by \cite{2011ApJ...732...61Z}. However, there are sufficient indications of a young age that the youth of this system is well-accepted. \cite{2015MNRAS.453.2378M} derive a moving group-independent age of 30 $\--$ 200 Myr, which is consistent with \cite{2005ApJS..159..141V} but with a much smaller uncertainty. This age remains consistent with updated studies of both the star and companion, such as \cite{2022AJ....163...50F} and \cite{2024AA...686A.294C}.} The [M/H] measurements by \cite{2018AJ....155..111L} using HARPS data is significantly higher than our work. (similar to 51 Eri and \cite{2017AJ....153...21L}). However, the values do agree if uncertainties are taken into account. The most recent parameters are by \cite{2020ApJ...898..119R} using $R$ $\sim$ 70,000 Keck HIRES spectra and agree almost perfectly with our measurements. \textcolor{black}{More recently, \cite{2024AA...686A.294C}  measured [C/H], [O/H], and [Fe/H] for HD 984. All three values agree with our measurements. Their C/O = 0.40 $\pm$ 0.20 for HD 984 also agrees with both our spectral fit and equivalent width C/O ratios.}

\begin{deluxetable}{cccccccccccc}
\tablecaption{Abundances for each ionic species for targets\label{tab:ewabundance}}
\tablewidth{0pt}
\tablehead{
\colhead{Ion} & \colhead{HR 8799} & \colhead{} & \colhead{51 Eri} & \colhead{} & \colhead{HD 984} & \colhead{} & \colhead{GJ 504} & \colhead{} & \colhead{HD 206893} & \colhead{} & \colhead{Solar values} \\
\colhead{} & \colhead{log(N)} & \colhead{\textcolor{black}{rms}} &  \colhead{log(N)} & \colhead{\textcolor{black}{rms}} & \colhead{log(N)} & \colhead{\textcolor{black}{rms}} & \colhead{log(N)} & \colhead{\textcolor{black}{rms}} & \colhead{log(N)} & \colhead{\textcolor{black}{rms}} & \colhead{}
}
\startdata
\colhead{C I} & \colhead{8.47} & \colhead{0.19} &  \colhead{8.41} & \colhead{0.16} & \colhead{8.47} & \colhead{0.09} & \colhead{8.78} & \colhead{0.19} & \colhead{8.55} & \colhead{0.16} & \colhead{8.43} \\
\colhead{O I} & \colhead{8.80} & \colhead{0.20} &  \colhead{8.68} & \colhead{0.17} & \colhead{8.79} & \colhead{0.24} & \colhead{9.16} & \colhead{0.16} & \colhead{8.71} & \colhead{0.15} & \colhead{8.69} \\
\colhead{Na I} & \colhead{5.79} & \colhead{0.08} &  \colhead{6.19} & \colhead{0.08} & \colhead{6.30} & \colhead{0.17} & \colhead{6.38} & \colhead{0.16} & \colhead{6.61} & \colhead{0.12} & \colhead{6.24} \\
\colhead{Mg I} & \colhead{7.26} & \colhead{0.27} &  \colhead{7.59} & \colhead{0.23} & \colhead{7.58} & \colhead{0.28} & \colhead{7.64} & \colhead{0.30} & \colhead{7.73} & \colhead{0.24} & \colhead{7.60} \\
\colhead{Si I} & \colhead{7.24} & \colhead{0.11} &  \colhead{7.61} & \colhead{0.18} & \colhead{7.54} & \colhead{0.12} & \colhead{7.71} & \colhead{0.11} & \colhead{7.52} & \colhead{0.08} & \colhead{7.51} \\
\colhead{Si II} & \colhead{7.26} & \colhead{0.22} &  \colhead{7.51} & \colhead{0.35} & \colhead{7.61} & \colhead{0.20} & \colhead{8.27\tablenotemark{*}} & \colhead{0.27} & \colhead{7.60} & \colhead{0.17} & \colhead{7.51} \\
\colhead{S I} & \colhead{6.90} & \colhead{0.09} &  \colhead{7.11} & \colhead{0.12} & \colhead{7.21} & \colhead{0.17} & \colhead{7.54} & \colhead{0.17} & \colhead{7.14} & \colhead{0.05} & \colhead{7.12}  \\
\colhead{Ca I} & \colhead{5.72} & \colhead{0.19} &  \colhead{6.36} & \colhead{0.18} & \colhead{6.43} & \colhead{0.25} & \colhead{6.56} & \colhead{0.31} & \colhead{6.47} & \colhead{0.20} & \colhead{6.34} \\
\colhead{Sc II} & \colhead{2.55} & \colhead{0.15} &  \colhead{3.08} & \colhead{0.15} & \colhead{3.23} & \colhead{0.17} & \colhead{3.61} & \colhead{0.16} & \colhead{3.12} & \colhead{0.19} & \colhead{3.15} \\
\colhead{Ti I} & \colhead{4.62} & \colhead{0.36} &  \colhead{5.04} & \colhead{0.23} & \colhead{5.00} & \colhead{0.29} & \colhead{5.13} & \colhead{0.19} & \colhead{4.89} & \colhead{0.15} & \colhead{4.95} \\
\colhead{Ti II} & \colhead{4.47} & \colhead{0.14} &  \colhead{4.93} & \colhead{0.20} & \colhead{5.00} & \colhead{0.21} & \colhead{5.19} & \colhead{0.17} & \colhead{4.88} & \colhead{0.14} & \colhead{4.95} \\
\colhead{Cr I} & \colhead{5.11} & \colhead{0.19} &  \colhead{5.63} & \colhead{0.21} & \colhead{5.56} & \colhead{0.29} & \colhead{5.84} & \colhead{0.20} & \colhead{5.57} & \colhead{0.18} & \colhead{5.64} \\
\colhead{Cr II} & \colhead{5.15} & \colhead{0.17} &  \colhead{5.56} & \colhead{0.13} & \colhead{5.51} & \colhead{0.21} & \colhead{5.91} & \colhead{0.19} & \colhead{5.44} & \colhead{0.21} & \colhead{5.64} \\
\colhead{Mn I} & \colhead{4.75} & \colhead{0.11} &  \colhead{5.48} & \colhead{0.11} & \colhead{5.46} & \colhead{0.45} & \colhead{5.64} & \colhead{0.22} & \colhead{5.42} & \colhead{0.23} & \colhead{5.43} \\
\colhead{Fe I} & \colhead{6.98} & \colhead{0.19} &  \colhead{7.52} & \colhead{0.21} & \colhead{7.55} & \colhead{0.23} & \colhead{7.63} & \colhead{0.26} & \colhead{7.51} & \colhead{0.19} & \colhead{7.50} \\
\colhead{Fe II} & \colhead{7.01} & \colhead{0.18} &  \colhead{7.54} & \colhead{0.17} & \colhead{7.57} & \colhead{0.24} & \colhead{7.75} & \colhead{0.23} & \colhead{7.48} & \colhead{0.19} & \colhead{7.50} \\
\colhead{Ni I} & \colhead{5.69} & \colhead{0.14} &  \colhead{6.13} & \colhead{0.21} & \colhead{6.46} & \colhead{0.23} & \colhead{6.38} & \colhead{0.20} & \colhead{6.34} & \colhead{0.25} &\colhead{6.22} \\
\colhead{Zn I} & \colhead{3.68} & \colhead{0.10} &  \colhead{4.41} & \colhead{0.17} & \colhead{4.42} & \colhead{0.30} & \colhead{4.69} & \colhead{0.08} & \colhead{4.47} & \colhead{0.28} & \colhead{4.56} \\
Y II & 1.67 & .10 &  2.47 & 0.21 & 2.25 & 0.04 & 2.41 & 0.15 & 2.20 & 0.05 & 2.21 \\
\enddata
\tablenotetext{*}{Abundance corresponds to equivalent width of unresolved blend}
\end{deluxetable}

\subsubsection{GJ 504} \label{subsub:discgj504}
 Extensive previous work has been done in determining the atmospheric parameters of GJ 504 as it is a G-type star fairly close to Earth ($d<$20pc). In addition, its low rotation velocity greatly aids in the analysis of its spectra. The most recent measurement was by \cite{2021AJ....161..134H} using HIRES spectra (\ref{tab:litcompare}); our measurements agree with this work within 2$\sigma$. Summarizing the results of other measurements in literature (e.g., \citealt{1993A&A...275..101E}; \citealt{2004A&A...418..551M}; \citealt{2005ApJS..159..141V}; \citealt{2007PASJ...59..335T}; \citealt{2010MNRAS.403.1368G}; \citealt{2012AA...541A..40M}; \citealt{2012A&A...542A..84D}; \citealt{2013ApJ...764...78R}; \citealt{2017AA...598A..19D}; \citealt{2017AJ....153...21L}; \citealt{2018A&A...614A..55A}), we have $T_\mathrm{eff}$ = 5995$\--$6234 K, $\log{g}$ = 4.15$\--$4.63 and $\mathrm{[M/H]}$  = 0.10$\--$0.28. Our stellar parameters fall well within this range. However, when we compare to \cite{2017AA...598A..19D}, who did a detailed abundance analysis in addition to measuring the atmospheric parameters, we find that our stellar atmospheric parameters agree with their values only at a 2$\sigma$ level. In addition, their [C/H] = -0.004, [O/H] = 0.030 is solar, while we obtain super-solar values for both elements using both spectral fitting and equivalent width. However, their C/O ratio = 0.56$^{+0.26}_{-0.18}$ does agree with our work. Among other elements, they get super-solar Mg (as opposed to our solar values) and solar S (super-solar in this work).

\subsubsection{HD 206893}
From Table \ref{tab:litcompare}, we see that all three atmospheric parameters agree with \cite{2022AA...667A..63Z} within 2$\sigma$. The metallicity also agrees with that obtained by \cite{2012AA...541A..40M}. In their paper on the companion HD 206893 B, \cite{2017AA...608A..79D} obtained host atmospheric parameters as well using FEROS spectroscopy, which agree quite well with our measurements. The recent discovery of a second planet \citep{doi:10.1051/0004-6361/202244727} makes it imperative to obtain accurate values for its individual elemental abundances to establish the formation and evolution of this planetary system. 

\subsection{Comparison between Spectral fitting and Equivalent width methods}
In the course of this work, we use both the spectral fitting and equivalent width methods to obtain the carbon and oxygen abundances, and hence the C/O ratio for our five targets. The results are summarized in Table \ref{tab:abundanceratio}. While the abundances obtained from both approaches agree well, each approach comes with its caveats. \par

The spectral fitting approach gives smaller uncertainties in abundance as compared to equivalent width. This occurs as most of our targets (except GJ 504) are fast rotators, which leads to the carbon and oxygen spectral lines blending with other neighboring lines. While there are several lines available for most of the other elements, the limited number of lines for carbon, oxygen, and sulfur necessitates the need to resolve the line blends to obtain the abundance. This is additionally compounded by the fact that the NIST database occasionally does not have line information on some important lines in a given blend, further compounding the errors. This is particularly an issue with the lines around the oxygen triplet at 7771--75 $\mathrm{\AA}$. As GJ 504 is a relatively slow rotator, the lines of interest are either quite resolved or blended with fewer neighboring lines, allowing us to measure the abundance of the line more accurately \textcolor{black}{in theory. However, the large uncertainty in $\log{g}$ for this target makes the uncertainties larger.}\par

\begin{deluxetable}{ccccccc}
\tablecaption{Abundance ratios for targets in this paper\label{tab:abundanceratio}}
\tablewidth{0pt}
\tablehead{\colhead{Abundance ratio}    & \colhead{HR 8799}     & \colhead{51 Eri}            & \colhead{HD 984}            & \colhead{GJ 504}            & \colhead{HD 206893}     & \colhead{Solar}  } 
\startdata
C/O (spectral fit) & 0.59 $\pm$ 0.11   & 0.54 $\pm$ 0.14   & 0.63 $\pm$ 0.14   &    0.54 $\pm$ 0.14  & 0.81 $\pm$ 0.14   & \multirow{2}{*}{0.55} \\
C/O (equiv. width)    & 0.47 $\pm$ 0.30   & 0.54 $\pm$ 0.29   & 0.48 $\pm$ 0.28   &   0.42 $\pm$ 0.24  & 0.69 $\pm$ 0.35    &                  \\ \hline
C/S   & 37.15 $\pm$ 17.99 & 19.95 $\pm$ 9.19  & 18.20 $\pm$ 8.06  & 17.38 $\pm$ 10.20  & 25.70 $\pm$ 9.92  & 20.42                 \\ 
O/S   & 79.43 $\pm$ 40.11 & 37.15 $\pm$ 17.80 & 38.02 $\pm$ 25.75 & 41.69 $\pm$ 22.41 & 37.15 $\pm$ 13.53  & 37.15                 \\ 
\enddata
\end{deluxetable}

While the issue of blends makes spectral fitting seem like the superior method, it ultimately relies upon synthetic atmospheric models. Even the best synthetic models are unable to take into account all factors that affect stellar atmospheric spectra. The $PHOENIX$ models that we use for our analysis do not model all the lines in the spectra perfectly. Thus, any abundances obtained using the spectral fitting method also include errors due to the atmospheric model not being a perfect fit to the spectra. The equivalent width method comes out superior in this regard, as it involves measuring the line strengths directly from the stellar spectra without resorting to any synthetic models. \par
Ultimately, it is difficult to obtain a consensus regarding the superior method for determining abundance. Using just the equivalent width method seems the superior option for slower rotators due to the relative absence of blends and avoiding the pitfalls of synthetic atmospheric models. However, for fast rotators (which comprise a substantial population of the directly imaged companion host stars), using both methods to obtain abundances and cross-verifying the results is advisable. This allows a secondary check of the drawbacks of either method of measuring abundances. 

\subsection{Metallicity and elemental abundance trends in current sample}
\begin{figure}
    \centering
    \includegraphics[width=1\linewidth]{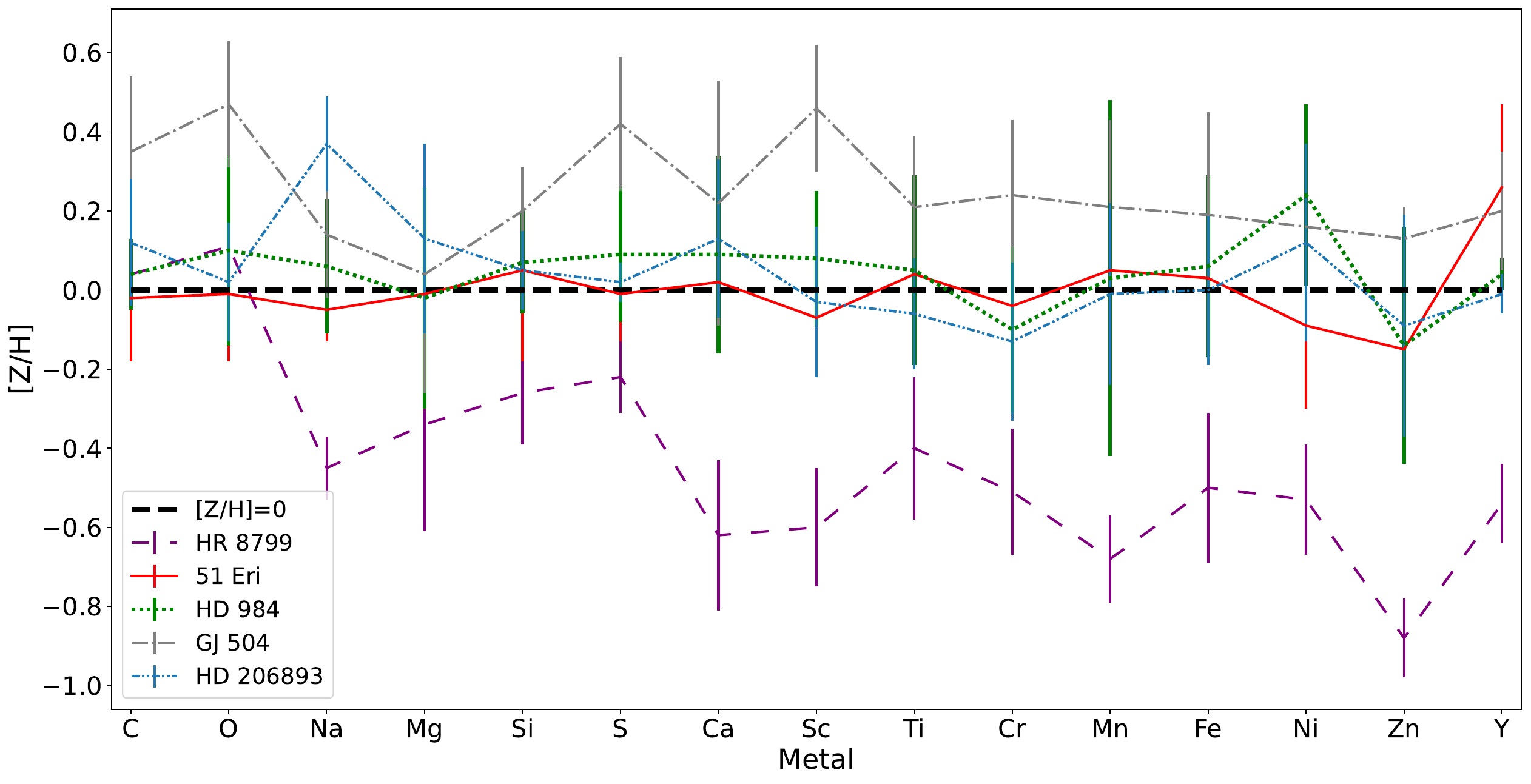}
    \caption{Abundances relative to solar for the 15 elements with abundances measured using the equivalent width method. The black dotted line denotes where the abundance of an element is solar. Relative abundances are shown for all five targets in this paper. We note that HR 8799 has consistently sub-solar abundances (except C and O), while GJ 504 has consistently super-solar average abundances (except Mg). For the other three host stars, the abundances of most elements are solar with no consistent trend regarding which element/s deviates from solar. Overall, we do not notice any unusual abundance patterns amongst the stars that host directly imaged planets that persist for the whole sample.}
    \label{fig:metalabundance}
\end{figure}

We do a small all-sample analysis of our 5 targets, looking at the deviation of the overall metallicity and the individual metal abundances when compared to solar values. The abundance of various species relative to solar values is encapsulated in Table \ref{tab:relabundance}, but we also plot the relative abundances in Figure \ref{fig:metalabundance} for each of our targets. \par

\begin{deluxetable*}{ccccccccccc}
\tablecaption{Metal abundance relative to solar values\label{tab:relabundance}}
\tablewidth{0pt}
\tablehead{
\colhead{Element} & \colhead{HR 8799} & \colhead{} & \colhead{51 Eri} & \colhead{} & \colhead{HD 984} & \colhead{} & \colhead{GJ 504} & \colhead{} & \colhead{HD 206893} & \colhead{} \\
\colhead{(X)} & \colhead{[X/H]} & \colhead{\textcolor{black}{$\Delta$[X/H]}} &  \colhead{[X/H]} & \colhead{\textcolor{black}{$\Delta$[X/H]}} & \colhead{[X/H]} & \colhead{\textcolor{black}{$\Delta$[X/H]}} & \colhead{[X/H]} & \colhead{\textcolor{black}{$\Delta$[X/H]}} & \colhead{[X/H]} & \colhead{\textcolor{black}{$\Delta$[X/H]}}
}
\startdata
\colhead{[M/H]\tablenotemark{a}} & \colhead{-0.57} & \colhead{0.08} &  \colhead{-0.01} & \colhead{0.11} & \colhead{-0.01} & \colhead{0.09} & \colhead{0.12} & \colhead{0.08} & \colhead{0.06} & \colhead{0.18} \\
\hline
\colhead{C} & \colhead{0.04} & \colhead{0.19} &  \colhead{-0.02} & \colhead{0.16} & \colhead{0.04} & \colhead{0.09} & \colhead{0.35} & \colhead{0.19} & \colhead{0.12} & \colhead{0.16} \\
\colhead{O} & \colhead{0.11} & \colhead{0.20} &  \colhead{-0.01} & \colhead{0.17} & \colhead{0.10} & \colhead{0.24} & \colhead{0.47} & \colhead{0.16} & \colhead{0.02} & \colhead{0.15} \\
\colhead{Na} & \colhead{-0.45} & \colhead{0.08} &  \colhead{-0.05} & \colhead{0.08} & \colhead{0.06} & \colhead{0.17} & \colhead{0.14} & \colhead{0.16} & \colhead{0.37} & \colhead{0.12}  \\
\colhead{Mg} & \colhead{-0.34} & \colhead{0.27} &  \colhead{-0.01} & \colhead{0.23} & \colhead{-0.02} & \colhead{0.28} & \colhead{0.04} & \colhead{0.30} & \colhead{0.13} & \colhead{0.24} \\
\colhead{Si} & \colhead{-0.26} & \colhead{0.13} &  \colhead{0.05} & \colhead{0.23} & \colhead{0.05} & \colhead{0.13} & \colhead{0.20\tablenotemark{b}} & \colhead{0.11} & \colhead{0.03} & \colhead{0.10}  \\
\colhead{S} & \colhead{-0.22} & \colhead{0.09} &  \colhead{-0.01} & \colhead{0.12} & \colhead{0.09} & \colhead{0.17} & \colhead{0.42} & \colhead{0.17} & \colhead{0.02} & \colhead{0.05} \\
\colhead{Ca} & \colhead{-0.62} & \colhead{0.19} &  \colhead{0.02} & \colhead{0.18} & \colhead{0.09} & \colhead{0.25} & \colhead{0.22} & \colhead{0.31} & \colhead{0.13} & \colhead{0.20} \\
\colhead{Sc} & \colhead{-0.60} & \colhead{0.15} &  \colhead{-0.07} & \colhead{0.15} & \colhead{0.08} & \colhead{0.17} & \colhead{0.46} & \colhead{0.16} & \colhead{-0.03} & \colhead{0.19}  \\
\colhead{Ti} & \colhead{-0.45} & \colhead{0.18} &  \colhead{0.02} & \colhead{0.20} & \colhead{0.05} & \colhead{0.24} & \colhead{0.21} & \colhead{0.18} & \colhead{-0.07} & \colhead{0.14} \\
\colhead{Cr} & \colhead{-0.50} & \colhead{0.16} &  \colhead{-0.06} & \colhead{0.14} & \colhead{-0.12} & \colhead{0.21} & \colhead{0.24} & \colhead{0.19} & \colhead{-0.18} & \colhead{0.20}  \\
\colhead{Mn} & \colhead{-0.68} & \colhead{0.11} &  \colhead{0.05} & \colhead{0.11} & \colhead{0.03} & \colhead{0.45} & \colhead{0.21} & \colhead{0.22} & \colhead{-0.01} & \colhead{0.23}  \\
\colhead{Fe} & \colhead{-0.51} & \colhead{0.19} &  \colhead{0.02} & \colhead{0.20} & \colhead{0.05} & \colhead{0.23} & \colhead{0.15} & \colhead{0.26} & \colhead{0.00} & \colhead{0.19}  \\
\colhead{Ni} & \colhead{-0.53} & \colhead{0.14} &  \colhead{-0.09} & \colhead{0.21} & \colhead{0.24} & \colhead{0.23} & \colhead{0.16} & \colhead{0.20} & \colhead{0.12} & \colhead{0.25} \\
\colhead{Zn} & \colhead{-0.88} & \colhead{0.10} &  \colhead{-0.15} & \colhead{0.17} & \colhead{-0.14} & \colhead{0.30} & \colhead{0.13} & \colhead{0.08} & \colhead{-0.09} & \colhead{0.28} \\
\colhead{Y} & \colhead{-0.54} & \colhead{0.10} &  \colhead{0.26} & \colhead{0.21} & \colhead{0.04} & \colhead{0.04} & \colhead{0.20} & \colhead{0.15} & \colhead{-0.01} & \colhead{0.05} \\
\enddata

\tablenotetext{a}{[M/H] values from spectral fitting for atmospheric parameters}
\tablenotetext{b}{Only Si \textsc{i} abundance used to calculate [Si/H]}
\end{deluxetable*}

HR 8799 has an overall metallicity 0.58 dex below solar; this is also reflected in the abundance of all other elements except C and O being sub-solar. On the other hand, GJ 504 has an overall metallicity 0.12 dex above solar. This is also consistent with the average abundance of all the measured elements (except Mg) ranging from 0.13$\--$0.47 dex above solar. 51 Eri, HD 984 and HD 206893 have a solar overall metallicity at the 1$\sigma$ level. When considering individual elemental abundances for these targets, most elements have solar abundances for these three targets. 51 Eri has super-solar Y abundance, HD 984 has super-solar Ni, and HD 206893 has super-solar Na abundance. \par
Looking at our sample of five targets as a whole, we do not notice any significant trends in metallicity or individual elemental abundances. Targets with multiple elements deviating from solar (HR 8799 and GJ 504) have non-solar overall metallicities themselves. For the three solar metallicity targets, only isolated elements show $>$1$\sigma$ deviation from solar. Thus based on this sample, we notice no trends between directly imaged planet hosts and their metallicities or elemental abundances. Even the overall metallicity is solar or slightly enhanced for all the sources (except HR 8799), indicating that directly imaged companion host stars do not need to be super metal-rich to host giant planets.  This is inconsistent with some predictions of massive planet formation via core accretion, such as \cite{2012A&A...541A..97M}, but not necessarily a clear indication for formation via gravitational instability.

\subsection{Abundance ratios}
\subsubsection{C/O Ratio}
As mentioned previously, the individual elemental abundances, and particularly, the elemental abundance ratios are quite important in constraining planet formation. The C/O ratio is one such diagnostic that has been measured for several directly imaged exoplanets. For a planet formed by gravitational instability, the C/O ratio of its atmosphere would match that of its host star. For a planet formed by the core-accretion process (both the pebble and planetesimal accretion scenarios), the exact C/O ratio relative to the host star would depend on the formation location of the planet relative to the snowlines of $\mathrm{CO}$, $\mathrm{CO_2}$, and $\mathrm{H_2O}$ (\citealt{2011ApJ...743L..16O}; \citealt{2019ARA&A..57..617M}).
In this work, we find solar C/O ratios using both methods for all targets except HD 206893. The spectral fit method gives a super-solar C/O (0.81 $\pm$ 0.14) for HD 206893 at a 2$\sigma$ level. The elevated C/O is due to the super-solar [C/H] for  HD 206893 ($\sim$5$\sigma$ confidence). \textcolor{black}{\cite{2011ApJ...743L..16O} and \cite{2016ApJ...833..203P} state that in most core accretion scenarios, planet C/O ratios would differ from stellar C/O ratios by $>$0.1 dex. We meet this requirement for all our targets using the spectral fit method.}\par

As mentioned previously, there are a number of ongoing efforts to measure C/O ratios among the directly imaged planet population. So far around fifteen planets from this population have C/O ratio measurements (\citealt{doi:10.1051/0004-6361/201832942}; \citealt{doi:10.1051/0004-6361/202244826}; \citealt{2023AJ....166...85H}), including all of the companions to the host stars in this work (51 Eri b, HR 8799 b$\--$e, HD 984 B, GJ 504 b, and HD 206893 B). In addition, various JWST programs are currently in progress to characterize the atmospheres and measure the C/O ratios of a number of directly imaged planets. \par

\begin{figure}
    \centering
    \includegraphics[width=1\linewidth]{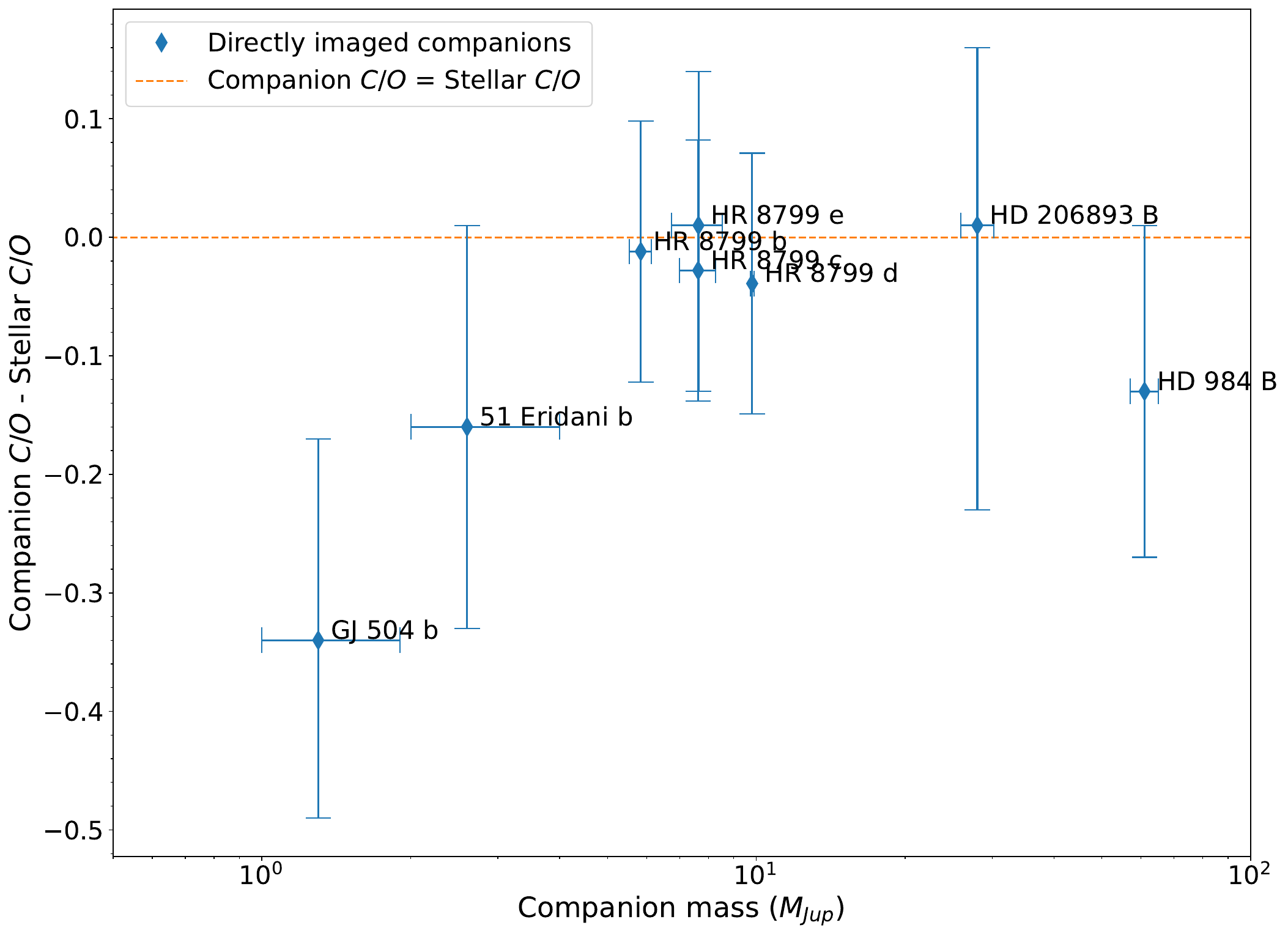}
    \caption{Companion C/O ratio relative to the stellar C/O ratio plotted against the companion mass in $M_\mathrm{Jup}$. Most companions effectively have stellar C/O ratios at the 1$\sigma$ level. Only GJ 504 b has a non-stellar (sub-stellar) C/O ratio, however, the C/O ratio of this companion is highly model-dependent and hence still uncertain. Stellar C/O ratios are from Table 4 and planetary parameters are from Table 6. We assume a mass of 2$\--$4$M_\mathrm{Jup}$ for 51 Eri b (\citealt{doi:10.1051/0004-6361/202244826}) and a C/O = $0.82^{+0.04}_{-0.19}$ for HD 206893 B from \cite{doi:10.1051/0004-6361/202140749}.}
    \label{fig:cocompare}
    
\end{figure}

Of the five targets studied in this work, two of them are known to host multiple companions: HR 8799 has four, while HD 206893 has two. The companion information from Section \ref{sub:targets} is again summarized in Table \ref{tab:companiontable}. We compare the available companion C/O ratios to the host star C/O in Figure \ref{fig:cocompare}. For this exercise, we use the C/O results from the method that gives better uncertainties for the host star. If both methods give similar uncertainties, we use the C/O value from the spectral fit method as it gives lower uncertainties for most targets. \par

Seven of the eight companions to our targets have stellar C/O ratios to within 1$\sigma$. GJ 504 b is the only companion with sub-stellar C/O. This suggests that most of these companions might have formed through the gravitational instability process, as this formation mechanism gives similar planetary and stellar C/O ratios. Theoretical modeling by \cite{2010Icar..207..503H} also predicted stellar compositions for planets in the HR 8799 system formed by gravitational instability. However, the large uncertainties indicate that other formation mechanisms cannot be ignored as of yet. In addition, the C/O reported for companions such as GJ 504 b and HD 206893 B is highly model-dependent, suggesting that making conclusions about their formation would be premature with current data. Rotational broadening for the early-type directly imaged hosts makes improving uncertainties on their C/O difficult. Hence, tighter bounds would be needed for planetary C/O measurements to make definitive conclusions about their formation. Observations using JWST might contribute significantly in this regard. \par

\begin{deluxetable}{ccccccc}
\tablecaption{Companion masses and C/O ratios\label{tab:companiontable}}
\tablewidth{0pt}
\tablehead{\colhead{Companion} & \colhead{Mass ($M_\mathrm{Jup}$)}    & \colhead{C/O Ratio} & \colhead{References}} 
\startdata
\multirow{2}{*}{51 Eri b} & $\leq$11 (Dynamical)   & \multirow{2}{*}{0.38 $\pm$ 0.09}  & \multirow{2}{*}{1, 2, 3, 4, 5, 6, 7, 8}\\
& 2$\--$9 (Evolutionary) & \\
\hline
HR 8799 b & 5.84 $\pm$ 0.3   & $0.578^{+0.004}_{-0.005}$ & {9, 10, 11} \\
HR 8799 c &  $7.63^{+0.64}_{-0.63}$ & 0.562 $\pm$ 0.004  & {9, 10, 11} \\
HR 8799 d &  9.81 $\pm$ 0.08  & $0.551^{+0.005}_{-0.004}$ & {9, 10, 11} \\
HR 8799 e &  $7.64^{+0.89}_{-0.91}$ & $0.60^{+0.07}_{-0.08}$  & {9, 10, 12} \\
\hline
HD 984 B & 61 $\pm$ 4   &  0.50 $\pm$ 0.01 & {13, 17} \\
\hline
\multirow{2}{*}{GJ 504 b} & $1.3^{+0.6}_{-0.3}$ (Young system) &  \multirow{2}{*}{$0.20^{+0.09}_{-0.06}$} & \multirow{2}{*}{14} \\
& $23.0^{+10}_{-9}$ (Old system) &  \\
\hline
HD 206893 B & $28.0^{+2.2}_{-2.1}$   & 0.65$\--$0.90 & {15, 16} \\
HD 206893 c & $12.7^{+1.2}_{-1.0}$    &   & {15} \\
\enddata
\tablerefs{(1) \cite{2020AJ....159....1D}, (2) \cite{2022MNRAS.509.4411D}, (3) \cite{doi:10.1126/science.aac5891}, (4) \citet{doi:10.1051/0004-6361/201629767}, (5) \cite{2019AJ....158...13N}, (6) \cite{doi:10.1051/0004-6361/202244826}, (7) \cite{2023MNRAS.525.1375W}, (8) \cite{2024arXiv240101468E}, (9) \cite{2022AJ....163...52S}, (10) \cite{doi:10.1051/0004-6361/202243862}, (11) \cite{2021AJ....162..290R}, (12) \cite{doi:10.1051/0004-6361/202038325}, (13) \cite{2022AJ....163...50F}, (14) \cite{doi:10.1051/0004-6361/201832942}, (15) \cite{doi:10.1051/0004-6361/202244727}, (16) \cite{doi:10.1051/0004-6361/202140749}, (17) \cite{2024AA...686A.294C}}
\end{deluxetable}

\subsubsection{Volatile-to-sulfur Ratios}
An increasing amount of evidence suggests that a planet's C/O ratio is insufficient to definitively interpret its formation history. Several studies on different aspects of the planet formation process in protoplanetary disks reveal that the C/O ratio is not uniquely correlated with the initial formation location or where a planet accreted most of its mass and hence provides limited information about the same (e.g., \citealt{2016ApJ...832...41M}; \citealt{2021ApJ...909...40T}; \citealt{2022ApJ...937...36P}). \cite{2023AJ....166...85H} demonstrate how the directly imaged planet population has solar C/O ratios. This necessitates using additional abundance ratios rather than just the C/O ratio to probe formation histories. \par
The possibility of sulfur detections in exoplanet atmospheres using JWST has initiated speculation that volatile-to-sulfur ratios (C/S and O/S) might be particularly promising. The high condensation temperature of sulfur implies that it should be present exclusively in refractories beyond $\sim$0.3 AU in the protoplanetary disk (e.g., \citealt{2011ApJ...738..141O}). The pebble accretion model of planet formation suggests that gas giants would become enriched in volatiles over refractories such as sulfur. This is because the pressure extrema induced by gas giants inhibits the inward migration and accretion of solids, thus giants formed by this mechanism should have elevated volatile-to-sulfur ratios (\citealt{2021A&A...654A..71S}; \citealt{2021A&A...654A..72S};  \citealt{2022A&A...659C...3S}). Meanwhile, planet formation by standard core accretion (through planetesimals) would give nearly stellar C/S and O/S ratios (\citealt{2022ApJ...937...36P}). Thus the volatile-to-sulfur ratios could be used to track the refractory-to-volatile ratios for the planet. \par
From Table \ref{tab:abundanceratio}, we see 51 Eri, HD 984, GJ 504, and HD 206893 all have solar volatile-to-sulfur ratios. HR 8799 being a $\lambda$ Bootis star is deficient in sulfur, but has solar to super-solar abundance for carbon and oxygen, leading to elevated average values of volatile-to-sulfur ratios. Unfortunately, the companions of none of these five targets have sulfur abundance measurements, making planet$\--$host star volatile-to-sulfur ratio comparisons currently impossible. We hope future observations with JWST will provide the missing measurements to enable such comparisons.

\subsubsection{Nitrogen elemental ratios}
In addition to carbon, oxygen, and sulfur, nitrogen (N) abundance is also a useful tracer for planet formation and migration pathways, in the form of C/N and N/O ratios (\citealt{2021ApJ...909...40T}; \citealt{2022ApJ...937...36P}). The bulk of the nitrogen is mostly in the form of the hypervolatile $\mathrm{N_2}$ with a significant fraction in the form of $\mathrm{NH_3}$ (\citealt{2018A&A...613A..14E}; \citealt{2021PhR...893....1O}). The snowline for $\mathrm{N_2}$ is farther away from the star compared to all the other C and O snowlines, thus most of the N is in the gas phase through most of the disk (\citealt{2021ApJ...909...40T}). This leads to the C and O sequestering in the solid phase faster than N with increasing orbital distance. The greater the deviations in the planetary C/N and N/O ratios relative to the stellar values, the greater the distance migrated by the planet. Moreover, considering both the C/N and N/O ratios could also allow us to distinguish between gas-dominated and solid-enriched atmospheric accretion (\citealt{2021ApJ...909...40T}). However, we do not measure nitrogen abundance in stellar atmospheres as part of our analysis for two reasons: (1) forbidden [NI] lines in the optical being too weak for abundance measurements and (2) the sufficiently strong NI lines and the NH vibration-rotation lines being present only in the infrared region (\citealt{2020A&A...636A.120A}; \citealt{2021A&A...653A.141A}). Additional spectral coverage in the infrared would be required to measure nitrogen abundance in the host stars.

\subsubsection{Other elemental ratios}
All other elements for which we measure stellar abundances (Na, Mg, Si, Ca, Sc, Ti, Cr, Mn, Fe, Ni, Zn, Y) are considered highly refractory, condensing into solids very close to the host star (\citealt{2003ApJ...591.1220L}, \citealt{2014A&A...562A..27T})and rarely detected in the gaseous atmospheres of directly imaged companions. However, refractories such as $\mathrm{Fe}$, $\mathrm{MgSiO_3}$ and $\mathrm{Mg_2SiO_4}$ are present in the atmosphere of some of these companions in the form of cloud condensates (e.g., \citealt{2006asup.book....1L}, \citealt{2012ApJ...754..135M}, \citealt{2023ApJ...946L...6M}). If we consider that the refractories in these clouds are present solely due to the sublimation of accreted pebbles/planetesimals, we would expect the ratio of the elements contained in clouds of different kinds to correspond to the ratio of the elements in the accreted rocks. Since both the host star and the pebbles/planetesimals formed from the same initial nebula, we would expect the abundance ratio of these refractory elements in the accreted bodies to be equal to the stellar value (\citealt{2015A&A...574A.138T}). This could inform the mass fraction of clouds composed of different condensates while modeling cloud properties in the atmospheres of these companions. \par 
While the measurement of the Mg/Si ratios for the host star would not be informative about the atmospheric gaseous composition, and hence the formation and migration history of their substellar companions, they could tell us about the mineral composition of the (potentially) rocky/icy cores of these gaseous companions. In addition, they would also be indicative of the mineral composition in the hitherto undiscovered terrestrial planets in these directly imaged systems. Having these measurements available when such hidden terrestrial planets are discovered via direct imaging or other methods would be hugely advantageous. \par
The abundance ratios for the host star, and hence, for the pebbles/planetesimals could also be used to inform the distribution of different kinds of minerals in the protoplanetary disk. \cite{2015A&A...580A..30T} propose that the Mg/Si ratio governs the silicate distribution within the protoplanetary disk and the distribution of silicon among different minerals depending on the Mg/Si ratio in the disk. Depending on the specific Mg/Si ratio, the Mg and Si could be incorporated into various minerals such as orthopyroxene ($\mathrm{MgSiO_3}$), olivine
($\mathrm{Mg_2SiO_4}$), feldspars ($\mathrm{CaAl_2Si_2O_8}$, $\mathrm{NaAlSi_3O_8}$) and oxides. Minerals such as olivine have been detected in the debris disk around $\mathrm{\beta}$Pictoris (\citealt{2012Natur.490...74D}). Depending on the exact elemental abundance ratio for the host star, one could search for signatures of specific minerals in a protoplanetary/debris disk. The ratios could also be used to inform the specific modeling of protoplanetary disks. \par 
However, there have been detections of refractory species in the atmospheres of directly imaged substellar objects: Na I and K I for VHS 1256$\--$1257b (\citealt{2023ApJ...946L...6M}); Na I, Mg I, K I, Ca I, Ti I$\--$Si I doublet, Mn I, Fe I, etc., for TWA 27A and TWA 28 (\citealt{2024arXiv240204230M}). In addition to the volatile-to-sulfur ratio, these elements could also be used to track the refractory-to-volatile elemental abundance ratio for the hotter ($T_{eq}$\>2,000K) directly imaged companions. For such companions, could also measure elemental abundance ratios involving these refractory elements; deviations from the corresponding stellar values might imply non-uniform composition of accreted pebbles/planetesimals or other unknown processes at play. \par

\section{Conclusions}
\label{sec:conclusion}
Abundance ratios can be a tracer for planet formation and evolution, with planet abundance ratios varying differently relative to the host star ratios depending on the formation mechanism and evolutionary processes undergone by the planets. While several directly imaged planets have C/O ratio measurements, HR 8799 is the only directly imaged companion host star with a measured C/O ratio. In this work, we measure the metallicities, individual abundances of 15 elements, and various abundance ratios for a sample of five stars with well-studied companions. \par

We use high-resolution optical spectra from the Levy spectrograph at Lick Observatory to estimate the abundances of carbon and oxygen and hence the C/O ratio using two methods $\--$ spectral template fitting and equivalent width measurement. The abundances are measured only using the equivalent width method for the other 13 elements (Na, Mg, Si, S, Ca, Sc, Ti, Cr, Mn, Fe, Ni, Zn, Y). We also use the sulfur abundance measurements to measure the C/S and O/S ratios for the host stars. \par

We first use the spectral fit method to determine the atmospheric parameters ($T_{eff}$, $\log{g}$, $\mathrm{[M/H]}$) for our host stars. The values we obtain are generally in agreement with those in the literature. These atmospheric parameters are subsequently used to obtain the custom \textit{PHOENIX}$\--$\textit{C/O} grid to fit for the carbon and oxygen abundances. The spectral fit method gives solar C/O ratios ($\sim$0.55) for 51 Eri, HR 8799, HD 984, and GJ 504. However, we obtain a super-solar C/O ratio for HD 206893 (0.81 $\pm$ 0.14) at a 2$\sigma$ confidence. \par
On the other hand, the equivalent width method gives solar C/O ratios for all five stars, albeit with the caveat that the C/O uncertainties are \textcolor{black}{up to $\sim$0.2 dex higher} compared to the spectral fit method. Despite this, the C/O obtained using the two methods agree for each of the targets in this paper. We also calculate the C/S and O/S ratios, which are found to be solar for all targets \textcolor{black}{(though the average value for HR 8799 is super-solar for both ratios)}. While the sulfur abundance (and hence the C/S and O/S ratios) have not been measured for any of the directly imaged companions, recent JWST detections of sulfur-bearing species in the atmosphere of hot Jupiters provide hope that similar detections could be possible in the atmospheres of the directly imaged companions using JWST. With approved JWST programs for companions around four of the five host stars in this work, these detections might happen quite soon. \par

We also compare the available C/O ratios for companions around our host stars with the stellar C/O measured in this work, with most companions effectively having similar C/O ratios to their host stars. GJ 504 b is found to be the only exception, with a sub-stellar C/O ratio. However, the large uncertainties \textcolor{black}{and the model-dependent C/O ratios for several companions, in addition to the complex nature of the C/O ratio as a formation tracer,} indicate that making conclusions about formation would be premature at this juncture. JWST observations could greatly assist in this as well by obtaining more precise planetary C/O measurements. \par

Abundances of elements heavier than carbon and oxygen do not show any significant trends among any of our directly imaged planet host stars. Overall metallicities are also solar or slightly super-solar for all targets (except HR 8799), indicating that directly imaged companion host stars do not need to be metal-rich to host giant planets. However, measurement of these abundances and other ratios such as Fe/Si and Mg/Si could possibly be used to constrain various elements of planet modeling such as clouds and internal structure. \par

The future steps would be to extend the analysis in this paper to more host stars from the directly imaged companion population, which will be the topic of future work. Performing a similar analysis for host stars of the transiting planet population would allow us to investigate the similarities and differences between these two planetary populations, not just due to formation and evolution, but also from a modeling perspective.

\section{Acknowledgements}
The authors would like to thank Brad Holden for his support in obtaining these observations. A.B. and Q.M.K acknowledge support by the National Aeronautics and Space Administration under Grants/Contracts/Agreements No. 80NSSC24K0210 issued through the Astrophysics Division of the Science Mission Directorate. S.P acknowledges support by NASA under award number 80GSFC21M0002.
This research has made use of the NASA Exoplanet Archive, which is operated by the California Institute
of Technology, under contract with the National Aeronautics and Space Administration under the Exoplanet Exploration Program.
Any opinions, findings, conclusions, and/or recommendations expressed in this paper are those of the author(s) and do not reflect the views of the National Aeronautics and Space Administration. The data presented herein were obtained at the Lick Observatory, which is operated by the University of California. We acknowledge that the land on which Lick Observatory is located is the unceded territory of the Ohlone (Costanoans), Tamyen, and Muwekma Ohlone tribes. This work used Bridges-2 (\citealt{osti_10299128}) at Pittsburgh Supercomputing Center through allocation PHY230140 from the Advanced Cyberinfrastructure Coordination Ecosystem: Services \& Support (ACCESS) program, which is supported by National Science Foundation grants \#2138259, \#2138286, \#2138307, \#2137603, and \#2138296. The research shown here acknowledges use of the Hypatia Catalog Database, an online compilation of stellar abundance data as described in Hinkel et al. (2014, AJ, 148, 54), which was supported by NASA's Nexus for Exoplanet System Science (NExSS) research coordination network and the Vanderbilt Initiative in Data-Intensive Astrophysics (VIDA). The majority of this work was conducted at the University of California, San Diego, which was built on the unceded territory of the Kumeyaay Nation, whose people continue to maintain their political sovereignty and cultural traditions as vital members of the San Diego community.

\bibliographystyle{aasjournal}
\bibliography{main}

\appendix
\restartappendixnumbering
\section{Atmospheric Parameter MCMC Fit Plots}

\begin{figure}[H]
    \centering
    \includegraphics[width=0.98\linewidth]{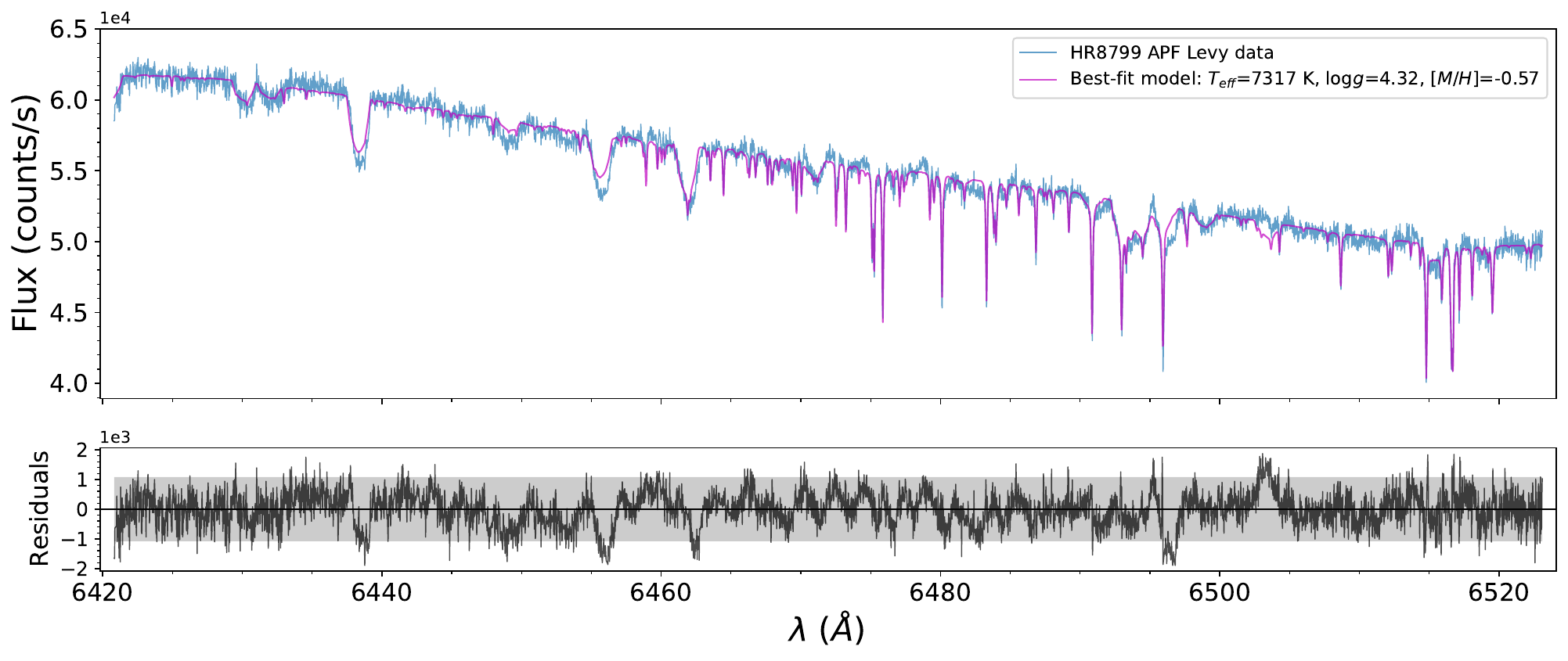}
    \caption{Best-fit PHOENIX model to the APF spectrum for the target HR 8799 (cyan), shown for APF order \#72. The parameters of the best-fit PHOENIX model are estimated by computing the median for the effective temperature ($T_\mathrm{eff}$), surface gravity ($\log{g}$), and the metallicity ($\mathrm{[M/H]}$) over seven retained runs after performing multiple multi-order MCMC runs. This model has $T_\mathrm{eff}$ = 7317 K, $\log{g}$ = 4.32, $\mathrm{[M/H]}$ = -0.57 (magenta). The residuals between the data and the model are plotted in black and other noise limits are shown in grey}
    \label{fig:schmeatic}
\end{figure}

\begin{figure}[H]
    \centering
    \includegraphics[width=0.98\linewidth]{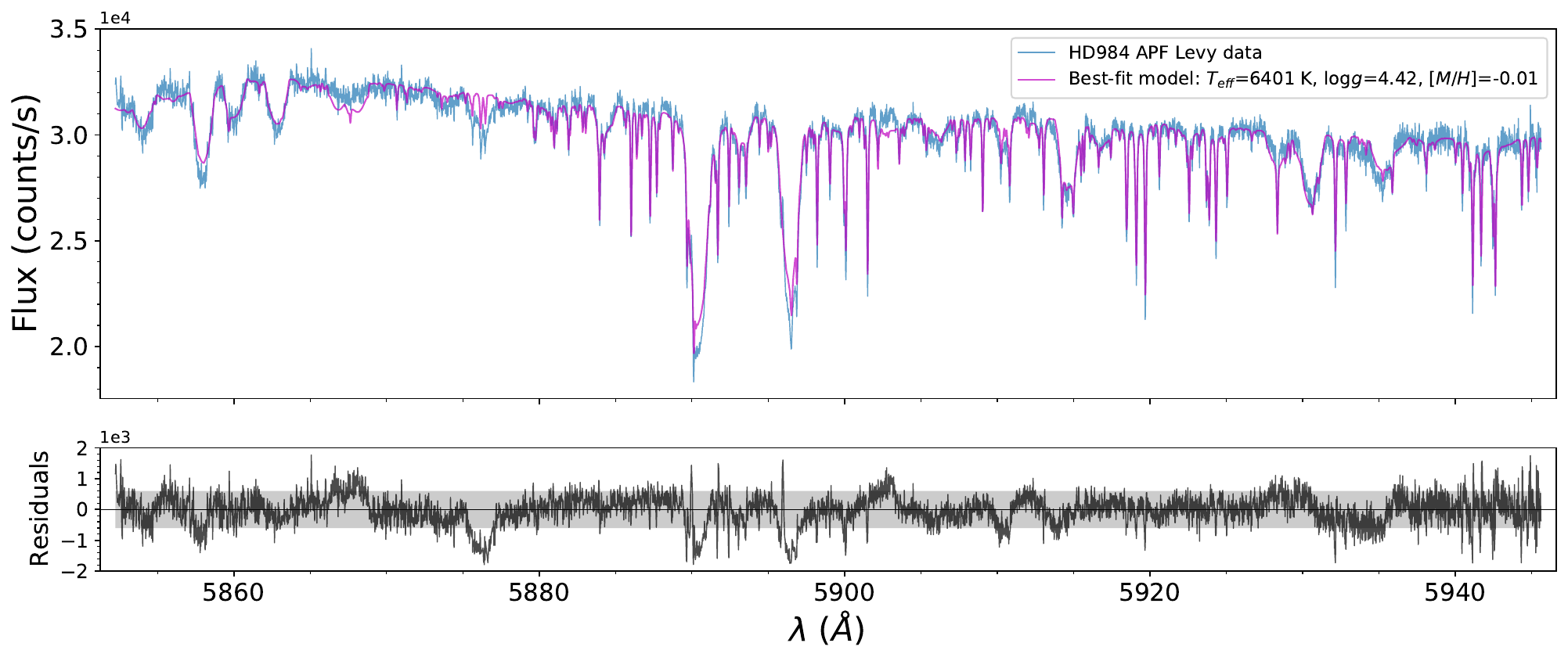}
    \caption{Best-fit PHOENIX model to the APF spectrum for the target HD 984 (cyan), shown for APF order \#79. The parameters of the best-fit PHOENIX model are estimated by computing the median for the effective temperature ($T_\mathrm{eff}$), surface gravity ($\log{g}$), and the metallicity ($\mathrm{[M/H]}$) over seven retained runs after performing multiple multi-order MCMC runs. This model has $T_\mathrm{eff}$ = 6401 K, $\log{g}$ = 4.42, $\mathrm{[M/H]}$ = -0.01 (magenta). The residuals between the data and the model are plotted in black and other noise limits are shown in grey}
    \label{fig:schmeatic}
\end{figure}

\begin{figure}[H]
    \centering
    \includegraphics[width=1.0\linewidth]{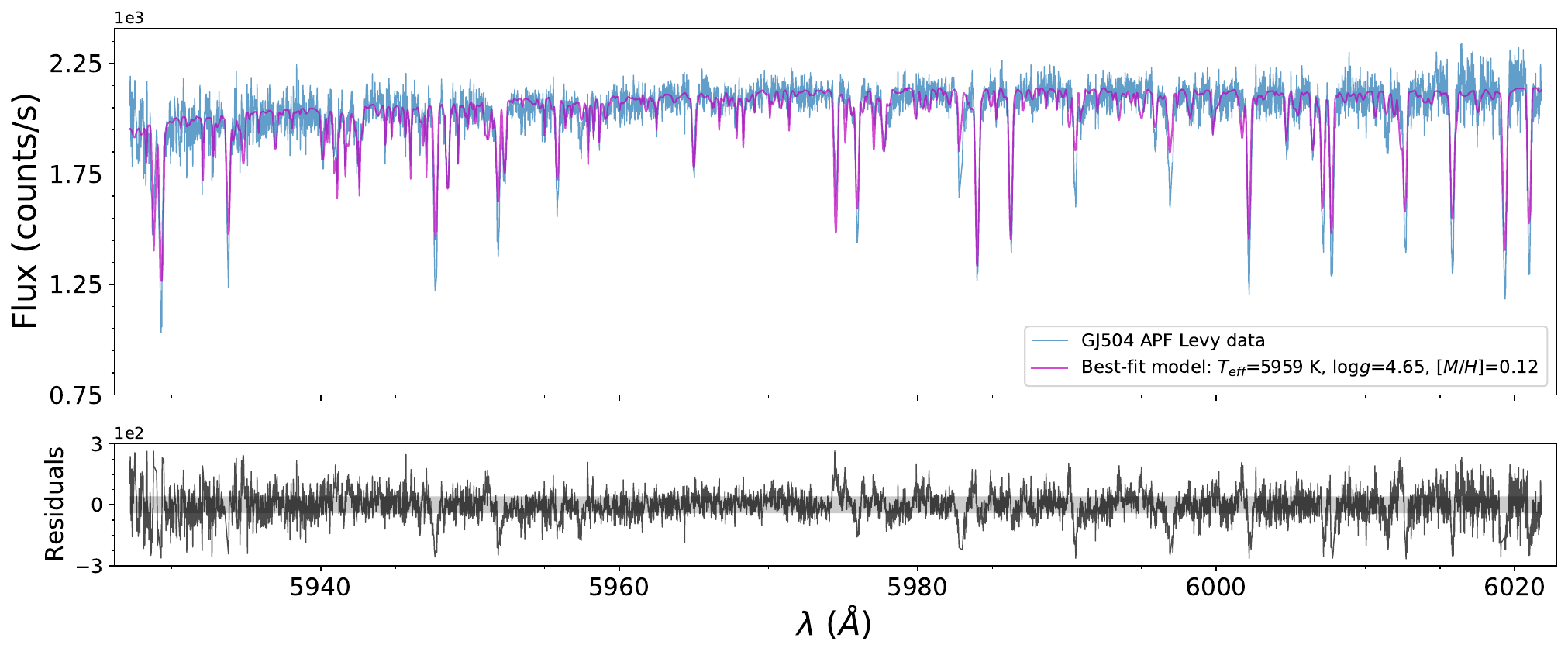}
    \caption{Best-fit PHOENIX model to the APF spectrum for the target GJ 504 (cyan), shown for APF order \#78. The parameters of the best-fit PHOENIX model are estimated by computing the median for the effective temperature ($T_\mathrm{eff}$), surface gravity ($\log{g}$), and the metallicity ($\mathrm{[M/H]}$) over seven retained runs after performing multiple multi-order MCMC runs. This model has $T_\mathrm{eff}$ = 5959 K, $\log{g}$ = 4.65, $\mathrm{[M/H]}$ = 0.12 (magenta). The residuals between the data and the model are plotted in black and other noise limits are shown in grey}
    \label{fig:schmeatic}
\end{figure}

\begin{figure}[H]
    \centering
    \includegraphics[width=1.0\linewidth]{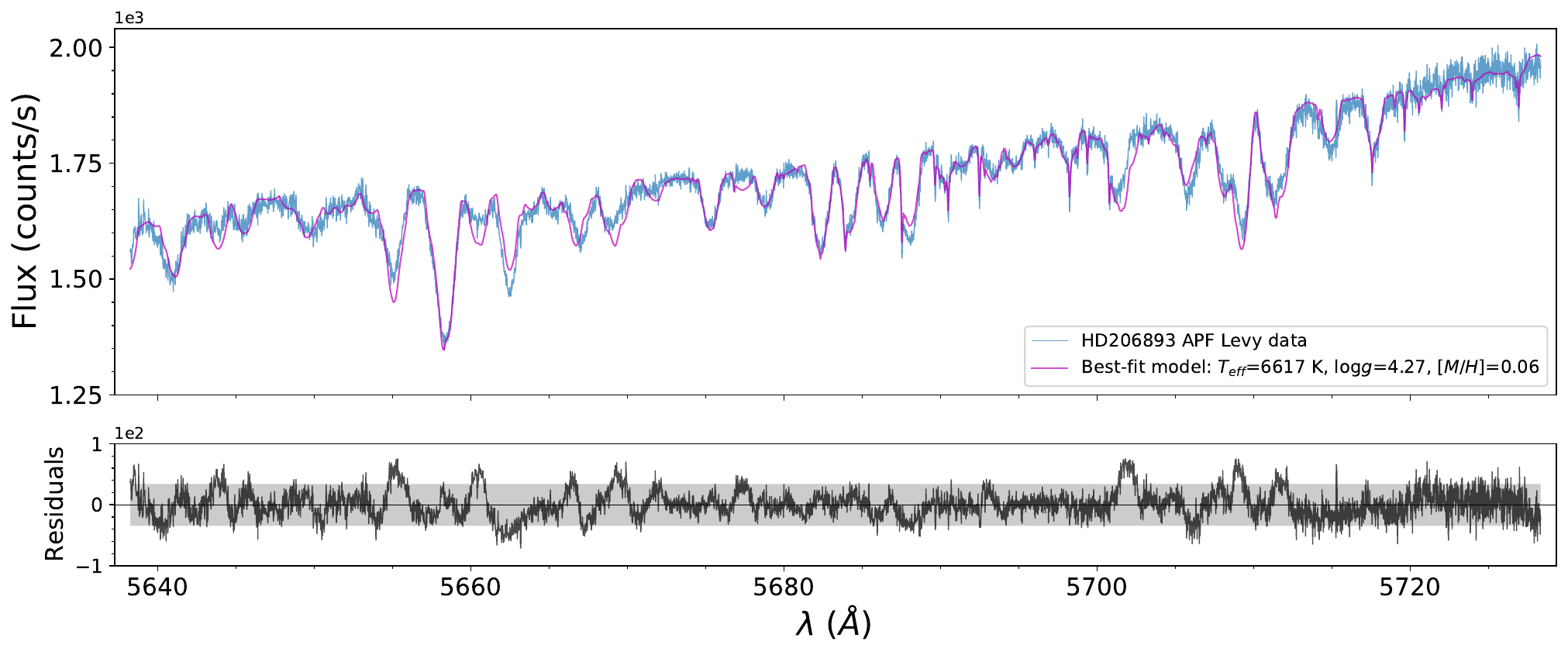}
    \caption{Best-fit PHOENIX model to the APF spectrum for the target HD 206893 (cyan), shown for APF order \#82. The parameters of the best-fit PHOENIX model are estimated by computing the median for the effective temperature ($T_\mathrm{eff}$), surface gravity ($\log{g}$), and the metallicity ($\mathrm{[M/H]}$) over 10 retained runs after performing multiple multi-order MCMC runs. This model has $T_\mathrm{eff}$ = 6617 K, $\log{g}$ = 4.27, $\mathrm{[M/H]}$ = 0.06 (magenta). The residuals between the data and the model are plotted in black and other noise limits are shown in grey}
    \label{fig:schmeatic}
\end{figure}

\begin{figure}[H]
    \centering
    \includegraphics[width=1.0\linewidth]{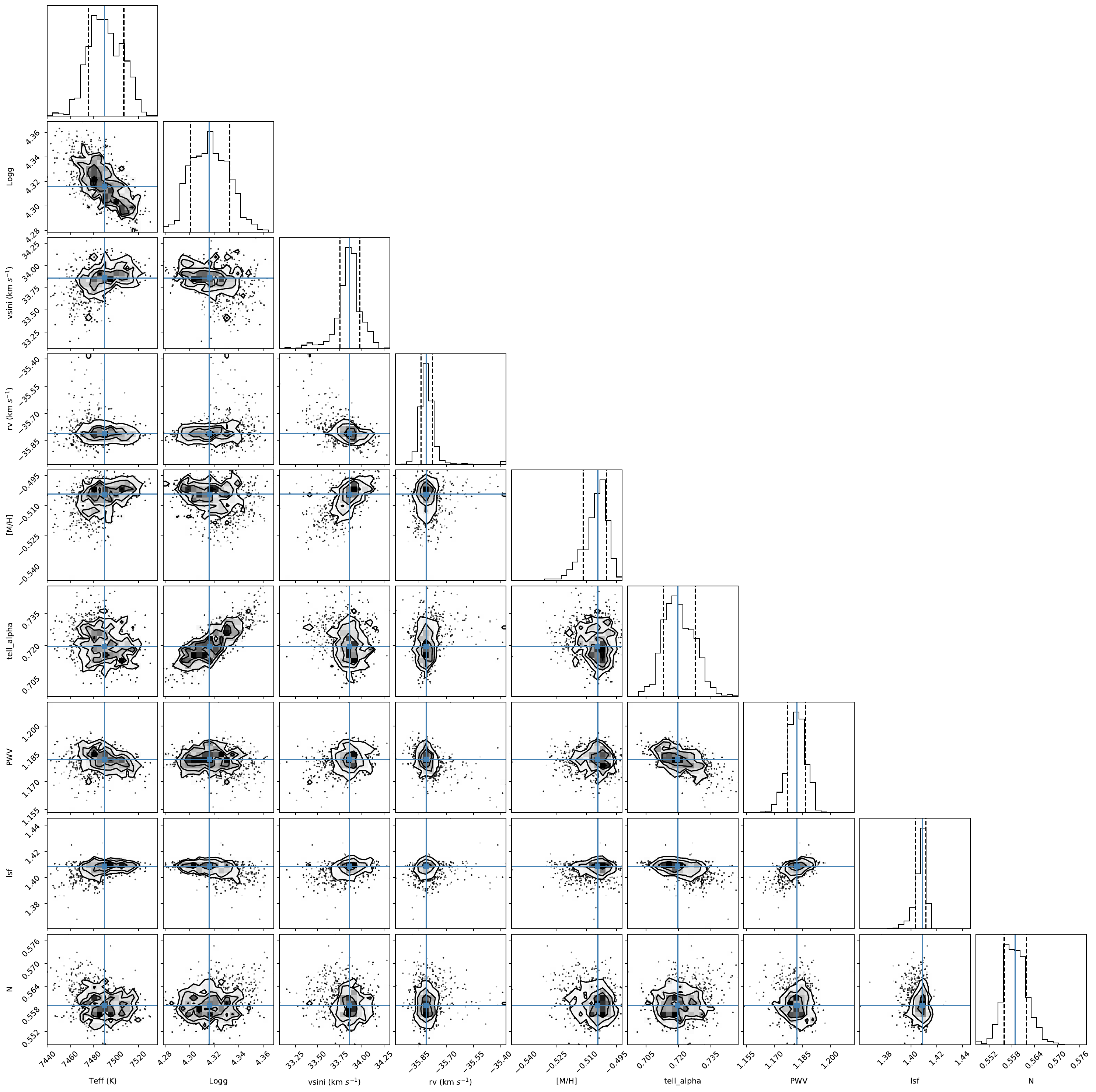}
    \caption{Corner plot for one of the retained runs for a multi-order fit of the PHOENIX grid to the spectrum of HR 8799. The marginalized posteriors are shown along the diagonal. The blue lines represent the 50 percentile, and the dotted lines represent the 16 and 84 percentiles. The subsequent covariances between all the parameters are in the corresponding 2-D histograms. This run gives a best-fit $T_\mathrm{eff}$ = 7490 K, $\log{g}$ = 4.32, $\mathrm{[M/H]}$ = -0.50.}
    \label{fig:schmeatic}
\end{figure}

\begin{figure}[H]
    \includegraphics[width=1.0\linewidth]{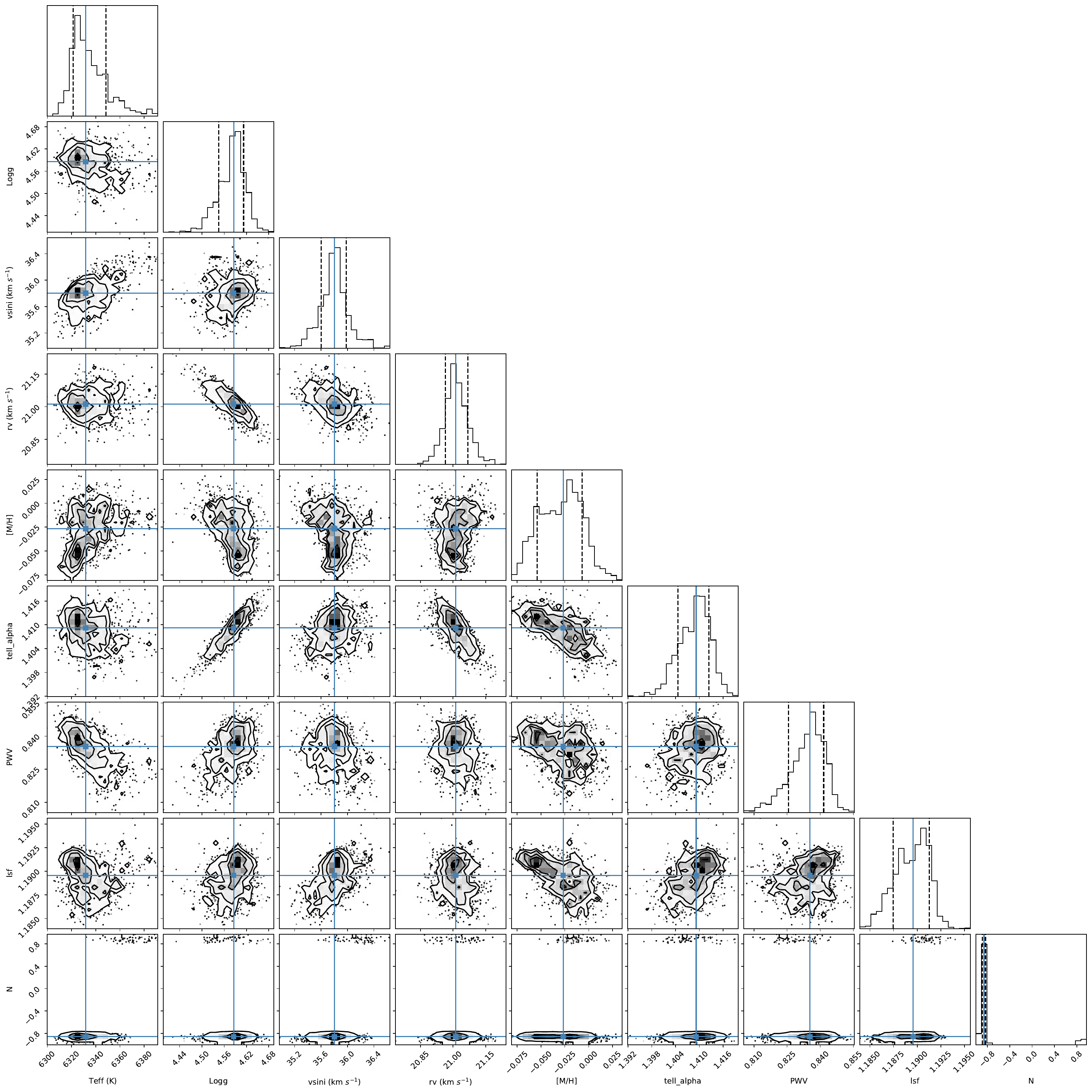}
    \caption{Corner plot for one of the retained runs for a multi-order fit of the PHOENIX grid to the spectrum of HD 984. The marginalized posteriors are shown along the diagonal. The blue lines represent the 50 percentile, and the dotted lines represent the 16 and 84 percentiles. The subsequent covariances between all the parameters are in the corresponding 2-D histograms. This run gives a best-fit $T_\mathrm{eff}$ = 6332 K, $\log{g}$ = 4.59, $\mathrm{[M/H]}$ = -0.03.}
    \label{fig:schmeatic}
\end{figure}

\begin{figure}[H]
    \centering
    \includegraphics[width=1.0\linewidth]{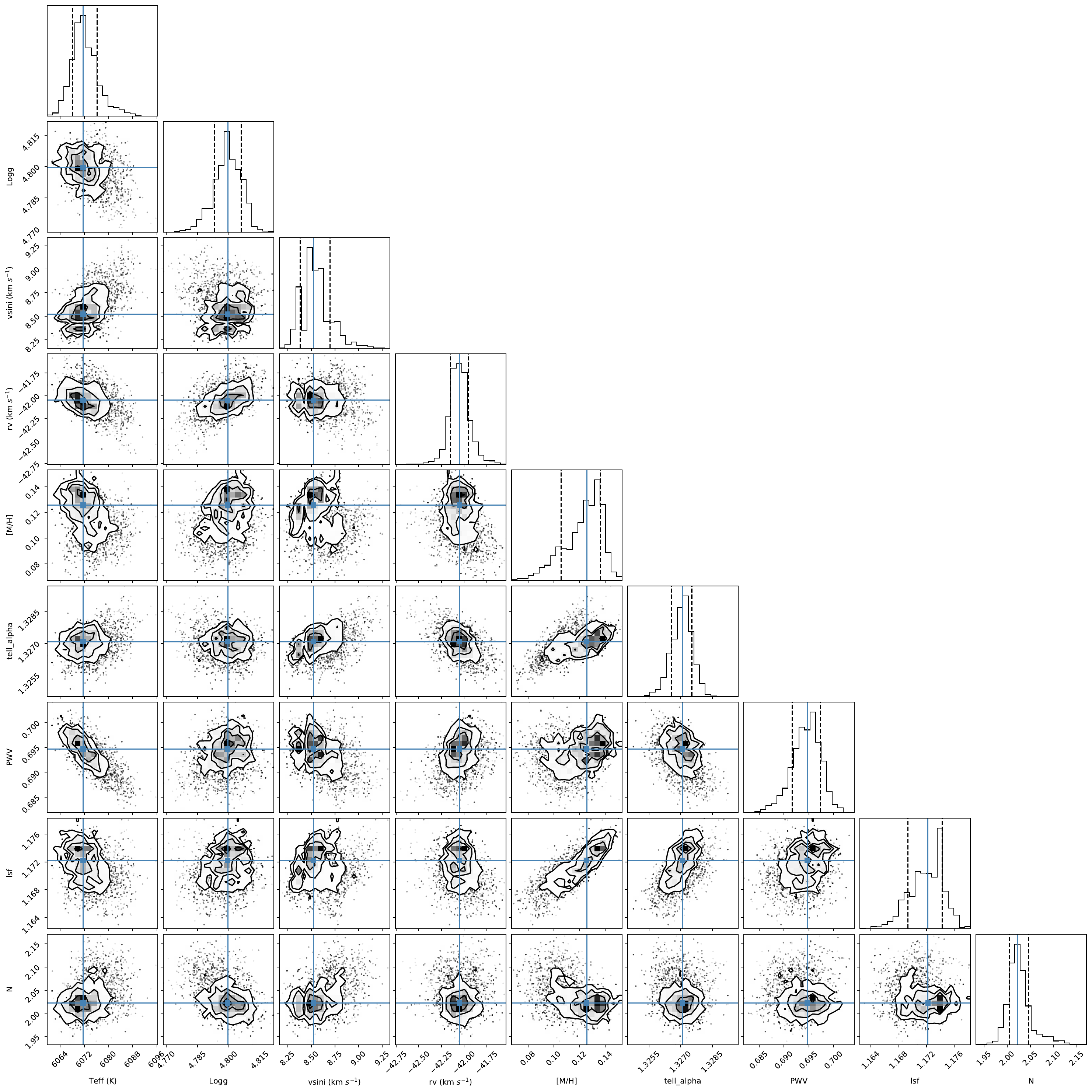}
    \caption{Corner plot for one of the retained runs for a multi-order fit of the PHOENIX grid to the spectrum of GJ 504. The marginalized posteriors are shown along the diagonal. The blue lines represent the 50 percentile, and the dotted lines represent the 16 and 84 percentiles. The subsequent covariances between all the parameters are in the corresponding 2-D histograms. This run gives a best-fit $T_\mathrm{eff}$ = 6072 K, $\log{g}$ = 4.80, $\mathrm{[M/H]}$ = 0.13.}
    \label{fig:schmeatic}
\end{figure}

\begin{figure}[H]
    \centering
    \includegraphics[width=1.0\linewidth]{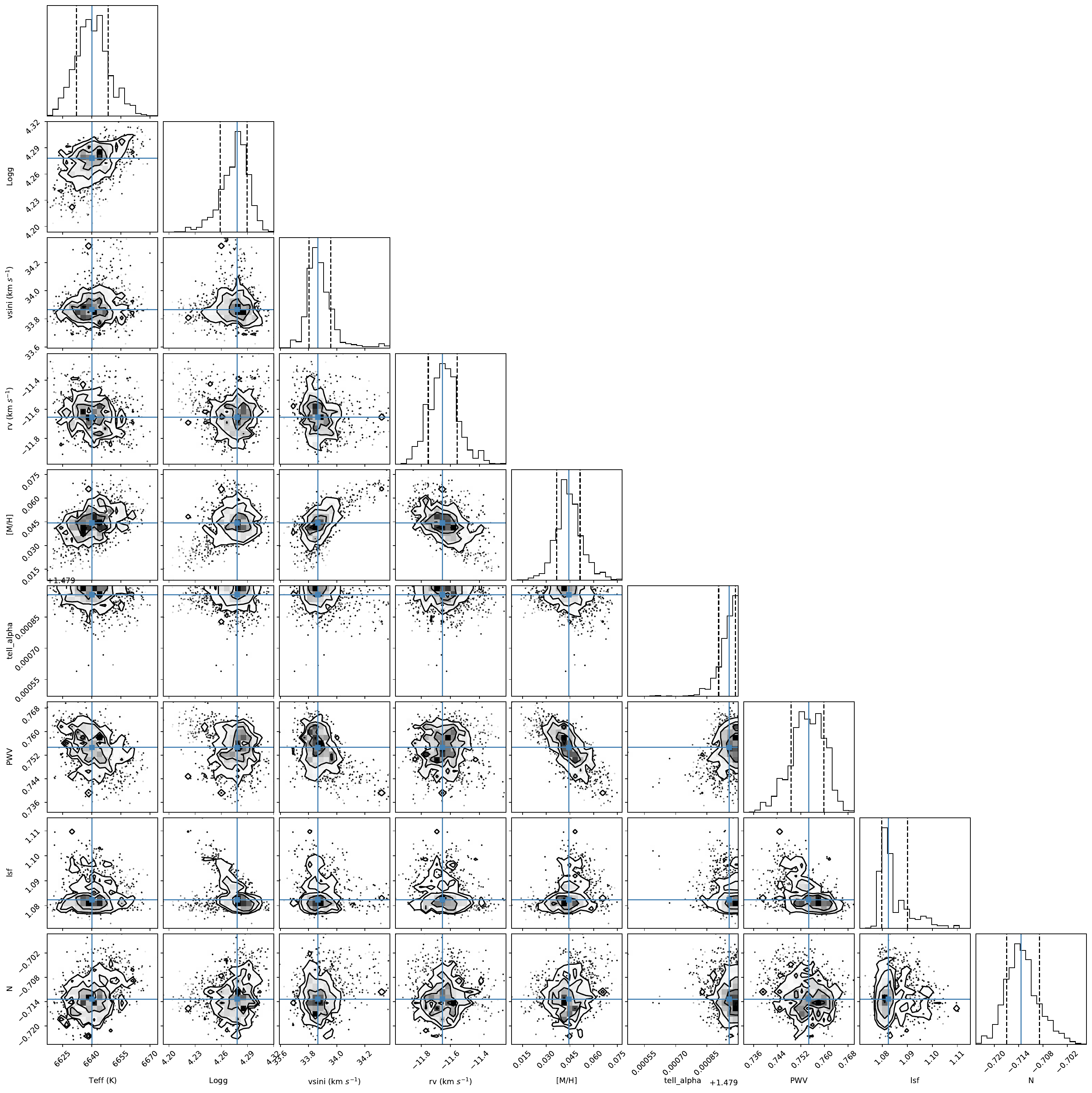}
    \caption{Corner plot for one of the retained runs for a multi-order fit of the PHOENIX grid to the spectrum of HD 206893. The marginalized posteriors are shown along the diagonal. The blue lines represent the 50 percentile, and the dotted lines represent the 16 and 84 percentiles. The subsequent covariances between all the parameters are in the corresponding 2-D histograms. This run gives a best-fit $T_\mathrm{eff}$ = 6640 K, $\log{g}$ = 4.28, $\mathrm{[M/H]}$ = 0.04.}
    \label{fig:schmeatic}
\end{figure}

\newpage

\section{Carbon and Oxygen Abundances MCMC Fit Plots}
\restartappendixnumbering

\begin{figure}[H]
    \centering
    \includegraphics[width=1.0\linewidth]{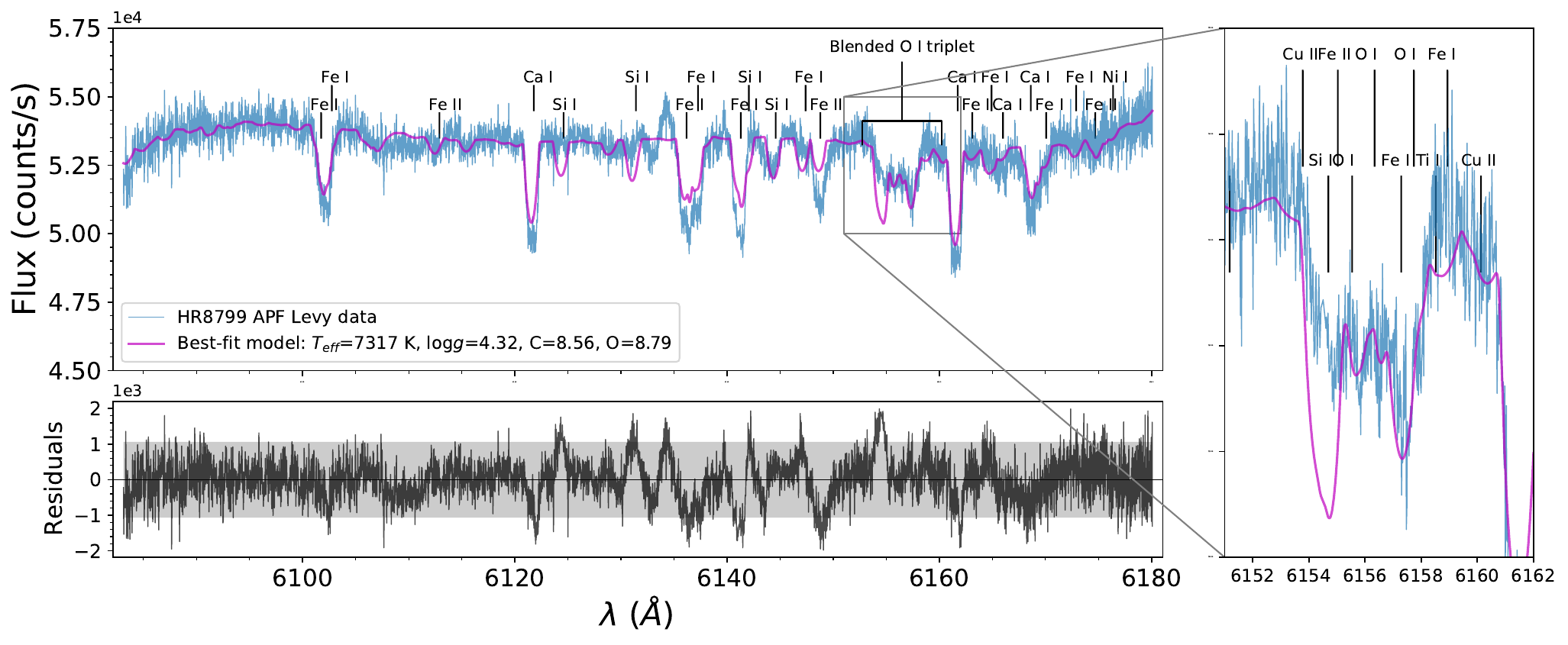}
    \caption{Best-fit \textit{PHOENIX}$\--$\textit{C/O} model to the APF spectrum for the target HR 8799 (cyan), shown for order \#76 containing an O I triplet feature at 6155$\--$6158 $\mathrm{\AA}$. The zoomed-in subplot on the right shows a $\sim$10$\mathrm{\AA}$ region around the blended feature. The best-fit \textit{PHOENIX}$\--$\textit{C/O} model has $T_\mathrm{eff}$ = 7317 K, $\log{g}$ = 4.32, $\log{\epsilon_C}$ = 8.56, $\log{\epsilon_O}$ = 8.79 (magenta). The residuals between the data and the model are plotted in black and other noise limits are shown in grey}
    \label{fig:schmeatic}
\end{figure}

\begin{figure}[H]
    \centering
    \includegraphics[width=1.0\linewidth]{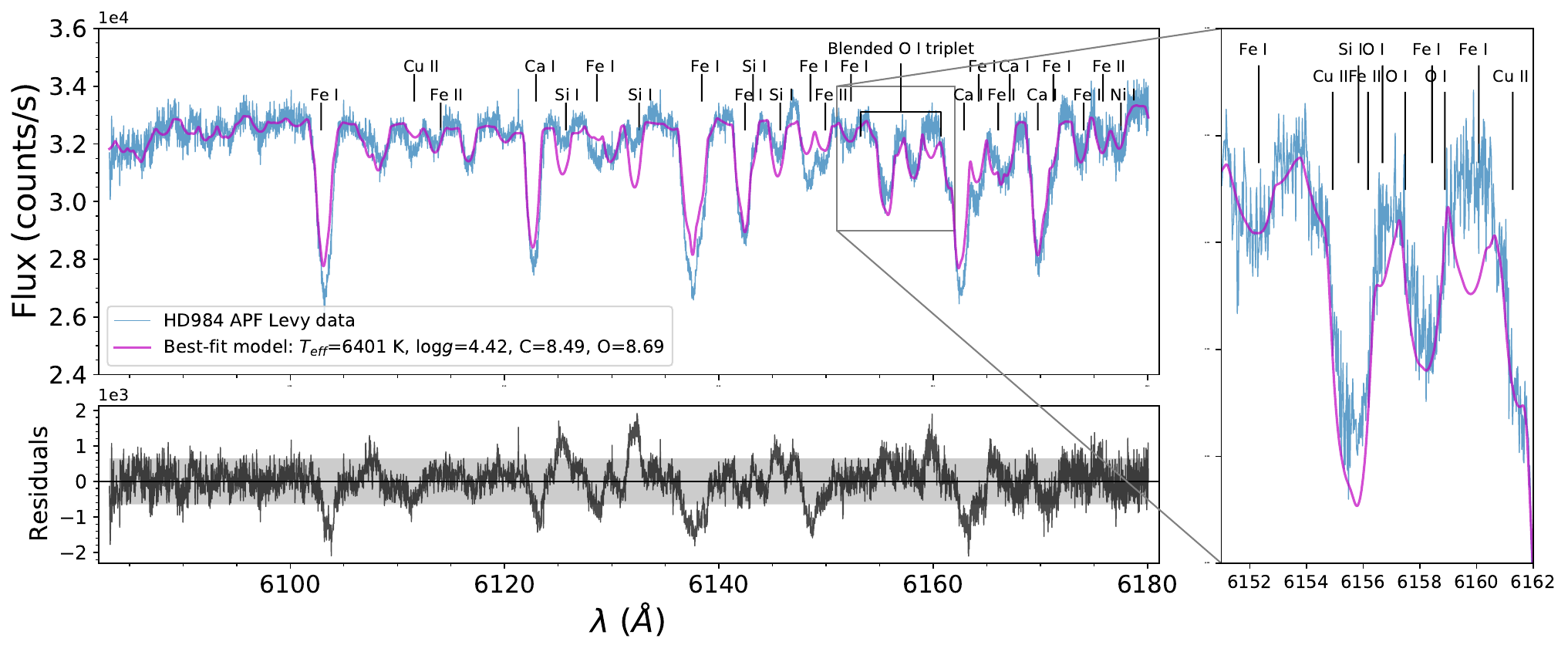}
    \caption{Best-fit \textit{PHOENIX}$\--$\textit{C/O} model to the APF spectrum for the target HD 984 (cyan), shown for order \#76 containing an O I triplet feature at 6155$\--$6158 $\mathrm{\AA}$. The zoomed-in subplot on the right shows a $\sim$10$\mathrm{\AA}$ region around the blended feature. The best-fit \textit{PHOENIX}$\--$\textit{C/O} model has $T_\mathrm{eff}$ = 6401 K, $\log{g}$ = 4.42, $\log{\epsilon_C}$ = 8.49, $\log{\epsilon_O}$ = 8.69 (magenta). The residuals between the data and the model are plotted in black and other noise limits are shown in grey}
    \label{fig:schmeatic}
\end{figure}

\begin{figure}[H]
    \centering
    \includegraphics[width=1.0\linewidth]{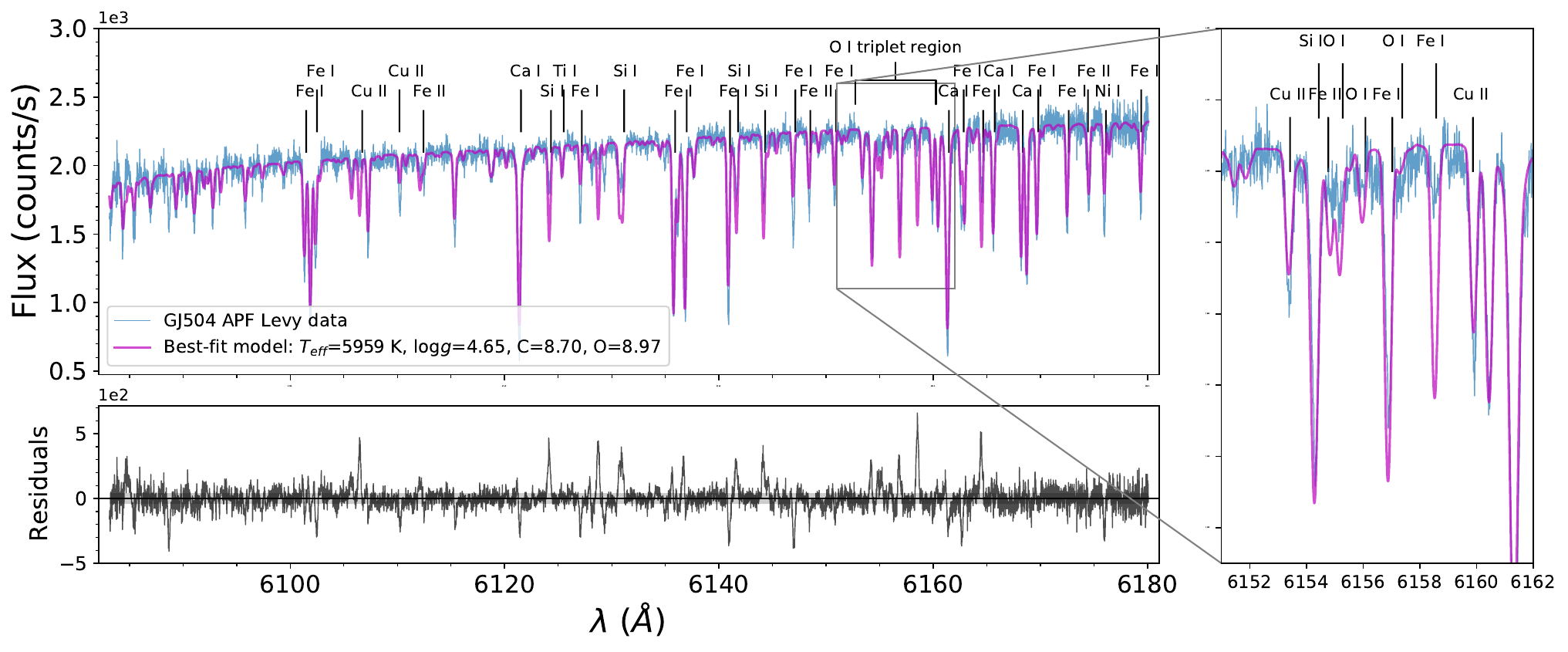}
    \caption{Best-fit \textit{PHOENIX}$\--$\textit{C/O} model to the APF spectrum for the target GJ 504 (cyan), shown for order \#76 containing an O I triplet feature at 6155$\--$6158 $\mathrm{\AA}$. The zoomed-in subplot on the right shows a $\sim$10$\mathrm{\AA}$ region around the slightly blended feature. The best-fit \textit{PHOENIX}$\--$\textit{C/O} model has $T_\mathrm{eff}$ = 5959 K, $\log{g}$ = 4.65, $\log{\epsilon_C}$ = 8.70, $\log{\epsilon_O}$ = 8.97 (magenta). The residuals between the data and the model are plotted in black and other noise limits are shown in grey}
    \label{fig:schmeatic}
\end{figure}

\begin{figure}[H]
    \centering
    \includegraphics[width=1.0\linewidth]{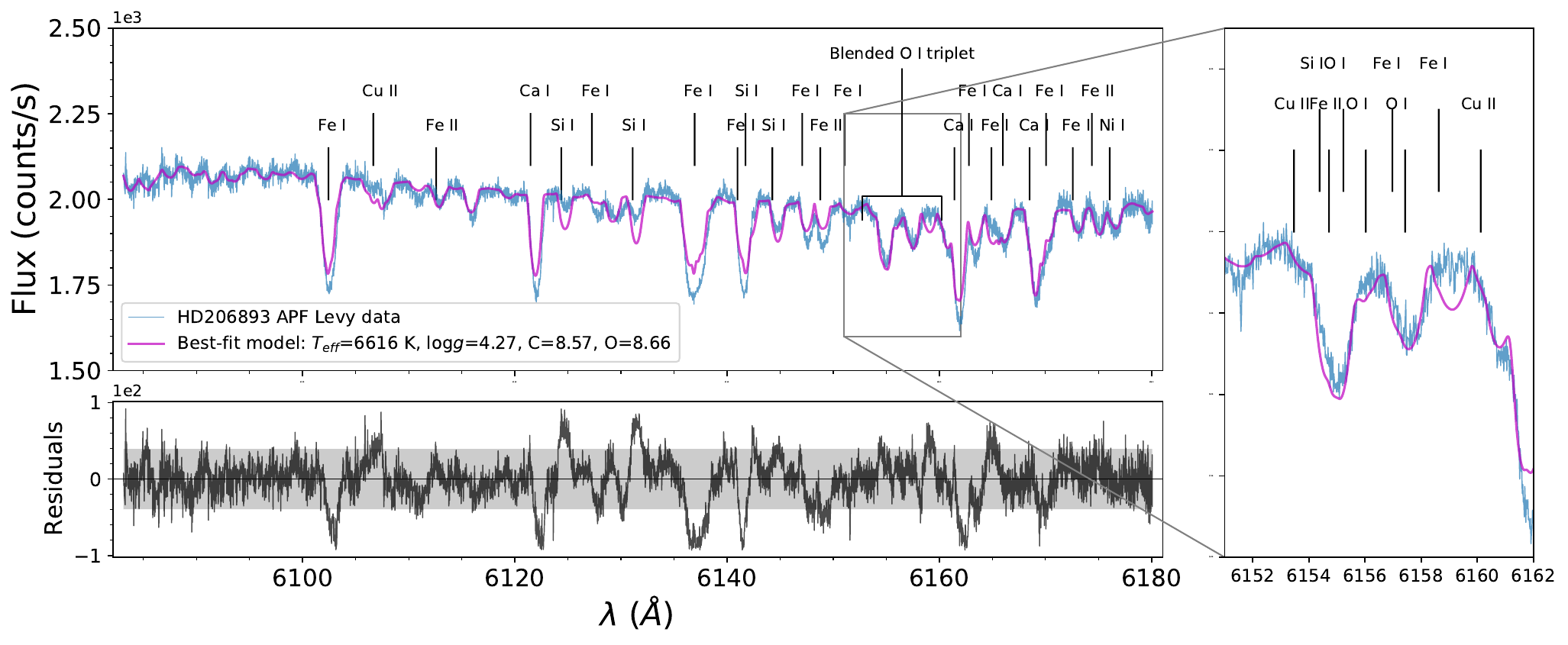}
    \caption{Best-fit \textit{PHOENIX}$\--$\textit{C/O} model to the APF spectrum for the target HD 206893 (cyan), shown for order \#76 containing an O I triplet feature at 6155$\--$6158 $\mathrm{\AA}$. The zoomed-in subplot on the right shows a $\sim$10$\mathrm{\AA}$ region around the blended feature. The best-fit \textit{PHOENIX}$\--$\textit{C/O} model has $T_\mathrm{eff}$ = 6617 K, $\log{g}$ = 4.27, $\log{\epsilon_C}$ = 8.57, $\log{\epsilon_O}$ = 8.66 (magenta). The residuals between the data and the model are plotted in black and other noise limits are shown in grey}
    \label{fig:schmeatic}
\end{figure}

\newpage
\begin{figure}[H]
    \centering
    \includegraphics[width=1.0\linewidth]{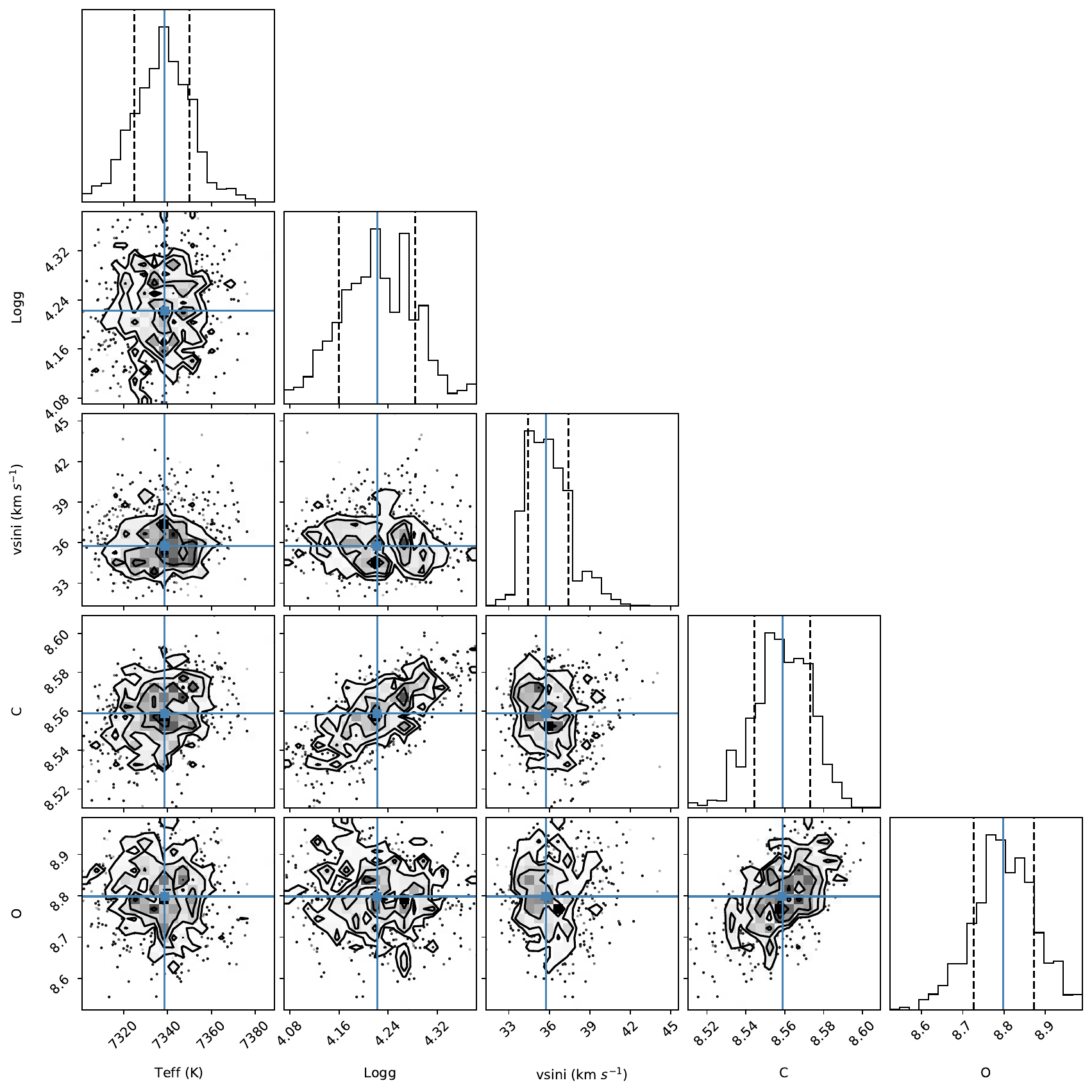}
    \caption{Corner plot for one of the retained runs for \textit{PHOENIX}$\--$\textit{C/O} grid fit to the spectrum of HR 8799. The marginalized posteriors are shown along the diagonal. The blue lines represent the 50 percentile, and the dotted lines represent the 16 and 84 percentiles. The subsequent covariances between all the parameters are in the corresponding 2-D histograms. This run is one of those performed with the carbon abundance fixed to 8.56 $\pm$ 0.04 (obtained in the carbon-only runs performed beforehand). We obtain a best-fit $\log{\epsilon_C}$ = 8.56, $\log{\epsilon_O}$ = 8.80 from this MCMC run.}
    \label{fig:schmeatic}
\end{figure}

\begin{figure}[H]
    \centering
    \includegraphics[width=1.0\linewidth]{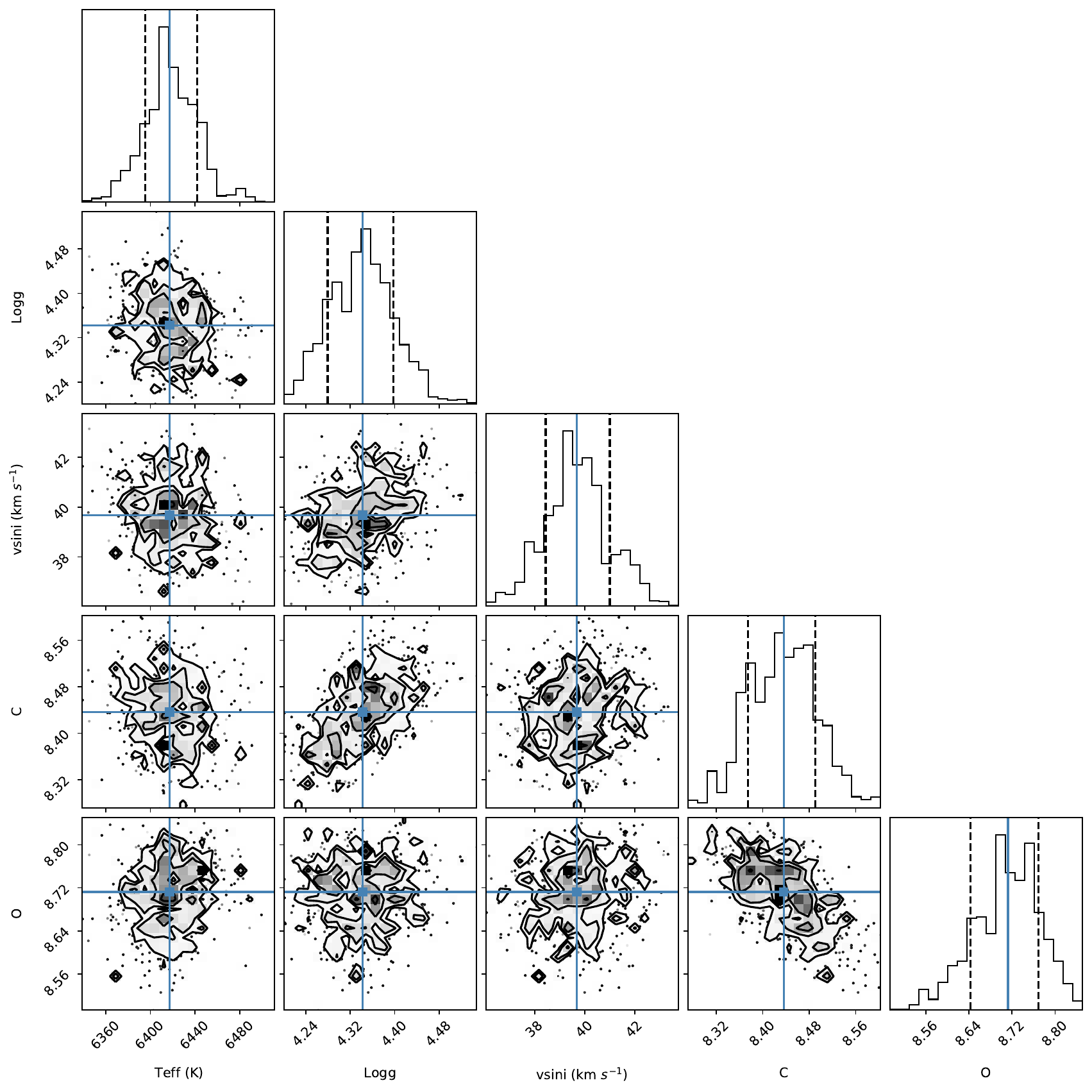}
    \caption{Corner plot for one of the retained runs for \textit{PHOENIX}$\--$\textit{C/O} grid fit to the spectrum of HD 984. The marginalized posteriors are shown along the diagonal. The blue lines represent the 50 percentile, and the dotted lines represent the 16 and 84 percentiles. The subsequent covariances between all the parameters are in the corresponding 2-D histograms. This run gives a best-fit $\log{\epsilon_C}$ = 8.44, $\log{\epsilon_O}$ = 8.71.}
    \label{fig:schmeatic}
\end{figure}

\begin{figure}[H]
    \centering
    \includegraphics[width=1.0\linewidth]{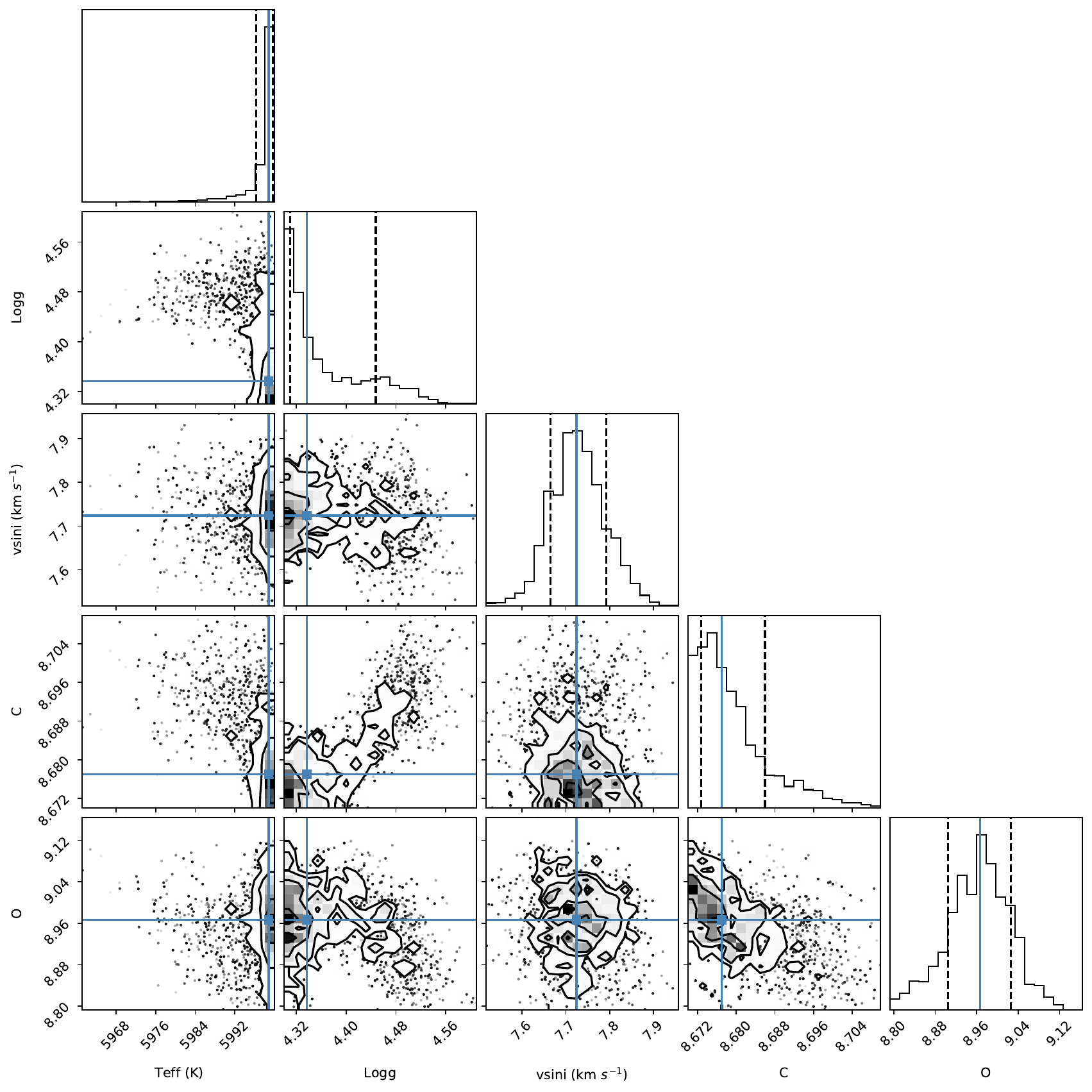}
    \caption{Corner plot for one of the retained runs for \textit{PHOENIX}$\--$\textit{C/O} grid fit to the spectrum of GJ 504. The marginalized posteriors are shown along the diagonal. The blue lines represent the 50 percentile, and the dotted lines represent the 16 and 84 percentiles. The subsequent covariances between all the parameters are in the corresponding 2-D histograms. This run is one of those performed with the carbon abundance fixed to 8.70 $\pm$ 0.03 (obtained in the carbon-only runs performed beforehand). We obtain a best-fit $\log{\epsilon_C}$ = 8.68, $\log{\epsilon_O}$ = 8.97 from this MCMC run.}
    \label{fig:schmeatic}
\end{figure}

\begin{figure}[H]
    \centering
    \includegraphics[width=1.0\linewidth]{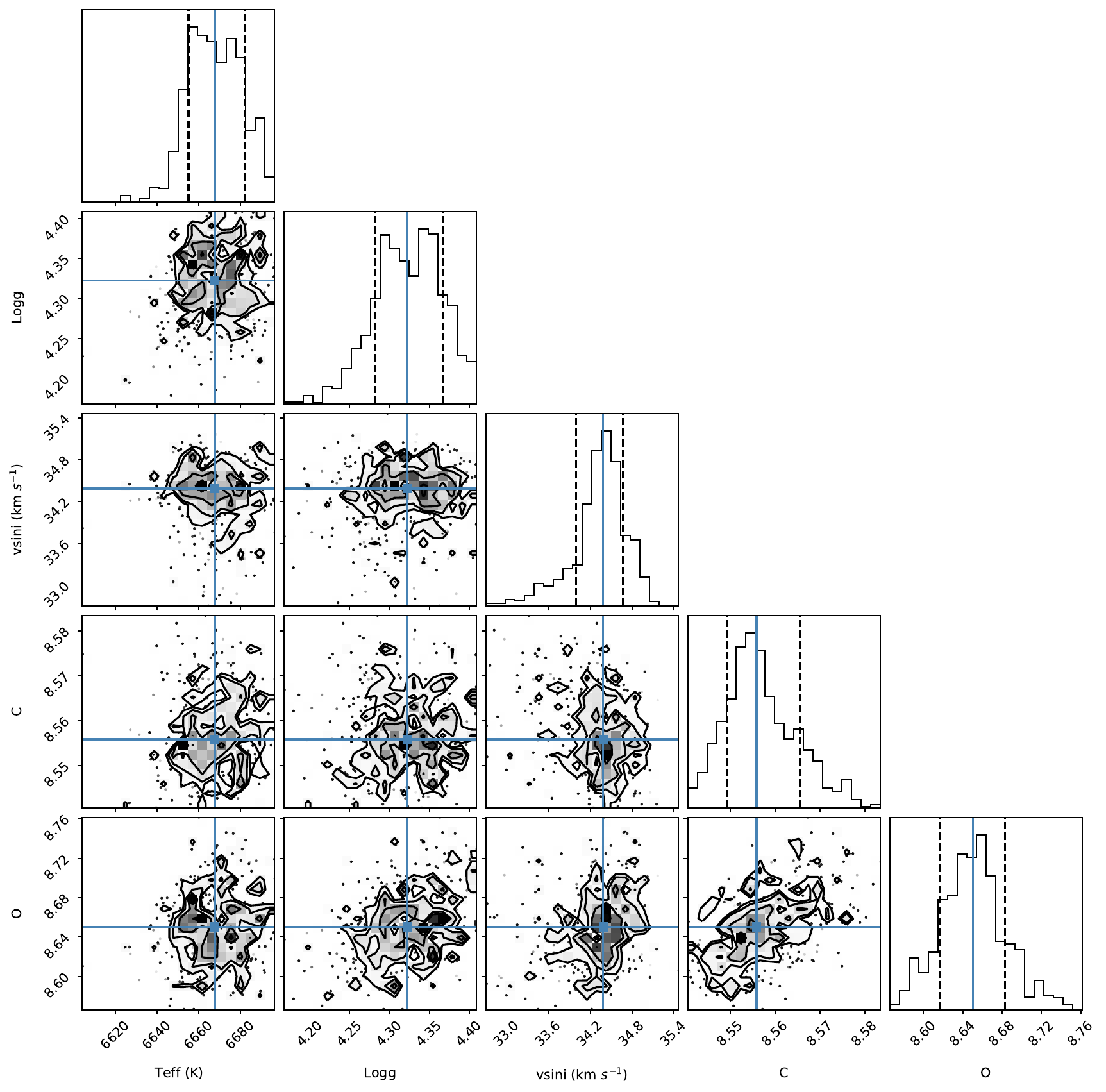}
    \caption{Corner plot for one of the retained runs for \textit{PHOENIX}$\--$\textit{C/O} grid fit to the spectrum of HD 206893. The marginalized posteriors are shown along the diagonal. The blue lines represent the 50 percentile, and the dotted lines represent the 16 and 84 percentiles. The subsequent covariances between all the parameters are in the corresponding 2-D histograms. This run is one of those performed with the carbon abundance fixed to 8.57 $\pm$ 0.03 (obtained in the carbon-only runs performed beforehand). We obtain a best-fit $\log{\epsilon_C}$ = 8.56, $\log{\epsilon_O}$ = 8.65 from this MCMC run.}
    \label{fig:schmeatic}
\end{figure}

\newpage

\section{Equivalent Widths}
\restartappendixnumbering

\startlongtable
\begin{deluxetable}{ccccccccc}
\tablecaption{Equivalent widths of elemental lines \label{tab:lineabundance}}
\tablewidth{0pt}
\tablehead{
\colhead{Ion} & \colhead{Wavelength} & \colhead{$\chi$} & \colhead{$\log{gf}$} & \colhead{HR 8799} & \colhead{51 Eri} & \colhead{HD 984} & \colhead{GJ 504} & \colhead{HD 206893} \\
\colhead{} & \colhead{(\AA)} & \colhead{(eV)} & \colhead{} & \colhead{(m\AA)} & \colhead{(m\AA)} & \colhead{(m\AA)} & \colhead{(m\AA)} & \colhead{(m\AA)}
}
\startdata
{C I}  & {4771.733} &  {7.488} &  {-1.87}  &  {64.5\tablenotemark{a}}   &   {105.0\tablenotemark{a}}   &  {214.1\tablenotemark{a}}    &  {76.9\tablenotemark{a}}     &  {182.3\tablenotemark{a}}   \\
{}     & {4932.026} & {7.685}  & {-1.66}   &  {60.0}    &   {}   &  {38.3\tablenotemark{a}}    &  {54.7\tablenotemark{a}}     &  {50.3\tablenotemark{a}}   \\
{}     & {5052.149} & {7.685}  & {-1.30}   &  {142.6\tablenotemark{a}}  &   {366.4\tablenotemark{a}}   &  {235.5\tablenotemark{a}}    &  {57.5}     &  {245.6\tablenotemark{a}}   \\
{}     & {5380.331} & {7.685}  &  {-1.62}  &  {82.9\tablenotemark{a}}   &   {266.2\tablenotemark{a}}   &  {160.7\tablenotemark{a}}    &  {41.4}     &  {}   \\
{}     & {6587.608} & {8.537}  &  {-1.00}  &  {68.6}    &   {67.5\tablenotemark{a}}   &  {57.4\tablenotemark{a}}    &  {41.2}     &  {59.8\tablenotemark{a}}   \\
{O I}  & {6300.304} &  {0.000} &  {-9.776}      &  {}    &   {}   &  {}    &  {6.2}     &  {}   \\
{}     & {6155.98}  &  {10.740}  &  {-1.011}    &  \multirow{3}{*}{111.9\tablenotemark{b}}    &   \multirow{3}{*}{164.5\tablenotemark{b}}    &  \multirow{3}{*}{192.4\tablenotemark{b}}   &  {24.0\tablenotemark{a}}     &  \multirow{3}{*}{192.4\tablenotemark{b}}   \\
{}     & {6156.77}  &  {10.740}  &  {-0.694}    &    &    &     &  {9.6}     &     \\
{}     & {6158.18}  &  {10.741}  &  {-0.409}    &    &    &     &  {14.8\tablenotemark{a}}     &     \\
{Na I} & {5682.633} & {2.102} & {-0.706} & {26.6} & {50.6} & {82.7} & {121.9} & {97.3} \\
{Mg I} & {4702.991} & {4.346} & {-0.440} & {119.2} & {190.5} & {287.6} & {355.1} & {225.4} \\
{}     & {4730.029} & {4.346} & {-2.347} & {} & {} & {} & {79.5} & {} \\
{}     & {5172.684} & {2.712} & {-0.393} & {} & {438.3} & {873.6} & {} & {508.4} \\
{}     & {5183.604} & {2.717} & {-0.167} & {272.6} & {397.9} & {927.0} & {1002.8} & {633.2} \\
{}     & {5528.405} & {4.346} & {-0.498} & {103.4} & {} & {239.8} & {310.3} & {208.4} \\
{}     & {5711.088} & {4.346} & {-1.724} & {16.7} & {59.0} & {} & {119.0} & {} \\
{}     & {6841.19}  & {5.75}  & {-1.720} & {} & {25.0} & {53.8} & {} & {} \\
{Si I} & {5690.425} & {4.930}  & {-1.870} & {} & {} & {37.5} & {60.7} & {36.9} \\
{}     & {5701.105} & {4.930}  & {-2.050} & {11.3} & {24.3} & {} & {43.4} & {} \\
{}     & {5708.397} & {4.954}  & {-1.470} & {29.6} & {} & {} & {94.7} & {} \\
{}     & {5772.145} & {5.082}  & {-1.750} & {10.9} & {21.8} & {50.7} & {65.0} & {31.9} \\
{}     & {6125.021} & {5.61}  & {-1.464} & {} & {} & {26.6} & {37.7} & {22.2} \\
{}     & {6142.487} & {5.62}  & {-1.295} & {} & {} & {} & {48.4} & {} \\
{}     & {6145.015} & {5.62}  & {-1.310} & {14.9} & {} & {43.0} & {55.3} & {37.9} \\
{}     & {6243.813} & {5.62}  & {-1.242} & {} & {} & {} & {68.3} & {} \\
{}     & {6244.468} & {5.62}  & {-1.093} & {} & {} & {} & {62.6} & {} \\
{}     & {6414.98}  & {5.87}  & {-1.035} & {} & {} & {42.5} & {79.8} & {46.3} \\
{}     & {6741.63}  & {5.98}  & {-1.428} & {} & {} & {} & {33.4} & {} \\
{}     & {6848.568} & {5.86}  & {-1.524} & {} & {} & {} & {} & {} \\
{}     & {7003.567} & {5.964} & {-0.754} & {} & {} & {} & {104.7} & {} \\
{Si II} & {6347.109} & {8.121} & {0.23}   & {94.9} & {} & {77.3} & {98.7} & {98.4} \\
{}     & {6371.371} & {8.121} & {-0.08}  & {64.6} & {} & {63.6} & {69.9} & {66.3} \\
{S I}  & {5706.110} & {7.870} & {-0.93}  &  {24.0\tablenotemark{a}} & {69.6\tablenotemark{a}} & {109.7\tablenotemark{a}} & {109.4\tablenotemark{a}} & {105.4\tablenotemark{a}} \\
{}     & {6046.040} & {7.868} & {-0.959}  & {10.3\tablenotemark{a}} & {19.5\tablenotemark{a}} & {14.1\tablenotemark{a}} & {22.3\tablenotemark{a}} & {16.0\tablenotemark{a}} \\
{}     & {6052.660} & {7.870} & {-0.672}  & {14.7\tablenotemark{a}} & {} & {} & {19.6\tablenotemark{a}} & {16.6\tablenotemark{a}} \\
{}     & {6743.580} & {7.866} & {-1.070}  & {} & {} & {} & {25.9\tablenotemark{a}} & {} \\
{Ca I}  & {4425.437} & {1.879} & {-0.358} & {63.5} & {108.2} & {150.3} & {177.6} & {141.8} \\
{}     & {4578.551} & {2.521} & {-0.558} & {15.4} & {} & {91.4} & {113.5} & {75.6} \\
{}     & {4585.655} & {2.526} & {-0.187} & {33.8} & {84.8} & {113.1} & {155.9} & {130.5} \\
{}     & {5581.965} & {2.523} & {-0.71}  & {21.6} & {50.7} & {81.1} & {118.0} & {81.5} \\
{}     & {5588.749} & {2.526} & {0.21}   & {84.2} & {} & {} & {206.9} & {132.7} \\
{}     & {5590.114} & {2.521} & {-0.71}  & {} & {} & {} & {132.0} & {79.5} \\
{}     & {5594.462} & {2.523} & {-0.05}  & {75.2} & {108.0} & {} & {236.0} & {159.8} \\
{}     & {5601.277} & {2.526} & {-0.69}  & {29.4} & {} & {} & {133.4} & {} \\
{}     & {5857.451} & {2.932} & {0.23}   & {56.7} & {92.0} & {158.6} & {184.2} & {128.3} \\
{}     & {6122.217} & {1.886} & {-0.315} & {79.0} & {119.8} & {177.3} & {243.4} & {146.4} \\
{}     & {6162.173} & {1.899} & {-0.089} & {77.8} & {120.6} & {} & {283.3} & {} \\
{}     & {6439.075} & {2.526} & {0.47}   & {91.9} & {132.2} & {171.0} & {243.4} & {151.5} \\
{}     & {6449.808} & {2.521} & {-0.55}  & {28.1} & {59.8} & {130.3} & {} & {100.0} \\
{}     & {6717.681} & {2.709} & {-0.61}  & {23.9} & {50.4} & {96.2} & {154.5} & {81.4} \\
{Sc II} & {5031.021} & {1.357} & {-0.40}   & {33.1} & {64.2} & {84.7} & {106.8} & {85.0} \\
{}     & {5526.790} & {1.768} & {0.02}   & {48.0} & {} & {} & {112.7} & {82.2} \\
{}     & {5667.149} & {1.500} & {-1.31}  & {} & {17.4} & {43.3} & {49.6} & {45.2} \\
{}     & {5669.042} & {1.500} & {-1.20}  & {} & {32.1} & {30.5} & {65.1} & {38.4} \\
{}     & {6245.637} & {1.507} & {-1.03}  & {} & {} & {} & {60.2} & {} \\
{}     & {6320.851} & {1.500} & {-1.82}  & {} & {} & {} & {20.5} & {} \\
{}     & {6604.601} & {1.357} & {-1.31}  & {} & {} & {42.9} & {60.3} & {26.4} \\
{Ti I} & {5022.866} & {0.826} & {-0.434} & {} & {} & {} & {86.9} & {} \\
{}     & {5024.844} & {0.818} & {-0.602} & {} & {} & {68.0\tablenotemark{a}} & {83.6} & {47.1\tablenotemark{a}} \\
{}     & {5039.955} & {0.021} & {-1.068} & {155.9\tablenotemark{a}} & {405.3\tablenotemark{a}} & {673.1\tablenotemark{a}} & {100.6} & {558.4\tablenotemark{a}} \\
{}     & {5064.652} & {0.048} & {-0.929} & {} & {} & {} & {104.3} & {} \\
{}     & {5210.384} & {0.048} & {-0.85}  & {18.2\tablenotemark{a}} & {34.4\tablenotemark{a}} & {87.1\tablenotemark{a}} & {114.8} & {60.1\tablenotemark{a}} \\
{}     & {5866.449} & {1.067} & {-0.840} & {} & {10.3} & {30.7} & {60.3} & {23.2} \\
{}     & {6258.099} & {1.443} & {-0.355} & {} & {} & {58.6\tablenotemark{a}} & {71.5} & {47.5\tablenotemark{a}} \\
{}     & {6261.096} & {1.430} & {-0.479} & {} & {} & {19.3} & {66.2} & {} \\
{Ti II} & {4450.482} & {1.084} & {-1.45}  & {84.0} & {} & {} & {} & {} \\
{}     & {4501.273} & {1.116} & {-0.75}  & {} & {} & {} & {168.4} & {} \\
{}     & {4563.761} & {1.221} & {-0.96}  & {} & {} & {} & {149.3} & {} \\
{}     & {4571.968} & {1.572} & {-0.53}  & {} & {} & {} & {187.3} & {} \\
{}     & {4779.985} & {2.048} & {-1.37}  & {46.9} & {73.7} & {87.0} & {84.3} & {86.5} \\
{}     & {4798.521} & {1.080} & {-2.43}  & {21.0} & {} & {74.0} & {} & {59.7} \\
{}     & {4805.085} & {2.061} & {-1.12}  & {65.4} & {81.9} & {} & {} & {} \\
{}     & {5154.068} & {1.566} & {-1.92}  & {} & {78.5} & {} & {96.5} & {} \\
{}     & {5185.913} & {1.893} & {-1.35}  & {42.7} & {53.8} & {103.1} & {85.0} & {76.8} \\
{}     & {5336.771} & {1.582} & {-1.70}  & {49.8} & {75.0} & {80.3} & {92.4} & {82.1} \\
{}     & {5381.022} & {1.566} & {-1.921} & {} & {68.6} & {73.7} & {80.8} & {} \\
{}     & {5418.751} & {1.582} & {-1.999} & {21.3} & {} & {44.4} & {63.4} & {54.7} \\
{}     & {5490.690} & {1.566} & {-2.65}  & {10.3} & {} & {31.3} & {34.6} & {23.7} \\
{Cr I} & {4646.148} & {1.03}  & {-0.71}  & {40.5} & {241.0\tablenotemark{a}} & {356.4\tablenotemark{a}} & {158.3} & {105.3} \\
{}     & {4652.152} & {1.004} & {-1.03}  & {34.6} & {49.8} & {155.8\tablenotemark{a}} & {118.5} & {129.6\tablenotemark{a}} \\
{}     & {5409.772} & {1.030} & {-0.72}  & {} & {} & {} & {174.4} & {} \\
{}     & {5702.310} & {3.450} & {-0.667} & {} & {} & {} & {30.0} & {} \\
{}     & {5783.060} & {3.320} & {-0.500} & {} & {} & {} & {41.2} & {} \\
{}     & {5783.850} & {3.320} & {-0.295} & {} & {} & {} & {53.8} & {} \\
{}     & {5787.920} & {3.320} & {-0.083} & {} & {} & {} & {57.7} & {} \\
{}     & {6330.090} & {0.940} & {-2.920} & {} & {} & {} & {29.1} & {} \\
{}     & {7400.250} & {2.900} & {-0.111} & {} & {} & {} & {123.9} & {} \\
{Cr II} & {4558.650} & {4.073} & {-0.66}  & {91.7} & {} & {96.0} & {113.8} & {109.2} \\
{}     & {4588.199} & {4.071} & {-0.64}  & {80.2} & {102.7} & {90.7} & {102.9} & {87.1} \\
{}     & {4634.073} & {4.072} & {-1.24}  & {49.6} & {} & {81.6} & {80.8} & {72.9} \\
{}     & {4812.337} & {3.864} & {-1.8}   & {22.8} & {41.4} & {36.2} & {48.6} & {31.9} \\
{}     & {5237.322} & {4.073} & {-1.16}  & {39.3} & {64.2} & {60.0} & {68.1} & {59.4} \\
{}     & {5308.408} & {4.071} & {-1.81}  & {} & {} & {19.0} & {39.5} & {22.4} \\
{}     & {5310.687} & {4.072} & {-2.28}  & {} & {} & {} & {22.2} & {} \\
{}     & {5313.581} & {4.073} & {-1.65}  & {23.9} & {48.7} & {36.6} & {50.7} & {40.5} \\
{}     & {5334.869} & {4.072} & {-1.562} & {21.6} & {} & {38.8} & {53.1} & {38.3} \\
{}     & {5407.604} & {3.827} & {-2.088} & {} & {} & {40.0} & {} & {19.8} \\
{}     & {5478.365} & {4.177} & {-1.908} & {12.6} & {} & {} & {} & {39.7} \\
{}     & {5508.606} & {4.156} & {-2.12}  & {} & {} & {} & {} & {11.2} \\
{Mn I} & {4754.042} & {2.282} & {-0.085} & {35.7} & {75.5} & {100.3} & {137.4} & {95.7} \\
{}     & {4783.427} & {2.298} & {0.042} & {42.4} & {85.0} & {144.6} & {165.6} & {114.9} \\
{}     & {5399.500} & {3.850} & {-0.287} & {} & {} & {} & {48.3} & {} \\
{}     & {5432.550} & {0.000} & {-3.795} & {} & {} & {} & {44.1} & {} \\
{}     & {6021.819} & {3.075} & {0.035}  & {} & {} & {59.8} & {121.3} & {56.0} \\
{Fe I} & {4404.750} & {1.557} & {-0.142} & {182.4} & {220.1} & {483.0} & {606.6} & {383.0} \\
{}     & {4466.551} & {2.832} & {-0.60}  & {92.1} & {114.4} & {159.3} & {191.3} & {146.6} \\
{}     & {4484.219} & {3.602} & {-0.72}  & {32.3} & {} & {98.1} & {122.4} & {88.9} \\
{}     & {4485.675} & {3.686} & {-1.00}  & {15.6} & {} & {70.7} & {108.5} & {63.7} \\
{}     & {4494.563} & {2.198} & {-1.136} & {87.3} & {126.9} & {176.8} & {227.5} & {135.1} \\
{}     & {4528.613} & {2.176} & {-0.822} & {104.8} & {119.3} & {} & {322.4} & {} \\
{}     & {4607.647} & {3.266} & {-1.545} & {16.0} & {39.1} & {83.8} & {103.7} & {71.5} \\
{}     & {4611.284} & {3.654} & {-0.670} & {40.4} & {65.7} & {115.6} & {155.3} & {104.0} \\
{}     & {4625.044} & {3.241} & {-1.35}  & {24.0} & {45.5} & {77.6} & {111.9} & {69.4} \\
{}     & {4632.911} & {1.608} & {-2.913} & {} & {} & {90.4} & {130.6} & {71.9} \\
{}     & {4643.463} & {3.654} & {-1.29}  & {10.4} & {44.4} & {71.5} & {111.1} & {73.2} \\
{}     & {4647.433} & {2.949} & {-1.30}  & {43.4} & {} & {} & {142.6} & {} \\
{}     & {4678.845} & {3.602} & {-0.66}  & {57.7} & {} & {} & {125.6} & {} \\
{}     & {4691.411} & {2.990} & {-1.45}  & {21.7} & {61.6} & {99.8} & {153.8} & {98.9} \\
{}     & {4707.272} & {3.241} & {-1.08}  & {43.1} & {} & {} & {137.9} & {116.6} \\
{}     & {4741.529} & {2.832} & {-2.00}  & {14.2} & {43.2} & {} & {94.5} & {} \\
{}     & {4768.319} & {3.686} & {-1.109} & {} & {} & {88.4} & {115.1} & {69.6} \\
{}     & {4802.881} & {3.642} & {-1.165} & {13.0} & {20.6} & {41.2} & {73.3} & {39.1} \\
{}     & {4903.308} & {2.882} & {-1.07}  & {67.7} & {102.5} & {131.2} & {173.0} & {103.3} \\
{}     & {4920.502} & {2.832} & {0.06}   & {131.7} & {} & {241.5} & {323.9} & {236.2} \\
{}     & {4966.087} & {3.332} & {-0.89}  & {46.8} & {88.6} & {124.0} & {171.1} & {96.5} \\
{}     & {4973.101} & {3.960} & {-0.95}  & {17.1} & {46.9} & {66.3} & {110.9} & {62.5} \\
{}     & {5049.819} & {2.279} & {-1.43}  & {56.8} & {} & {138.0} & {175.8} & {112.8} \\
{}     & {5065.014} & {4.256} & {-0.134} & {58.2} & {96.4} & {} & {215.0} & {} \\
{}     & {5096.998} & {4.283} & {-0.277} & {42.6} & {} & {} & {149.5} & {} \\
{}     & {5121.641} & {4.283} & {-0.81}  & {10.5} & {} & {69.2} & {113.7} & {54.8} \\
{}     & {5133.681} & {4.178} & {0.14}   & {77.6} & {94.2} & {151.2} & {185.5} & {124.3} \\
{}     & {5159.050} & {4.283} & {-0.81}  & {11.6} & {} & {64.9} & {95.7} & {51.1} \\
{}     & {5242.491} & {3.634} & {-0.84}  & {28.6} & {71.4} & {99.9} & {107.5} & {87.3} \\
{}     & {5250.645} & {2.198} & {-2.05}  & {26.8} & {68.4} & {} & {130.3} & {99.5} \\
{}     & {5253.641} & {3.283} & {-1.68}  & {} & {} & {60.1} & {92.3} & {47.9} \\
{}     & {5281.790} & {3.038} & {-1.01}  & {54.5} & {} & {135.4} & {177.8} & {109.7} \\
{}     & {5302.299} & {3.283} & {-0.88}  & {50.8} & {85.1} & {117.8} & {172.2} & {107.4} \\
{}     & {5324.178} & {3.211} & {-0.24}  & {99.3} & {134.6} & {192.4} & {308.4} & {165.5} \\
{}     & {5341.023} & {1.608} & {-2.06}  & {} & {119.8} & {} & {202.2} & {} \\
{}     & {5353.373} & {4.103} & {-0.84}  & {20.8} & {58.0} & {79.5} & {126.9} & {67.1} \\
{}     & {5367.479} & {4.415} & {0.35}   & {74.3} & {115.5} & {143.5} & {176.5} & {116.9} \\
{}     & {5369.958} & {4.371} & {0.35}   & {86.6} & {} & {168.9} & {204.6} & {149.0} \\
{}     & {5371.489} & {0.958} & {-1.644} & {118.1} & {} & {211.3} & {305.5} & {177.1} \\
{}     & {5383.369} & {4.312} & {0.50}   & {91.8} & {133.4} & {160.9} & {213.4} & {135.2} \\
{}     & {5389.479} & {4.415} & {-0.40}  & {40.7} & {51.5} & {96.4} & {136.0} & {85.7} \\
{}     & {5393.167} & {3.241} & {-0.91}  & {57.4} & {90.2} & {107.5} & {184.1} & {102.2} \\
{}     & {5400.502} & {4.371} & {-0.15}  & {40.5} & {77.1} & {96.1} & {168.1} & {87.9} \\
{}     & {5405.774} & {0.990} & {-1.844} & {95.4} & {} & {197.4} & {264.2} & {151.9} \\
{}     & {5410.910} & {4.473} & {0.28}   & {70.8} & {150.5} & {} & {188.3} & {} \\
{}     & {5415.192} & {4.386} & {0.51}   & {84.3} & {128.7} & {156.3} & {209.3} & {134.0} \\
{}     & {5424.069} & {4.320} & {0.52}   & {} & {185.4} & {205.0} & {230.7} & {171.1} \\
{}     & {5434.523} & {1.011} & {-2.121} & {81.2} & {110.1} & {155.0} & {191.9} & {128.8} \\
{}     & {5445.042} & {4.386} & {-0.01}  & {53.1} & {} & {111.1} & {151.5} & {102.3} \\
{}     & {5466.390} & {4.371} & {-0.63}  & {21.2} & {46.6} & {67.4} & {112.6} & {67.0} \\
{}     & {5473.900} & {4.154} & {-0.76}  & {22.9} & {41.0} & {85.6} & {119.5} & {70.7} \\
{}     & {5565.704} & {4.607} & {-0.285} & {27.2} & {60.1} & {94.9} & {119.5} & {74.0} \\
{}     & {5569.618} & {3.417} & {-0.53}  & {61.2} & {89.2} & {134.2} & {179.9} & {117.1} \\
{}     & {5572.841} & {3.396} & {-0.31}  & {77.6} & {113.3} & {182.4} & {265.0} & {158.7} \\
{}     & {5576.090} & {3.430} & {-1.01}  & {38.4} & {64.6} & {98.8} & {145.7} & {94.0} \\
{}     & {5581.965} & {2.523} & {-0.710} & {} & {116.2} & {} & {} & {} \\
{}     & {5586.756} & {3.368} & {-0.21}  & {88.5} & {} & {188.8} & {278.5} & {152.5} \\
{}     & {5633.975} & {4.991} & {-0.27}  & {15.5} & {43.0} & {55.9} & {87.4} & {52.2} \\
{}     & {5705.981} & {4.607} & {-0.53}  & {23.6} & {45.4} & {57.5} & {109.6} & {68.4} \\
{}     & {5762.990} & {4.209} & {-0.46}  & {51.1} & {106.5} & {115.5} & {135.8} & {107.4} \\
{}     & {5816.367} & {4.548} & {-0.69}  & {21.9} & {28.8} & {78.5} & {102.5} & {63.1} \\
{}     & {5859.578} & {4.548} & {-0.398} & {33.3} & {} & {76.5} & {106.6} & {56.4} \\
{}     & {6024.049} & {4.548} & {-0.11}  & {32.5} & {63.0} & {95.2} & {140.5} & {88.3} \\
{}     & {6055.092} & {4.733} & {-0.46}  & {25.1} & {43.4} & {76.3} & {97.7} & {64.8} \\
{}     & {6065.482} & {2.608} & {-1.53}  & {60.6} & {71.6} & {98.8} & {159.4} & {102.7} \\
{}     & {6191.558} & {2.433} & {-1.60}  & {68.1} & {86.2} & {137.8} & {185.1} & {120.7} \\
{}     & {6213.429} & {2.223} & {-2.65}  & {7.1} & {36.7} & {56.0} & {111.3} & {47.0} \\
{}     & {6219.279} & {2.198} & {-2.434} & {15.6} & {51.6} & {81.8} & {122.7} & {66.9} \\
{}     & {6230.726} & {2.559} & {-1.281} & {63.2} & {99.2} & {149.0} & {179.1} & {110.2} \\
{}     & {6232.639} & {3.654} & {-1.271} & {16.2} & {} & {76.9} & {110.4} & {54.3} \\
{}     & {6252.554} & {2.404} & {-1.687} & {28.0} & {82.2} & {100.7} & {156.4} & {82.7} \\
{}     & {6265.131} & {2.176} & {-2.550} & {15.3} & {47.2} & {66.3} & {122.1} & {50.8} \\
{}     & {6335.328} & {2.198} & {-2.23}  & {} & {} & {82.0} & {144.1} & {78.8} \\
{}     & {6336.823} & {3.686} & {-1.05}  & {} & {} & {90.1} & {148.6} & {75.4} \\
{}     & {6393.602} & {2.433} & {-1.61}  & {52.8} & {90.1} & {134.8} & {171.7} & {100.0} \\
{}     & {6400.000} & {3.602} & {-0.52}  & {54.0} & {95.5} & {147.0} & {236.1} & {125.0} \\
{}     & {6411.647} & {3.654} & {-0.82}  & {42.3} & {72.5} & {105.1} & {171.2} & {101.0} \\
{}     & {6419.942} & {4.733} & {-0.25}  & {35.0} & {} & {91.8} & {127.5} & {69.4} \\
{}     & {6421.349} & {2.279} & {-2.027} & {39.8} & {} & {102.6} & {158.8} & {87.5} \\
{}     & {6430.844} & {2.176} & {-2.005} & {36.6} & {77.8} & {101.9} & {183.0} & {88.9} \\
{}     & {6633.746} & {4.558} & {-0.78}  & {17.6} & {} & {} & {100.5} & {71.4} \\
{}     & {6663.437} & {2.424} & {-2.478} & {22.7} & {54.1} & {83.7} & {148.8} & {74.2} \\
{}     & {6677.989} & {2.692} & {-1.47}  & {60.6} & {88.1} & {115.9} & {163.9} & {113.7} \\
{Fe II} & {4416.830} & {2.778} & {-2.61} & {} & {} & {} & {127.9} & {} \\
{}     & {4472.929} & {2.844} & {-3.43}  & {25.0} & {65.6} & {92.9} & {} & {95.9} \\
{}     & {4508.288} & {2.855} & {-2.21}  & {103.4} & {151.5} & {125.0} & {142.5} & {128.0} \\
{}     & {4515.339} & {2.844} & {-2.48}  & {88.6} & {125.6} & {119.8} & {139.0} & {110.0} \\
{}     & {4520.224} & {2.807} & {-2.61}  & {84.2} & {100.5} & {114.3} & {} & {114.5} \\
{}     & {4522.634} & {2.844} & {-2.03}  & {118.4} & {133.4} & {} & {} & {191.2} \\
{}     & {4541.524} & {2.855} & {-3.05}  & {56.4} & {121.5} & {} & {110.0} & {} \\
{}     & {4576.340} & {2.844} & {-3.04}  & {57.2} & {84.7} & {81.9} & {90.0} & {78.9} \\
{}     & {4620.521} & {2.828} & {-3.29}  & {30.5} & {73.4} & {65.2} & {82.3} & {61.5} \\
{}     & {4629.339} & {2.807} & {-2.38}  & {98.0} & {153.7} & {124.6} & {136.0} & {122.4} \\
{}     & {4635.316} & {5.956} & {-1.65}  & {12.1} & {} & {44.2} & {} & {} \\
{}     & {4731.453} & {2.891} & {-3.37}  & {46.3} & {77.0} & {} & {90.8} & {} \\
{}     & {4923.922} & {2.891} & {-1.21}  & {178.1} & {247.2} & {292.6} & {257.7} & {212.4} \\
{}     & {5197.577} & {3.230} & {-2.10}  & {87.6} & {125.2} & {157.4} & {121.2} & {117.3} \\
{}     & {5362.869} & {3.199} & {-2.739} & {70.3} & {91.7} & {102.1} & {140.6} & {107.0} \\
{}     & {5425.257} & {3.199} & {-3.36}  & {} & {} & {54.5} & {61.5} & {} \\
{}     & {5427.826} & {6.724} & {-1.664} & {} & {} & {} & {} & {} \\
{}     & {6084.111} & {3.199} & {-3.97}  & {} & {23.7} & {} & {41.4} & {17.7} \\
{}     & {6113.322} & {3.221} & {-4.30}  & {} & {} & {13.4} & {} & {14.9} \\
{}     & {6147.741} & {3.889} & {-2.721} & {} & {} & {62.5} & {} & {55.1} \\
{}     & {6149.258} & {3.889} & {-2.90}  & {43.1} & {} & {46.0} & {62.9} & {49.8} \\
{}     & {6369.462} & {2.891} & {-4.36}  & {10.5} & {} & {14.0} & {46.9} & {26.0} \\
{}     & {6416.919} & {3.892} & {-2.85}  & {22.8} & {46.3} & {48.4} & {75.3} & {52.7} \\
{}     & {6432.680} & {2.891} & {-3.74}  & {21.8} & {46.9} & {52.2} & {94.9} & {50.1} \\
{Ni I} & {4714.408} & {3.380} & {0.23}   & {55.1} & {96.4} & {158.7} & {175.3} & {133.4} \\
{}     & {4752.415} & {3.658} & {-0.69}  & {} & {} & {71.4} & {80.0} & {65.1} \\
{}     & {4829.016} & {3.542} & {-0.33}  & {25.7} & {41.0} & {110.5} & {108.4} & {89.5} \\
{}     & {4831.169} & {3.606} & {-0.41}  & {19.2} & {39.7} & {84.3} & {102.8} & {62.7} \\
{}     & {4904.407} & {3.542} & {-0.17}  & {} & {} & {} & {100.7} & {60.7} \\
{}     & {4980.166} & {3.606} & {-0.11}  & {23.0} & {} & {101.4} & {126.0} & {80.4} \\
{}     & {5115.389} & {3.834} & {-0.11}  & {22.4} & {39.0} & {71.8} & {93.1} & {69.2} \\
{}     & {6643.629} & {1.676} & {-2.30}  & {} & {} & {80.3} & {116.5} & {45.2} \\
{}     & {6767.768} & {1.826} & {-2.17}  & {11.6} & {} & {69.9} & {106.6} & {76.7} \\
{Zn I} & {4680.134} & {4.006} & {-0.815} & {} & {} & {64.2} & {} & {60.8} \\
{}     & {4722.153} & {4.030} & {-0.338} & {14.5} & {51.2} & {59.4} & {93.9} & {58.0} \\
{}     & {4810.528} & {4.078} & {-0.140} & {18.3} & {49.4} & {73.3} & {98.1} & {71.2} \\
{Y II} & {4883.684} & {1.084} & {0.07}   & {31.0} & {} & {78.2} & {89.8} & {76.7} \\
{}     & {4900.120} & {1.033} & {-0.09}  & {35.7} & {80.2} & {} & {} & {} \\
{}     & {5087.416} & {1.084} & {-0.17}  & {20.6} & {57.5} & {68.7} & {68.9} & {65.5} \\
{}     & {5200.406} & {0.992} & {-0.57}  & {14.2} & {} & {49.3} & {53.0} & {44.3} \\
\enddata
\tablenotetext{a}{EW of blended line reported}
\tablenotetext{b}{Oxygen triplet present as a single blended feature}
\end{deluxetable}

\section{Validation of analysis procedure}
\label{sec:test}
\restartappendixnumbering

\textcolor{black}{In order to validate our analysis procedure, we test our methods on the star HIP 25278 ($v\sin{i}\sim$15 $\mathrm{kms^{-1}}$). This target has previously measured abundances of C, O, and S from the Hypatia catalog (\citealt{2014AJ....148...54H}). The spectra for HIP 25278 were obtained from archival APF data originally taken to identify potential targets for the Gemini Planet Imager Exoplanet Survey (GPIES). After data reduction and blaze subtraction, we performed the first set of forward modeling runs to determine the stellar atmospheric parameters. The priors are given in table \ref{tab:hip25278atmos}.}

\begin{deluxetable}{cccccccccc}
\tablecaption{Forward modeling prior ranges for stellar parameter determination for HIP 25278 \label{tab:hip25278atmos}}
\tablewidth{0pt}
\tablehead{\colhead{Target}    & \colhead{$T_\mathrm{eff}$ (K)}    & \colhead{$\log{g}$ (cgs)}      & \colhead{$\mathrm{[M/H]}$}     & \colhead{$v\sin{i}$ (km\,s$^{-1}$)}    & \colhead{$RV$ (km\,s$^{-1}$)} & \colhead{$\alpha$} & \colhead{$pwv$} & \colhead{LSF} & \colhead{Noise factor (N)} }
\startdata
HIP 25278 & (6000, 7000) & (3.5, 5.0) & (-1.0,1.0) & {(0,100)} & {(-100,100)} & {(0.0, 2.0)} & {(0.5, 5.0)} & {(0.1, 3.0)} & {(0, 7.5e6)} \\
\enddata
\end{deluxetable}

\begin{figure}[H]
    \centering
    \includegraphics[width=1.0\linewidth]{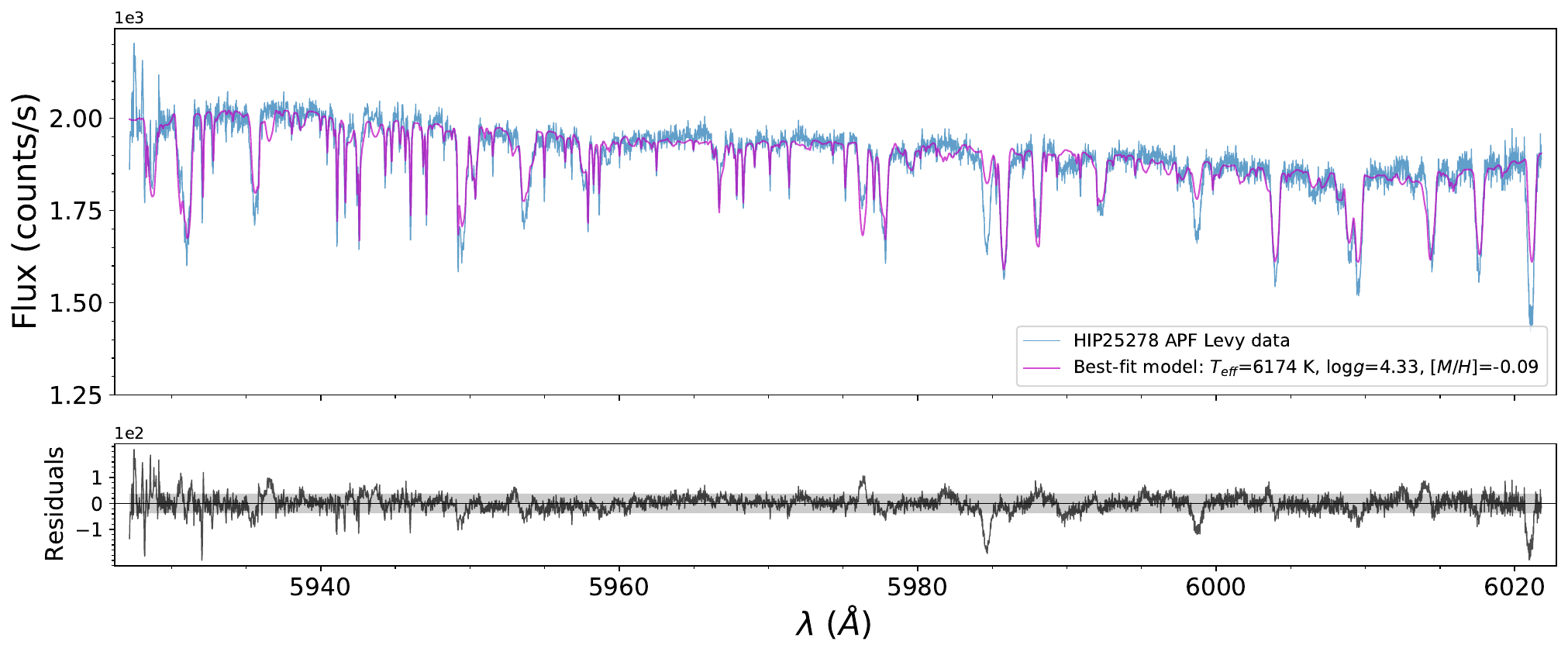}
    \caption{Best-fit PHOENIX model to the APF spectrum for the target HIP 25278 (cyan), shown for APF order \#78. The parameters of the best-fit PHOENIX model are estimated by computing the median for the effective temperature ($T_\mathrm{eff}$), surface gravity ($\log{g}$), and the metallicity ($\mathrm{[M/H]}$) over seven retained runs after performing multiple multi-order MCMC runs. This model has $T_\mathrm{eff}$ = 6174 K, $\log{g}$ = 4.33, $\mathrm{[M/H]}$ = -0.09 (magenta). The residuals between the data and the model are plotted in black and other noise limits are shown in grey}
    \label{fig:hip25278atmos}
\end{figure}

\begin{figure}[H]
    \centering
    \includegraphics[width=1.0\linewidth]{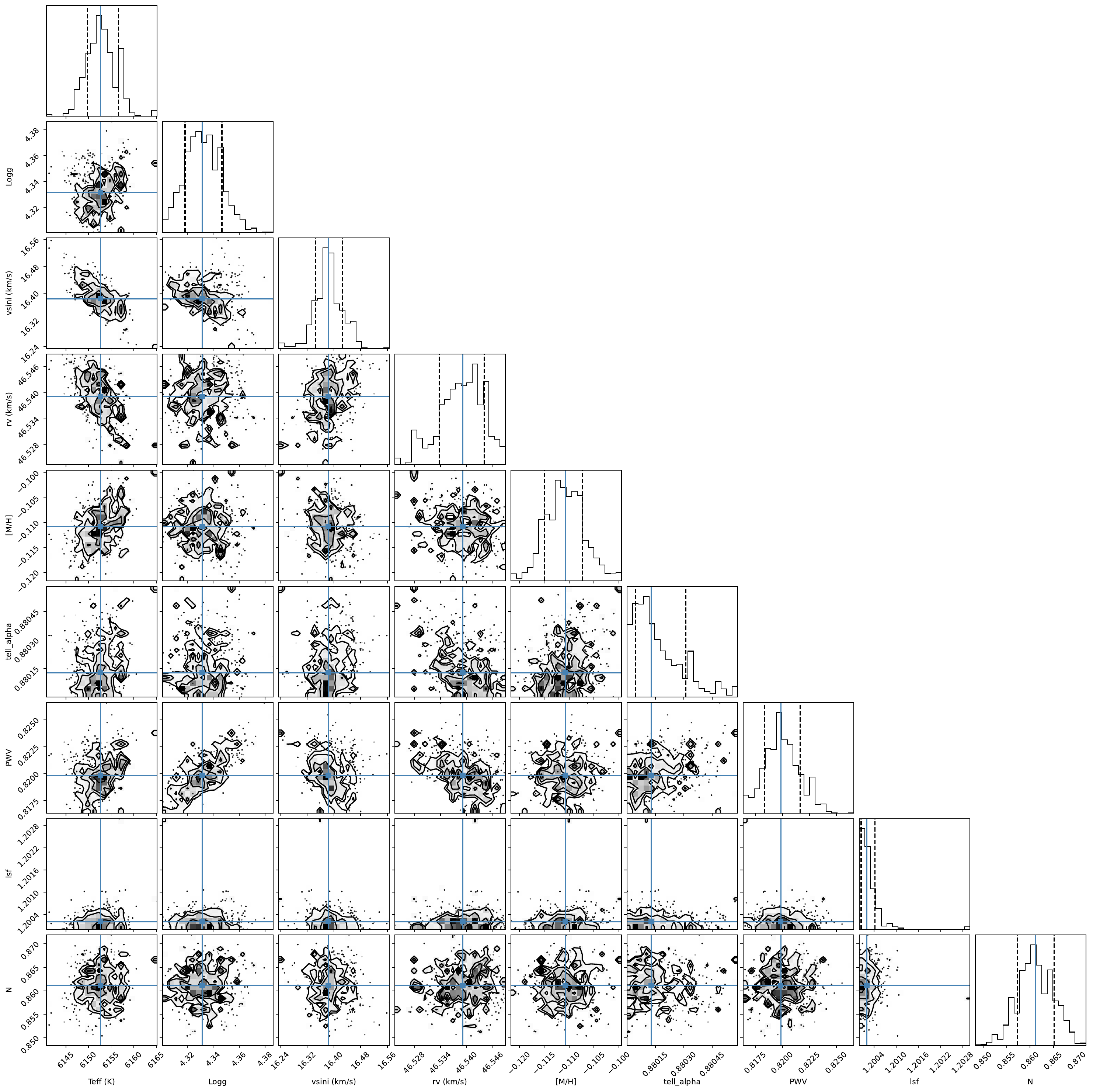}
    \caption{Corner plot for one of the retained runs for a multi-order fit of the PHOENIX grid to the spectrum of HD 206893. The marginalized posteriors are shown along the diagonal. The blue lines represent the 50 percentile, and the dotted lines represent the 16 and 84 percentiles. The subsequent covariances between all the parameters are in the corresponding 2-D histograms. This run gives a best-fit $T_\mathrm{eff}$ = 6153 K, $\log{g}$ = 4.33, $\mathrm{[M/H]}$ = -0.11.}
    \label{fig:hip25278atmoscorner}
\end{figure}

\textcolor{black}{Determining the atmospheric parameters for this target involved 20 two-part MCMC runs, each with 100 walkers and 2000 steps, with the first 1500 steps discarded as burn-in. Of these, five runs were discarded due to the values of $T_\mathrm{eff}$ and/or $\log{g}$ hitting the edge of the grid. Two runs were discarded due to poor fits to some spectral orders. Four runs were discarded due to the walkers attempting to converge a new minima toward the end of 2000 steps. For the retained nine runs, we compute the median and standard deviation for our fitted parameters and obtain $T_\mathrm{eff}$ = 6173 $\pm$ 173 K, $\log{g}$ = 4.33 $\pm$ 0.22, $\mathrm{[M/H]}$ = -0.09 $\pm$ 0.10. Figure \ref{fig:hip25278atmos} and figure \ref{fig:hip25278atmoscorner} show the best-fit stellar atmospheric model and an example corner plot for one of the atmospheric runs, respectively. Our atmospheric parameters agree with those reported in the Hypatia catalog (\cite{2014AJ....148...54H}; Table \ref{tab:hip25278lit}).} \par

\begin{figure}
    \centering
    \includegraphics[width=1.0\linewidth]{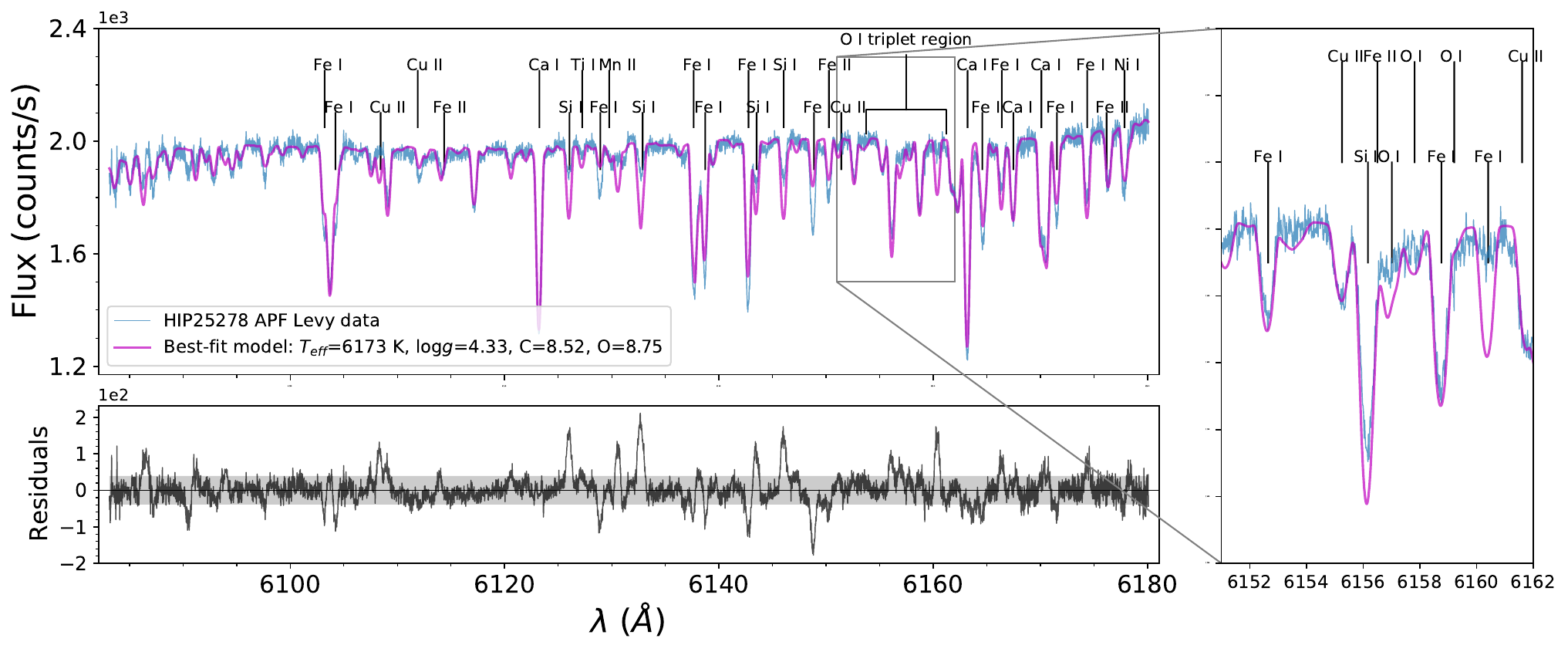}
    \caption{Best-fit \textit{PHOENIX}$\--$\textit{C/O} model to the APF spectrum for the target HIP 25278 (cyan), shown for order \#76 containing an O I triplet feature at 6155$\--$6158 $\mathrm{\AA}$. The zoomed-in subplot on the right shows a $\sim$10$\mathrm{\AA}$ region around the blended feature. The best-fit \textit{PHOENIX}$\--$\textit{C/O} model has $T_\mathrm{eff}$ = 6173 K, $\log{g}$ = 4.33, $\log{\epsilon_C}$ = 8.52, $\log{\epsilon_O}$ = 8.75 (magenta). The residuals between the data and the model are plotted in black and other noise limits are shown in grey. While the model disagrees with our spectra, especially around 6130, 6145, and 6160 $\mathrm{\AA}$, this disagreement is due to the model Fe I lines not agreeing with our spectra. This is expected as we do not use the \textit{PHOENIX}$\--$\textit{C/O} grid with the target metallicity ($[M/H]$ = -0.09) to fit for the C and O abundances.}
    \label{fig:hip25278spec76}
\end{figure}

\begin{figure}
    \centering
    \includegraphics[width=1.0\linewidth]{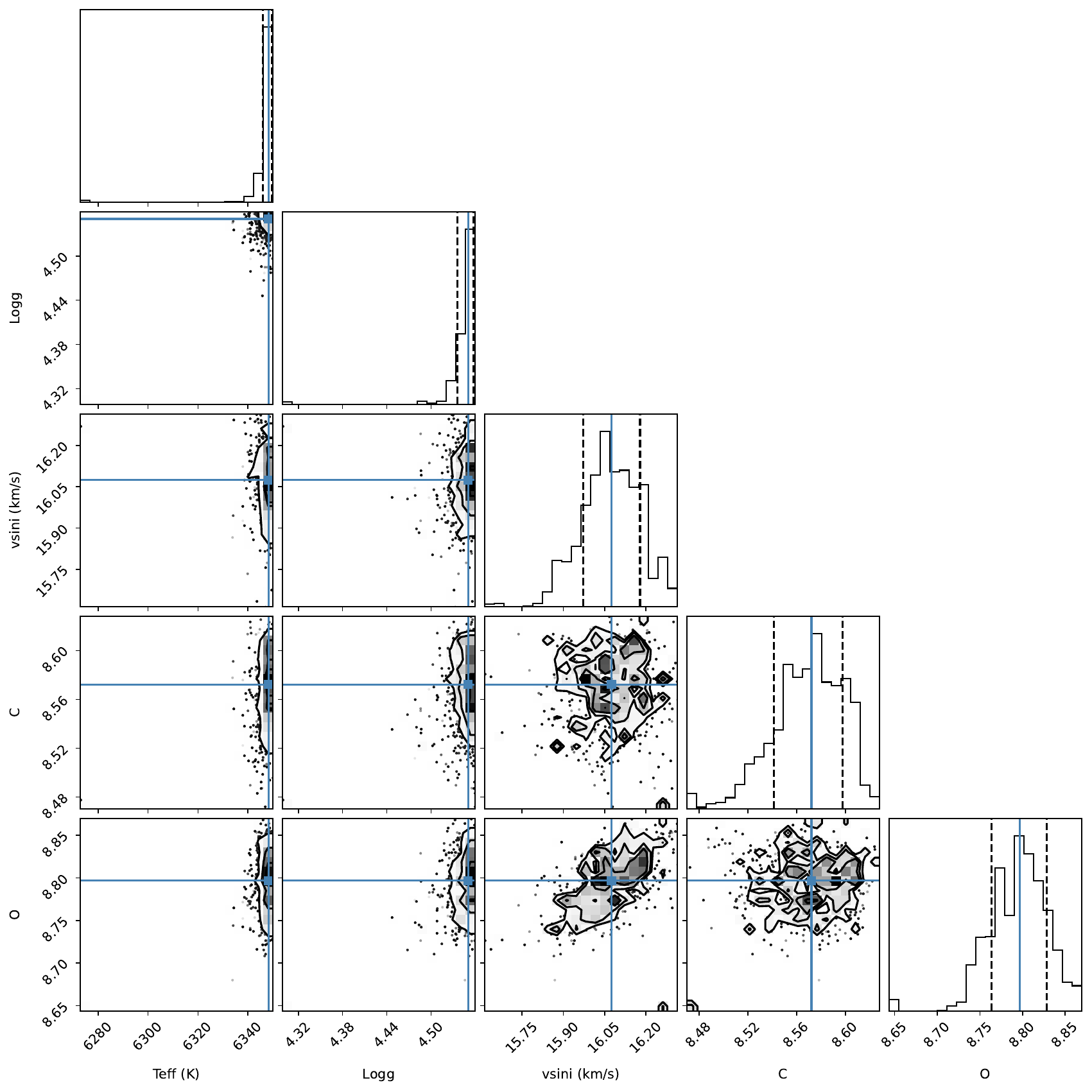}
    \caption{Corner plot for one of the retained runs for \textit{PHOENIX}$\--$\textit{C/O} grid fit to the spectrum of HIP 25278. The marginalized posteriors are shown along the diagonal. The blue lines represent the 50 percentile, and the dotted lines represent the 16 and 84 percentiles. The subsequent covariances between all the parameters are in the corresponding 2-D histograms. This run gives a best-fit $\log{\epsilon_C}$ = 8.57, $\log{\epsilon_O}$ = 8.80.}
    \label{fig:hip25278cocorner}
\end{figure}

 \textcolor{black}{We use the \textit{PHOENIX}$\--$\textit{C/O} grid to fit for the carbon and oxygen lines. Estimation of the carbon and oxygen abundance involved splicing a $\sim$20$\mathrm{\AA}$ region around the carbon and oxygen spectral lines and fitting all of the carbon and oxygen orders simultaneously. A smaller region was chosen as we do not have the \textit{PHOENIX}$\--$\textit{C/O} grid with the overall metallicity of the target. Twelve multi-order runs were performed with three runs discarded due to poor convergence of the walkers for the oxygen abundance and two runs discarded due to the walkers hitting the edge of the grid for carbon/or oxygen. Computing the median and standard deviation of the fitted parameters over the remaining seven runs, we obtain [C/H] = 0.09 $\pm$ 0.06 and [O/H] = 0.06 $\pm$ 0.09, giving a spectral fit C/O = 0.59 $\pm$ 0.15. The carbon abundance agrees with the Hypatia catalog at the 1.5$\sigma$ level and our oxygen abundance is in good agreement with their values (Table \ref{tab:hip25278lit}). The C/O = 0.44 $\pm$ 0.06 from the Hypatia catalog also agrees with our values. Figures \ref{fig:hip25278spec76} and \ref{fig:hip25278cocorner} show the best-fit model to our spectral data for the order with the O I triplet feature at 6155$\--$6158 $\mathrm{\AA}$ and an example corner plot for one of the C/O runs, respectively.} \par 

\textcolor{black}{We computed the carbon, oxygen, and sulfur abundance for this target using the equivalent width method. This gave us [C/H] = 0.09 $\pm$ 0.17, [O/H] = 0.04 $\pm$ 0.17, and [S/H] = 0.11 $\pm$ 0.09. All three abundances are in excellent agreement with those from the Hypatia Catalog (refer Table \ref{tab:hip25278lit}). The equivalent width C/O = 0.49 $\pm$ 0.27 agrees with the spectral fit value as well as the Hypatia catalog (C/O = 0.44 $\pm$ 0.06).} 

\textcolor{black}{Our analysis of the star HIP 25278 with previously measured abundances in the Hypatia Catalog (\citealt{2014AJ....148...54H}) using the same procedure used for the science targets in this paper gives similar abundance values for C, O, and S to those given in the catalog. Only our spectral fit carbon abundance is in slight disagreement with the catalog value (agrees only at a 1.5$\sigma$ level) due to the use of solar metallicity \textit{PHOENIX}$\--$\textit{C/O} grid instead of a grid with the exact target metallicity. This indicates that we can trust the results of the measurement procedures outlined in this paper as applied to our science targets.}

\begin{deluxetable}{c|c|cccccc}
\tablecaption{Comparison of HIP 25278 stellar parameters and abundances with literature \label{tab:hip25278lit}}
\tablewidth{0pt}
\tablehead{\colhead{Target}    & \colhead{Work}     & \colhead{$T_\mathrm{eff}$ (K)}    & \colhead{$\log{g}$ (cgs)}     & \colhead{[M/H]}     & \colhead{[C/H]} & \colhead{[O/H]} & \colhead{[S/H]}
}
\startdata
\multirow{3}{*}{HIP 25278} & \multirow{2}{*}{This work} & \multirow{2}{*}{6173 $\pm$ 173} & \multirow{2}{*}{4.33 $\pm$ 0.22} & \multirow{2}{*}{-0.09 $\pm$ 0.10} & {0.09 $\pm$ 0.06} \tablenotemark{a} & {0.06 $\pm$ 0.09} \tablenotemark{a} &\\ 
  & & & & &  0.02 $\pm$ 0.17 \tablenotemark{b} & 0.07 $\pm$ 0.17 \tablenotemark{b} & 0.11 $\pm$ 0.09 \tablenotemark{b}\\
  & \cite{2014AJ....148...54H} & {6184} & {4.38}  & {0.05 $\pm$ 0.18}  & {-0.02 $\pm$ 0.04} & {0.08 $\pm$ 0.04} & {0.11 $\pm$ 0.09}  \\ \hline
\enddata
\tablenotetext{a}{Abundances obtained using spectral fitting}
\tablenotetext{b}{Abundances obtained using equivalent width}
\end{deluxetable}

\end{document}